\documentclass[12pt,twoside, a4paper]{article}
\def\pd{\partial}
\def\mc{\mathcal}

\usepackage[dvips]{graphicx}
\usepackage{amssymb}
\usepackage{amssymb,amsmath}
\usepackage{graphicx}
\usepackage{dsfont}
\usepackage{caption}
\usepackage{subcaption}
\usepackage{mathtools}
\usepackage{verbatim}
\usepackage{graphicx}
\usepackage{multirow}
\usepackage[outline]{contour}
\usepackage{xcolor,colortbl}
\input{epsf.sty}  
\begin{document}
\begin{center}
\Large{\textbf{Superconformal interfaces from 5D $N=4$ gauged supergravity}}
\end{center}
\vspace{1 cm}
\begin{center}
\large{\textbf{Parinya Karndumri}}
\end{center}
\begin{center}
String Theory and Supergravity Group, Department
of Physics, Faculty of Science, Chulalongkorn University, 254 Phayathai Road, Pathumwan, Bangkok 10330, Thailand \\
E-mail: parinya.ka@hotmail.com \vspace{1 cm}
\end{center}
\begin{abstract}
We find a large class of supersymmetric Janus solutions from five-dimensional $N=4$ gauged supergravity coupled to five vector multiplets with $SO(2)_D\times SO(3)\times SO(3)$ gauge group. The gauged supergravity admits four supersymmetric $AdS_5$ vacua, two $N=4$ with $SO(2)_D\times SO(3)\times SO(3)$ and $SO(2)_D\times SO(3)_{\textrm{diag}}$ symmetric $AdS_5$ vacua and two $N=2$ with $SO(2)_{\textrm{diag}}\times SO(3)$ and $SO(2)_{\textrm{diag}}$ symmetric ones. In a truncation to three vector multiplets, the gauge group is given by $SO(2)\times SO(3)\times SO(3)$, and the resulting gauged supergravity admits only two $N=4$ supersymmetric $AdS_5$ vacua with $SO(2)\times SO(3)\times SO(3)$ and $SO(2)\times SO(3)_{\textrm{diag}}$ residual symmetries. By considering the $SO(2)_{\textrm{diag}}$ invariant sector within this truncation, we find a number of supersymetric Janus interfaces between the two $N=4$ vacua on both sides as well as an RG-flow interface between $SO(2)\times SO(3)\times SO(3)$ and $SO(2)\times SO(3)_{\textrm{diag}}$ symmetric vacua on each side. By repeating the analysis in the full $SO(2)_D\times SO(3)\times SO(3)$ gauged supergravity, we find Janus solutions interpolating between the aforementioned four supersymmetric $AdS_5$ vacua as well as examples of multi-Janus interfaces between these vacua.  
\end{abstract}
\newpage
%%%%%%%%%%%%%%%%%%%%%%%%%%%%%%%%%%%%%%%%%%%%%%%%%%%%%%%%%%%%%%%%%%%%%%%%%%%%%%%%%%%%%%%%%%%%%%%%%%%%%%%%%%%%%%%%%%%%%%%%%%%%%%%%%%%%%%%%%
\section{Introduction}
Conformal defects play an important role in the study of quantum field theories, string theory and condensed matter physics, see \cite{defect_review} for a comprehensive review. Since the original proposal of the AdS/CFT correspondence \cite{maldacena,Gubser_AdS_CFT,Witten_AdS_CFT}, various aspects of conformal defects within strongly-coupled superconformal field theories (SCFTs) can be explored using holographic duals in gauged supergravity theories. Of particular interest is a class of defects with co-dimension one called conformal interfaces. In the gravity dual, $(d-1)$-dimensional interfaces within $d$-dimensional SCFTs are described by Janus solutions in $(d+1)$-dimensional gauged supergravity and take the form of $AdS_d$-sliced domain walls with two asymptotic $AdS_{d+1}$ geometries. Far from these asymptotic geometries, the solutions preserve a $(d-1)$-dimensional conformal symmetry which is a subgroup of the conformal symmetry of the bulk $d$-dimensional SCFT.  
\\
\indent In this paper, we are interested in supersymmetric Janus solutions describing conformal interfaces within four-dimensional SCFTs. The first example for this type of solutions has been found in \cite{Bak_Janus} which breaks all supersymmetry, see also \cite{5D_Janus_CK} for a supersymmetric solution. These solutions describe conformal interfaces within $N=4$ SYM dual to $AdS_5\times S^5$ geometry in type IIB theory with different values of the gauge coupling constant on the two sides of the interfaces  \cite{Freedman_Janus,DHoker_Janus,Witten_Janus,Freedman_Holographic_dCFT}. A supersymmetric Janus solution can also be obtained from $N=8$ maximal gauged supergravity in five dimensions with $SO(6)$ gauge group \cite{5D_Janus_Suh}. More general Janus solutions have also appeared both in type IIB theory \cite{5D_Janus_DHoker1,5D_Janus_DHoker2,5D_Janus_DHoker3} and in maximal $SO(6)$ gauged supergravity in five dimensions uplifted to ten dimensions \cite{Bobev_5D_Janus1,Bobev_5D_Janus2,RG-interface_Gauntlett}, see \cite{6D_Janus}-\cite{3D_Janus4} for examples of Janus solutions in other dimensions.   
\\
\indent In the present work, we will look for new Janus solutions in half-maximal $N=4$ gauged supergravity in five dimensions. Janus solutions within this gauged supergravity are less known compared to the maximal case. As pointed out in \cite{5D_N4_curved_DW}, this type of solutions does not exist in pure $N=4$ gauged supergravity with $SU(2)\times U(1)$ gauge group \cite{Romans_5DN4}. Therefore, in order to obtain Janus solutions from $N=4$ gauged supergravity, we need to consider matter-coupled $N=4$ supergravity. The first examples of these solutions have appeared recently in \cite{5D_N4_Janus}, see also \cite{Janus_5D_ISO3} for Janus solutions interpolating between $N=2$ non-conformal phases. The solutions describe superconformal interfaces in $N=2$ and $N=1$ SCFTs in four dimensions and can be embedded in M-theory. The $N=4$ gauged supergravity considered in \cite{Janus_5D_ISO3} with $SO(2)\times ISO(3)$ gauge group admits only one supersymmetric $AdS_5$ vacuum, and it turns out that Janus solutions interpolating between $AdS_5$ vacua on both sides of the interfaces do not exist. On the other hand, there are two supersymmetric $AdS_5$ vacua with $N=2$ and $N=4$ supersymmetries in the $N=4$ gauged supergravity with $SO(2)_D\times SO(3)$ gauge group studied in \cite{5D_N4_Janus}. In this case, there are both Janus solutions interpolating between the same $AdS_5$ vacuum, dual to conformal interfaces, and solutions interpolating between different $AdS_5$ vacua, dual to RG-flow interfaces between two different SCFTs.     
\\
\indent As anticipated in \cite{RG-interface_Gauntlett}, it is interesting to consider the situation in which there are more than two conformal fixed points connected by holographic RG flows. In this case, there can exist RG flows from a UV SCFT to two SCFTs in the IR, and it could be of particular interest to study Janus solutions in this framework. In order to achieve this, the gauged supergravity must admit more than two $AdS_5$ vacua. We will then consider $N=4$ gauged supergravity with $SO(2)_D\times SO(3)\times SO(3)$ gauge group first considered in \cite{5D_flowII}. This can be obtained from $N=4$ gauged supergravity coupled to five vector multiplets with $\mathbb{R}^+\times SO(5,5)$ global symmetry. The $SO(2)_D$ is a diagonal subgroup of the $SO(2)$ subgroup of the $SO(5)_R\sim USp(4)_R$ R-symmetry and an $SO(2)\subset SO(5)$ from the symmetry of the matter multiplets. 
\\
\indent There are four supersymmetric $AdS_5$ vacua, two $N=4$ vacua with $SO(2)_D\times SO(3)\times SO(3)$ and $SO(2)_D\times SO(3)_{\textrm{diag}}$ symmetries together with two $N=2$ vacua with $SO(2)_{\textrm{diag}}\times SO(3)$ and $SO(2)_{\textrm{diag}}$ symmetries. It has also been shown in \cite{5D_flowII} that there exists an RG flow solution interpolating between the two $N=4$ vacua and the $N=2$ vacuum with $SO(2)_{\textrm{diag}}$ symmetries. Accordingly, we will look for Janus solutions in this $N=4$ gauged supergravity as well as a truncation to $N=4$ gauged supergravity coupled to three vector multiplets with $SO(2)\times SO(3)\times SO(3)$ gauge group considered in \cite{5D_N4_flow_Davide} and \cite{5D_N4_flow}. The latter only admits two $N=4$ $AdS_5$ vacua with $SO(2)\times SO(3)\times SO(3)$ and $SO(2)\times SO(3)_{\textrm{diag}}$ symmetries. It should be pointed out that both of the $N=4$ gauged supergravities considered in this paper have no known higher-dimensional origin to date. However, the results found here could hopefully still lead to some insights into various aspects of conformal interfaces within four-dimensional $N=1$ and $N=2$ SCFTs dual respectively to $N=2$ and $N=4$ $AdS_5$ vacua.  
\\
\indent The paper is organized as follows. In section \ref{N4_SUGRA},
we review five-dimensional $N=4$ gauged supergravity coupled to five vector multiplets and $SO(2)_D\times SO(3)\times SO(3)$ gauge group. Relevant BPS equations for obtaining supersymmetric Janus solutions are derived from the analysis of supersymmetry conditions in section \ref{Janus_BPS}. In section \ref{Janus_solutions}, the four supersymmetric $AdS_5$ vacua are reviewed, and a number of Janus solutions in $N=4$ gauged supergravities with both $SO(2)\times SO(3)\times SO(3)$ and $SO(2)_D\times SO(3)\times SO(3)$ gauge groups is given. We end the paper by giving some conclusions and comments in section \ref{conclusion}. 
%%%%%%%%%%%%%%%%%%%%%%%%%%%%%%%%%%%%%%%%%%%%%%%%%%%%%%%%%%%%%%%%%%%%%%%%%%%%%%%%%%%%%%%%%%%%%%%%%%%%%%%%%%%%%%%%%%%%%%%%%%%%%%%%%%%%%%%%%
\section{Five-dimensional $N=4$ gauged supergravity with $SO(2)_D\times SO(3)\times SO(3)$ gauge group}\label{N4_SUGRA}
In this section, we briefly review five-dimensional $N=4$ gauged supergravity coupled to vector multiplets. We will mainly focus on finding supersymmetric solutions with only the metric and scalar fields non-vanishing and refer to the original construction in \cite{N4_gauged_SUGRA} and \cite{5D_N4_Dallagata} for more details. At the end of this section, we will consider a particular case of $SO(2)_D\times SO(3)\times SO(3)$ gauge group obtained from $N=4$ supergravity coupled to five vector multiplets. 

\subsection{$N=4$ gauged supergravity in five dimensions}
The $N=4$ supergravity multiplet consists of the graviton $e^{\hat{\mu}}_\mu$, four gravitini $\psi_{\mu i}$, six vectors $(A_\mu^0,A_\mu^m)$, four spin-$\frac{1}{2}$ fields $\chi_i$ and one real scalar $\Sigma$, the dilaton. The only matter multiplet in this case is given by vector multiplets. Each of these multiplets consists of a vector field $A_\mu$, four gaugini $\lambda_i$ and five scalars $\phi^m$. Throughout the paper, space-time and tangent space indices are denoted respectively by $\mu,\nu,\ldots =0,1,2,3,4$ and
$\hat{\mu},\hat{\nu},\ldots=0,1,2,3,4$. The fundamental representation of $SO(5)_R\sim USp(4)_R$
R-symmetry is described respectively by $m,n=1,\ldots, 5$ for $SO(5)_R$ and $i,j=1,2,3,4$ for $USp(4)_R$. The vector multiplets are labeled by indices $a,b=1,\ldots, n$ with $n$ being the number of vector multiplets. 
\\
\indent Since we are only interested in supersymmetric Janus solutions involving only the metric and scalars as previously mentioned, we will set all the remaining fields to zero. The dilaton and $5n$ scalar fields from the vector multiplets parametrize the scalar manifold $\mathbb{R}^+\times SO(5,n)/SO(5)\times SO(n)$ with the first factor corresponding to the dilaton. The $SO(5,n)/SO(5)\times SO(n)$ factor can be described by a coset representative $\mc{V}_M^{\phantom{M}A}$ transforming under the global $G=SO(5,n)$ and the local $H=SO(5)\times SO(n)$ by left and right multiplications, respectively. The global $SO(5,n)$ and local $SO(5)\times SO(n)$ indices are denoted by $M,N,\ldots=1,2,\ldots, 5+n$ and $A,B,\ldots=1,2,\ldots, 5+n$, respectively. The latter can be split into $A=(m,a)$ resulting in 
\begin{equation}
\mc{V}_M^{\phantom{M}A}=(\mc{V}_M^{\phantom{M}m},\mc{V}_M^{\phantom{M}a}).
\end{equation}
\indent Gaugings of $N=4$ supergravity are characterized by the embedding tensor consisting of three components $\xi^{M}$, $\xi^{MN}=\xi^{[MN]}$ and $f_{MNP}=f_{[MNP]}$. 
We are interested in gaugings with $\xi^M=0$ since these are the only gaugings that admit supersymmetric $AdS_5$ vacua \cite{AdS5_N4_Jan}. With $\xi^{M}=0$, the resulting gauge groups are entirely embedded in $SO(5,n)$. In order to define consistent gaugings, the embedding tensor needs to satisfy a set of quadratic constraints of the form
\begin{equation}
f_{R[MN}{f_{PQ]}}^R=0\qquad \textrm{and}\qquad {\xi_M}^Qf_{QNP}=0\, .\label{QC}
\end{equation}
In particular, the gauge generators in the fundamental representation of $SO(5,n)$ can be written as
\begin{equation}
{(X_M)_N}^P=-{f_M}^{QR}{(t_{QR})_N}^P={f_{MN}}^P\quad \textrm{and}\quad {(X_0)_N}^P=-\xi^{QR}{(t_{QR})_N}^P={\xi_N}^P
\end{equation}
with ${(t_{MN})_P}^Q=\delta^Q_{[M}\eta_{N]P}$ being $SO(5,n)$ generators. We note that the definition of $\xi^{MN}$ and $f_{MNP}$ includes the gauge coupling constants. In addition, $SO(5,n)$ indices $M,N,\ldots$ are lowered and raised by the invariant tensor $\eta_{MN}$ and its inverse $\eta^{MN}$ with $\eta_{MN}=\textrm{diag}(-1,-1,-1,-1,-1,1,\ldots,1)$
\\
\indent The bosonic Lagrangian of a general gauged $N=4$ supergravity coupled to vector multiplets can be written as
\begin{eqnarray}
e^{-1}\mc{L}&=&\frac{1}{2}R-\frac{3}{2}\Sigma^{-2}\pd_\mu \Sigma \pd^\mu \Sigma +\frac{1}{16} \pd_\mu M_{MN}\pd^\mu
M^{MN}-V\label{Lar}
\end{eqnarray}
with $e$ being the vielbein determinant and  
\begin{equation}
M_{MN}=\mc{V}_M^{\phantom{M}m}\mc{V}_N^{\phantom{M}m}+\mc{V}_M^{\phantom{M}a}\mc{V}_N^{\phantom{M}a}\, .
\end{equation}
The scalar potential reads
\begin{eqnarray}
V&=&\frac{1}{4}\left[f_{MNP}f_{QRS}\Sigma^{-2}\left(\frac{1}{12}M^{MQ}M^{NR}M^{PS}-\frac{1}{4}M^{MQ}\eta^{NR}\eta^{PS}
\right.\right.\nonumber \\ & &\left.
+\frac{1}{6}\eta^{MQ}\eta^{NR}\eta^{PS}\right) +\frac{1}{4}\xi_{MN}\xi_{PQ}\Sigma^4(M^{MP}M^{NQ}-\eta^{MP}\eta^{NQ})
\nonumber \\ & &\left.
+\frac{\sqrt{2}}{3}f_{MNP}\xi_{QR}\Sigma M^{MNPQR}\right]
\end{eqnarray}
where $M^{MN}$ is the inverse of $M_{MN}$. $M^{MNPQR}$ is defined as
\begin{equation} 
M^{MNPQR}=\eta^{MM'}\eta^{NN'}\eta^{PP'}\eta^{QQ'}\eta^{RR'}M_{M'N'P'Q'R'}
\end{equation}
with
\begin{equation}
M_{MNPQR}=\epsilon_{mnpqr}\mc{V}_{M}^{\phantom{M}m}\mc{V}_{N}^{\phantom{M}n}
\mc{V}_{P}^{\phantom{M}p}\mc{V}_{Q}^{\phantom{M}q}\mc{V}_{R}^{\phantom{M}r}\, .
\end{equation}
\indent Supersymmetric solutions can be obtained from the BPS equations which in turn arise from setting supersymmetry transformations of fermions to zero. In the present case, the fermionic supersymmetry transformations are given by
\begin{eqnarray}
\delta\psi_{\mu i} &=&D_\mu \epsilon_i+\frac{i}{\sqrt{6}}\Omega_{ij}A^{jk}_1\gamma_\mu\epsilon_k,\\
\delta \chi_i &=&-\frac{\sqrt{3}}{2}i\Sigma^{-1} \pd_\mu
\Sigma\gamma^\mu \epsilon_i+\sqrt{2}\Omega_{ij}A_2^{kj}\epsilon_k,\\
\delta \lambda^a_i&=&i\Omega^{jk}({\mc{V}_M}^a\pd_\mu
{\mc{V}_{ij}}^M)\gamma^\mu\epsilon_k+\sqrt{2}\Omega_{ij}A_{2}^{akj}\epsilon_k\,
.
\end{eqnarray}
The fermion shift matrices are defined by
\begin{eqnarray}
A_1^{ij}&=&-\frac{1}{\sqrt{6}}\left(\sqrt{2}\Sigma^2\Omega_{kl}{\mc{V}_M}^{ik}{\mc{V}_N}^{jl}\xi^{MN}+\frac{4}{3}\Sigma^{-1}{\mc{V}^{ik}}_M{\mc{V}^{jl}}_N{\mc{V}^P}_{kl}{f^{MN}}_P\right),\nonumber
\\
A_2^{ij}&=&\frac{1}{\sqrt{6}}\left(\sqrt{2}\Sigma^2\Omega_{kl}{\mc{V}_M}^{ik}{\mc{V}_N}^{jl}\xi^{MN}-\frac{2}{3}\Sigma^{-1}{\mc{V}^{ik}}_M{\mc{V}^{jl}}_N{\mc{V}^P}_{kl}{f^{MN}}_P\right),\nonumber
\\
A_2^{aij}&=&-\frac{1}{2}\left(\Sigma^2{{\mc{V}_M}^{a}\mc{V}_N}^{ij}\xi^{MN}-\sqrt{2}\Sigma^{-1}\Omega_{kl}{\mc{V}_M}^a{\mc{V}_N}^{ik}{\mc{V}_P}^{jl}f^{MNP}\right).
\end{eqnarray}
The coset representative of the form $\mc{V}_M^{\phantom{M}ij}$ is defined in terms of ${\mc{V}_M}^m$ and $SO(5)$ gamma matrices ${\Gamma_{mi}}^j$ as
\begin{equation}
{\mc{V}_M}^{ij}=\frac{1}{2}{\mc{V}_M}^{m}\Gamma^{ij}_m
\end{equation}
with $\Gamma^{ij}_m=\Omega^{ik}{\Gamma_{mk}}^j$. Similarly, the inverse ${\mc{V}_{ij}}^M$ can be written as
\begin{equation}
{\mc{V}_{ij}}^M=\frac{1}{2}{\mc{V}_m}^M(\Gamma^{ij}_m)^*=\frac{1}{2}{\mc{V}_m}^M\Gamma_{m}^{kl}\Omega_{ki}\Omega_{lj}\,
.
\end{equation}
As in \cite{5D_flowII}, we use the following representation of ${\Gamma_{mi}}^j$ matrices
\begin{eqnarray}
\Gamma_1&=&-\sigma_2\otimes \sigma_2,\qquad \Gamma_2=\mathbb{I}_2\otimes \sigma_1,\qquad \Gamma_3=\mathbb{I}_2\otimes \sigma_3,\nonumber\\
\Gamma_4&=&\sigma_1\otimes \sigma_2,\qquad \Gamma_5=\sigma_3\otimes \sigma_2
\end{eqnarray}
with $\sigma_i$, $i=1,2,3$, being the Pauli matrices. We also note that raising and lowering of $i,j,\ldots$ indices by $\Omega^{ij}$ and $\Omega_{ij}$ are related to complex conjugation. The explicit form of $\Omega_{ij}$ is given by
\begin{equation}
\Omega_{ij}=\Omega^{ij}=i\sigma_2\otimes \mathbb{I}_2\, .
\end{equation}
\indent Finally, the covariant derivative on $\epsilon_i$ is given by
\begin{equation}
D_\mu \epsilon_i=\pd_\mu \epsilon_i+\frac{1}{4}\omega_\mu^{ab}\gamma_{ab}\epsilon_i+{Q_{\mu i}}^j\epsilon_j
\end{equation}
with the composite connection defined by
\begin{equation}
{Q_{\mu i}}^j={\mc{V}_{ik}}^M\pd_\mu {\mc{V}_M}^{kj}\, .
\end{equation}

\subsection{$N=4$ gauged supergravity with $SO(2)_D\times SO(3)\times SO(3)$ gauge group}
We now consider the case of $n=5$ vector multiplets with $SO(2)_D\times SO(3)\times SO(3)$ gauge group. The corresponding embedding tensor is given by 
\begin{eqnarray}
\xi^{MN}&=&g_1(\delta^M_1\delta^N_2-\delta^M_2\delta^N_1)-g_2(\delta^M_{10}\delta^N_9-\delta^M_9\delta^N_{10}),\\ 
f_{\tilde{m}+2,\tilde{n}+2,\tilde{p}+2}&=&h_1\epsilon_{\tilde{m}\tilde{n}\tilde{p}},\qquad \tilde{m},\tilde{n},\tilde{p}=1,2,3,\\
f_{\tilde{a}+5,\tilde{b}+5,\tilde{c}+5}&=&h_2\epsilon_{\tilde{a}\tilde{b}\tilde{c}},\qquad \tilde{a},\tilde{b},\tilde{c}=1,2,3
\end{eqnarray} 
with $g_1$, $g_2$, $h_1$ and $h_2$ being gauge coupling constants. We see that the $SO(2)_D$ is a diagonal subgroup of $SO(2)_R\subset SO(5)_R$ and $SO(2)\subset SO(5)$ generated by the $SO(5,5)$ generators $t_{12}$ and $t_{9,10}$. This factor is gauged by the vector field $A^0_\mu$ in the supergravity multiplet. The remaining $SO(3)\times SO(3)$ factor corresponds to $SO(3)_R\subset SO(5)_R$ and $SO(3)\subset SO(5)$, commuting subgroups to the aforementioned $SO(2)\times SO(2)$. The two $SO(3)$ factors are gauged by the vector fields $A^{3,4,5}_\mu$ and $A^{6,7,8}_\mu$, respectively. The remaining vector fields $A^{1,2}_\mu$ and $A^{9,10}_\mu$ must be dualized to four massive two-form fields \cite{N4_gauged_SUGRA,5D_N4_Dallagata}. We also point out that for $h_2=0$ or $g_2=0$, the gauging results in $SO(2)_D\times SO(3)$ or $SO(2)\times SO(3)\times SO(3)$ gauge groups considered in \cite{5D_N4_flow_Davide} and \cite{5D_N4_flow}, respectively. 
\\
\indent As shown in \cite{5D_flowII}, this gauged supergravity admits four supersymmetric $AdS_5$ vacua. Two of these vacua preserve $N=4$ supersymmetry with $SO(2)_D\times SO(3)\times SO(3)$ and $SO(2)_D\times SO(3)_{\textrm{diag}}$ with $SO(3)_{\textrm{diag}}$ being the diagonal subgroup of $SO(3)\times SO(3)$. The other two vacua preserve $N=2$ supersymmetry with  $SO(2)_{\textrm{diag}}\times SO(3)$ and $SO(2)_{\textrm{diag}}$ symmetries. In these two residual symmetries, the first $SO(2)_{\textrm{diag}}$ is a diagonal subgroup of $SO(2)_D$ and the $SO(2)$ subgroup of the first $SO(3)$ factor in the gauge group while the second $SO(2)_{\textrm{diag}}$ is the diagonal subgroup of $SO(2)_D\times SO(2)\times SO(2)$ with $SO(2)\times SO(2)\subset SO(3)\times SO(3)$. 
\\
\indent A number of holographic RG flows interpolating between these $AdS_5$ vacua has been found in \cite{5D_flowII}. In addition, black string solutions interpolating between these $AdS_5$ vacua and near horizon geometries of the form $AdS_3\times \Sigma$ for $\Sigma$ being a Riemann surface have also been considered in \cite{5D_N4_black_stringII}. In the present paper, we will consider supersymmetric Janus solutions interpolating between these four supersymmetric $AdS_5$ vacua. These solutions would holographically describe conformal interfaces between the dual $N=1$ and $N=2$ SCFTs. As in \cite{5D_flowII} and \cite{5D_N4_black_stringII}, the presence of many $AdS_5$ vacua leads to a very rich structure of solutions.
%%%%%%%%%%%%%%%%%%%%%%%%%%%%%%%%%%%%%%%%%%%%%%%%%%%%%%%%%%%%%%%%%%%%%%%%%%%%%%%%%%%%%%%%%%%%%%%%%%%%%%%%%%%%%%%%%%%%%%%%%%%%%%%%%%%%%%%%%
\section{BPS equations for supersymmetric Janus interfaces}\label{Janus_BPS}
In this section, we will analyze the BPS conditions arising from setting supersymmetry transformations of fermions to zero. Janus solutions are described by an $AdS_4$-sliced domain wall with the metric ansatz given by 
\begin{equation} 
ds^2=e^{2A(r)}ds^2_{AdS_4}+dr^2
\end{equation}
in which $ds^2_{AdS_4}$ is the metric on $AdS_4$ with radius $\ell$. 
\\
\indent To make the analysis more traceable, we will consider a truncation to $SO(2)_{\textrm{diag}}$ singlet scalars. This reduces the total number of $25+1$ scalars to a more managable number. Recall that $SO(2)_{\textrm{diag}}$ is the diagonal subgroup of $SO(2)_D\times SO(2)\times SO(2)$ with $SO(2)\times SO(2)\subset SO(3)\times SO(3)$. This $SO(2)_{\textrm{diag}}$ is generated by a linear combination of $SO(5,5)$ generators $t_{12}$, $t_{45}$, $t_{78}$ and $t_{9,10}$ under which there are nine singlet scalars from $SO(5,5)/SO(5)\times SO(5)$ coset, see \cite{5D_N4_black_stringII} for more detail. The non-compact generators of $SO(5,5)$ are given explicitly by
\begin{equation}
Y_{ma}=t_{m,a+5},\qquad m=1,2,\ldots, 5,\qquad a=1,2,\ldots, 5\, .
\end{equation}
The nine singlet scalars correspond to the following non-compact generators
\begin{eqnarray}
& &\hat{Y}_1=Y_{12}+Y_{23},\qquad \hat{Y}_2=Y_{42}+Y_{53},\qquad \hat{Y}_3=Y_{14}+Y_{25},\nonumber \\
& &\hat{Y}_4=Y_{44}+Y_{55},\qquad \hat{Y}_5=Y_{13}-Y_{22},\qquad \hat{Y}_6=Y_{43}-Y_{52},\nonumber \\
& &\hat{Y}_7=Y_{51}-Y_{24},\qquad \hat{Y}_8=Y_{45}-Y_{54},\qquad  \hat{Y}_9=Y_{31}\, . 
\end{eqnarray}
The coset representative can then be written as
\begin{equation}
\mc{V}=e^{\phi_1\hat{Y}_1}e^{\phi_2\hat{Y}_2}e^{\phi_3\hat{Y}_3}e^{\phi_4\hat{Y}_4}e^{\phi_5\hat{Y}_5}
e^{\phi_6\hat{Y}_6}e^{\phi_7\hat{Y}_7}e^{\phi_8\hat{Y}_8}e^{\phi_9\hat{Y}_9}\, .\label{coset_rep}
\end{equation}
In order to consistently truncate out all the gauge fields, we need to set 
\begin{equation}
\phi_5=\phi_6=\phi_7=\phi_8=0 
\end{equation}
which leads to all vanishing Yang-Mills currents.
\\
\indent Even within the $SO(2)_{\textrm{diag}}$ invariant truncation, the scalar potential still takes a highly complicated form wihch we refrain from giving its explicit form here. However, it can be verified that the aforementioned four supersymmetric $AdS_5$ vacua are critical points of the resulting scalar potential. For convenience, we will also present the locations of all these vacua on the scalar manifold. At all of these vacua, we have $\phi_1=\phi_3=0$ and $\phi_9=\phi_2$. The four vacua are given as follows:
\begin{itemize}
\item Critical point I: $N=4$ $AdS_5$ vacuum with $SO(2)_D\times SO(3)\times SO(3)$ symmetry
\begin{eqnarray}
\phi_2=\phi_4=0,\qquad \Sigma=-\left(\frac{h_1}{\sqrt{2}g_1}\right)^{\frac{1}{3}},\qquad \frac{1}{L}=-\left(\frac{g_1h_1^2}{4\sqrt{2}}\right)^{\frac{1}{3}}\, .
\end{eqnarray}
\item Critical point II: $N=4$ $AdS_5$ vacuum with $SO(2)_D\times SO(3)_{\textrm{diag}}$ symmetry
\begin{eqnarray}
& & \phi_4=0,\qquad \phi_2=\frac{1}{2}\ln\left[\frac{h_2-h_1}{h_2+h_1}\right],\nonumber \\
& & \Sigma=-\left(\frac{h_1h_2}{\sqrt{2}g_1\sqrt{h_2^2-h_1^2}}\right)^{\frac{1}{3}},\qquad \frac{1}{L}=\left[\frac{g_1h_1^2h_2^2}{4\sqrt{2}(h_1^2-h_2^2)}\right]^{\frac{1}{3}}\, .
\end{eqnarray}
\item Critical point III: $N=2$ $AdS_5$ vacuum with $SO(2)_{\textrm{diag}}\times SO(3)$ symmetry
\begin{eqnarray}
& &\phi_2=0,\qquad \phi_4=\frac{1}{2}\ln\left[\frac{g_2-4g_1+2\sqrt{4g_1^2-2g_1g_2-2g_2^2}}{3g_2}\right],\nonumber \\
& & \Sigma=\left(\frac{\sqrt{2}h_1}{g_2}\right)^{\frac{1}{3}},\qquad \frac{1}{L}=\frac{g_2-g_1}{3}\left(\frac{h_1}{\sqrt{2}g_2^2}\right)^{\frac{1}{3}}\, .
\end{eqnarray}
\item Critical point IV: $N=2$ $AdS_5$ vacuum with $SO(2)_{\textrm{diag}}$ symmetry
\begin{eqnarray}
& &\phi_2=\frac{1}{2}\ln\left[\frac{h_2-h_1}{h_2+h_1}\right],\qquad\phi_4=\frac{1}{2}\ln\left[\frac{g_2-4g_1+2\sqrt{4g_1^2-2g_1g_2-2g_2^2}}{3g_2}\right],\nonumber \\
& & \Sigma=\left(\frac{\sqrt{2}h_1h_2}{g_2\sqrt{h_2^2-h_1^2}}\right)^{\frac{1}{3}},\qquad \frac{1}{L}=\frac{g_2-g_1}{3}\left[\frac{h_1^2h_2^2}{\sqrt{2}g_2^2(h_2^2-h_1^2)}\right]^{\frac{1}{3}}\, .
\end{eqnarray}
\end{itemize}
For later convenience, we have also labeled these $AdS_5$ vacua as critical points I, II, III and IV. The $AdS_5$ radius $L$ is related to the cosmological constant $V_0$, the value of the scalar potential at the critical point, by
\begin{equation}
L=\sqrt{-\frac{6}{V_0}}\, .
\end{equation}
It is also useful to point out that we can rescale the dilaton such that critical point I is located at $\Sigma=1$. This amounts to setting $g_1=-\frac{h_1}{\sqrt{2}}$ leading to the $AdS_5$ radius $L=\frac{2}{h_1}$. 
\\
\indent We now analyze the supersymmetry conditions arising from setting supersymmetry transformations of fermionic fields to zero. With $\phi_5=\phi_6=\phi_7=\phi_8=0$, the $A_1^{ij}$ tensor takes a diagonal form
\begin{equation}
A^{ij}_1=\textrm{diag}(\alpha,\beta,\alpha^*,\beta).\label{A1_tensor}
\end{equation}
We see that $A^{ij}_1$ has one real eigenvalue $\beta$ and one complex eigevalue $\alpha=\alpha_1+i\alpha_2$. The real and imaginary parts of $\alpha$ are given by
\begin{eqnarray}
\alpha_1&=&\frac{1}{16\sqrt{3}\Sigma}\left\{4\sinh2\phi_1\sinh\phi_2\sinh\phi_3\sinh2\phi_4(\sqrt{2}h_2\sinh\phi_9+g_1\Sigma^3)-(3\right.\nonumber \\
& &-\cosh2\phi_4)\cosh2\phi_3\left[2\sqrt{2}h_2\sinh^2\phi_1\sinh\phi_9+(g_1-2g_2+g_1\cosh2\phi_1)\Sigma^3\right]\nonumber \\
& &+\cosh^2\phi_4\left[8\sqrt{2}h_1\cosh^2\phi_2\cosh\phi_9+4\sqrt{2}h_2\sinh\phi_9(\cosh^2\phi_1\cosh2\phi_2\right.\nonumber \\
& &\left.\left.\phantom{\sqrt{1}}-\cosh2\phi_1)-4\Sigma^3(g_2+g_1\cosh2\phi_1-g_1\cosh2\phi_2\sinh^2\phi_1)\right]\right\},\\
\alpha_2&=&-\frac{1}{4\sqrt{3}\Sigma}\left\{2\sqrt{2}h_2\sinh\phi_9(\cosh\phi_3\cosh\phi_4\sinh2\phi_1\sinh\phi_2\right.\nonumber \\
& &+\sinh^2\phi_1\sinh2\phi_2\sinh\phi_4)+\Sigma^3\left[2g_1\cosh\phi_3\cosh\phi_4\sinh2\phi_1\sinh\phi_2\right.\nonumber \\
& &\left.\phantom{\sqrt{1}}\left.+(g_1-2g_2+g_1\cosh2\phi_1)\sinh2\phi_3\sinh\phi_4\right]\right\}
\end{eqnarray}
while the real eigenvalue $\beta$ takes the form
\begin{eqnarray}
\beta&=&-\frac{1}{64\sqrt{6}\Sigma}\left\{4h_2\sinh\phi_9\left[\cosh2(\phi_1+\phi_2)+\cosh2\phi_2(2+4\cosh^2\phi_1\cosh2\phi_4)\right.\right.\nonumber \\
& &-8+\cosh2(\phi_1-\phi_2) +8\cosh2\phi_3\cosh^2\phi_4\sinh^2\phi_1\nonumber \\
& &\left.-8\cosh2\phi_1\sinh^2\phi_4+8\sinh2\phi_1\sinh\phi_2\sinh\phi_3\sinh2\phi_4 \right]\nonumber \\
& &+64h_1\cosh^2\phi_2\cosh^2\phi_4\cosh\phi_9+8\sqrt{2}g_2\Sigma^3(\cosh2\phi_4+2\cosh2\phi_3\cosh^2\phi_4\nonumber \\
& &-3)-\sqrt{2}g_1\Sigma^3\left[16+2\cosh2(\phi_1-\phi_2)-4\cosh2\phi_2+2\cosh2(\phi_1+\phi_2)\right.\nonumber \\
& &-4\cosh2(\phi_1-\phi_4)+\cosh2(\phi_1-\phi_2-\phi_4)-2\cosh2(\phi_2-\phi_4) \nonumber \\ 
& &+\cosh2(\phi_1+\phi_2-\phi_4)+8\cosh2\phi_3\cosh^2\phi_4+\cosh2(\phi_1-\phi_2+\phi_4)\nonumber \\
& &+8\cosh2\phi_1(1+\cosh2\phi_3\cosh^2\phi_4)-4\cosh2(\phi_1+\phi_4)-2\cosh2(\phi_2+\phi_4)\nonumber \\
& &\left.\left. +\cosh2(\phi_1+\phi_2+\phi_4)+16\sinh2\phi_1\sinh\phi_2\sinh\phi_3\sinh2\phi_4 \right]\right\}.
\end{eqnarray}
\indent We see that the structure of $A_1^{ij}$ tensor is the same as that considered in \cite{5D_N4_Janus} and \cite{Janus_5D_ISO3}. Accordingly, we will closely follow the analysis of the BPS equations given in \cite{5D_N4_Janus} and \cite{Janus_5D_ISO3}. In particular, as pointed out in \cite{5D_N4_Janus}, the real eigenvalue $\beta$ does not lead to supersymmetric Janus solutions. Therefore, in the following analysis, we will consider only the complex eigenvalue $\alpha$. As a result, the supersymmetry corresponding to the Killing spinors $\epsilon_2$ and $\epsilon_4$ is broken. We also note that, with non-vanishing $\phi_1$ and $\phi_3$, $\alpha$ does not lead to a superpotential in terms of which the scalar potential can be written, see also \cite{5D_N4_Janus} and \cite{Janus_5D_ISO3}. 
\\
\indent We first consider the supersymmetry transformation of $\chi_i$. Setting $\epsilon_2=\epsilon_4=0$, we find that the conditions $\delta\chi_i=0$ give 
\begin{equation}
\Sigma'\gamma_{\hat{r}}\epsilon_1=\mc{A}\epsilon_3\qquad \textrm{and}\qquad \Sigma'\gamma_{\hat{r}}\epsilon_3=\mc{A}^*\epsilon_1\, .\label{Sigma_eq1}
\end{equation}
The complex function $\mc{A}=\mc{A}_1+i\mc{A}_2$ has real and imaginary parts given by
\begin{eqnarray}
\mc{A}_1&=&\frac{1}{3}\cosh\phi_3\left\{h_2\sinh\phi_9(\cosh\phi_4\sinh2\phi_1\sinh\phi_2+2\sinh^2\phi_1\sinh\phi_3\sinh\phi_4)\right.\nonumber \\
& &+\sqrt{2}\Sigma^3\left[(2g_2-g_1-g_1\cosh2\phi_1)\sinh\phi_3\sinh\phi_4\right.\nonumber \\
& &\left.\left.-g_1\cosh\phi_4\sinh2\phi_1\sinh\phi_2\right]\right\}
\end{eqnarray}
and
\begin{eqnarray}
\mc{A}_2&=&-\frac{1}{3}h_1\cosh^2\phi_2\cosh^2\phi_4\cosh\phi_9-\frac{1}{12}h_2\sinh\phi_9\left\{\cosh2\phi_3\sinh^2\phi_1\times \right. \nonumber \\
& &(\cosh2\phi_4-3)+2\sinh2\phi_1\sinh\phi_2\sinh\phi_3\sinh2\phi_4-2\cosh^2\phi_4(\cosh2\phi_1
\nonumber \\
& &\left.-\cosh^2\phi_1\cosh2\phi_2)\right\}+\frac{1}{12\sqrt{2}}\Sigma^3\left\{(g_1-2g_2+g_1\cosh2\phi_1)\times \right.\nonumber \\
& &(\cosh2\phi_4-3)\cosh2\phi_3-4\cosh^2\phi_4(g_2+g_1\cosh2\phi_1\nonumber \\
& &\left.-g_1\cosh2\phi_2\sinh^2\phi_1)+4g_1\sinh2\phi_1\sinh\phi_2\sinh\phi_3\sinh2\phi_4 \right\}.
\end{eqnarray}
\indent Follow \cite{5D_N4_Janus}, we find that the two conditions in \eqref{Sigma_eq1} lead to a BPS equation for $\Sigma$ of the form
\begin{equation}
{\Sigma'}^2=|\mc{A}|^2=\mc{A}_1^2+\mc{A}_2^2
\end{equation}
and a projector on $\epsilon_1$ and $\epsilon_3$ given by 
\begin{equation}
\gamma_{\hat{r}}\epsilon_1=\eta \frac{\mc{A}}{|\mc{A}|}\epsilon_3\qquad \textrm{and}\qquad \gamma_{\hat{r}}\epsilon_3=\eta \frac{\mc{A}^*}{|\mc{A}|}\epsilon_1\label{gamma_r_proj}
\end{equation}
for $\eta=\pm 1$.
\\
\indent The supersymmetry conditions from $\delta \lambda^a_i$ give rise to three different sets of equations. The first one takes the form
\begin{eqnarray}
\left[\cosh\phi_2\sinh\phi_3\sinh\phi_4\phi_1'+\cosh\phi_4\phi_2'+i\cosh\phi_2\cosh\phi_3\phi_1'\right]\gamma_{\hat{r}}\epsilon_3=\mc{B}\epsilon_1
\end{eqnarray}
and 
\begin{eqnarray}
\left[\cosh\phi_2\sinh\phi_3\sinh\phi_4\phi_1'+\cosh\phi_4\phi_2'-i\cosh\phi_2\cosh\phi_3\phi_1'\right]\gamma_{\hat{r}}\epsilon_1=\mc{B}^*\epsilon_3
\end{eqnarray}
with $\mc{B}=\mc{B}_1+i\mc{B}_2$ and 
\begin{eqnarray}
\mc{B}_1&=&\frac{1}{4}\Sigma^{-1}\cosh\phi_2\cosh\phi_3\sinh2\phi_1(2h_2\sinh\phi_9-\sqrt{2}g_1\Sigma^3),\nonumber \\
\mc{B}_2&=&\frac{1}{2}\Sigma^{-1}\cosh\phi_2\left[2\cosh\phi_4\sinh\phi_2(h_1\cosh\phi_9+h_2\cosh^2\phi_1\sinh\phi_9)\right.\nonumber \\
& &-\sqrt{2}g_1\Sigma^3\sinh\phi_1(\cosh\phi_4\sinh\phi_1\sinh\phi_2+\cosh\phi_1\sinh\phi_3\sinh\phi_4)\nonumber \\
& &\left.+h_2\sinh2\phi_1\sinh\phi_3\sinh\phi_4\sinh\phi_9 \right].
\end{eqnarray}
Using the projector \eqref{gamma_r_proj}, we find the BPS equations for $\phi_1$ and $\phi_2$ of the form
\begin{eqnarray}
\cosh\phi_2\cosh\phi_3\phi'_1&=&\eta \frac{\textrm{Im}(\mc{A}\mc{B})}{|\mc{A}|},\\ 
\cosh\phi_4\phi'_2&=&\frac{\eta}{|\mc{A}|}\left[\textrm{Re}(\mc{A}\mc{B})-\tanh\phi_3\sinh\phi_4\textrm{Im}(\mc{A}\mc{B})\right].
\end{eqnarray}
\indent The second set of equations from $\delta\lambda^a_i$ is given by
\begin{equation}
\left[\sinh\phi_2\sinh\phi_3\phi_1'-\phi_4'-i(\cosh\phi_3\sinh\phi_2\sinh\phi_4\phi_1'+\cosh\phi_4\phi_3')\right]\gamma_{\hat{r}}\epsilon_3=\mc{C}\epsilon_1
\end{equation}
and
\begin{equation}
\left[\sinh\phi_2\sinh\phi_3\phi_1'-\phi_4'+i(\cosh\phi_3\sinh\phi_2\sinh\phi_4\phi_1'+\cosh\phi_4\phi_3')\right]\gamma_{\hat{r}}\epsilon_1
=\mc{C}\epsilon_3
\end{equation}
with $\mc{C}=\mc{C}_1+i\mc{C}_2$ and
\begin{eqnarray}
\mc{C}_1&=&\frac{1}{2}\Sigma^{-1}\cosh\phi_3\left\{h_2\sinh\phi_9(2\cosh\phi_4\sinh^2\phi_1\sinh\phi_3\right.\nonumber \\
& &+\sinh2\phi_1\sinh\phi_2\sinh\phi_4) -\sqrt{2}\Sigma^{3}\left[(g_1\cosh^2\phi_1-g_2)\cosh\phi_4\sinh\phi_3\right.\nonumber \\
& &\left.\left.+g_1\cosh\phi_1\sinh\phi_1\sinh\phi_2\sinh\phi_4\right]\right\},\nonumber \\
\mc{C}_2&=&\frac{1}{2}h_1\Sigma^{-1}\cosh^2\phi_2\cosh\phi_9\sinh2\phi_4+\frac{1}{2}\Sigma^{-1}h_2\sinh\phi_9\left[\cosh^2\phi_4\sinh2\phi_1\times \right.\nonumber \\
& &\sinh\phi_2\sinh\phi_3+\sinh2\phi_1\sinh\phi_2\sinh\phi_3\sinh^2\phi_4\nonumber \\
& &\left.+\sinh2\phi_4(\cosh^2\phi_1\sinh^2\phi_2+\sinh^2\phi_1\sinh^2\phi_3)\right].
\end{eqnarray}
By a similar analysis, we obtain the BPS equations for $\phi_3$ and $\phi_4$ as follows
\begin{eqnarray}
\cosh\phi_4\phi_3'&=&-\frac{\eta}{|\mc{A}|}\left[\textrm{Im}(\mc{A}\mc{C})+\tanh\phi_2\sinh\phi_4\textrm{Im}(\mc{A}\mc{B})\right],\\
\phi_4'&=&-\frac{\eta}{|\mc{A}|}\left[\textrm{Re}(\mc{A}\mc{C})+\tanh\phi_2\tanh\phi_3\textrm{Im}(\mc{A}\mc{B})\right].
\end{eqnarray}
\indent The last set of equations from $\delta\lambda^a_i$ reads
\begin{equation}
\phi_9'\gamma_{\hat{r}}\epsilon_1=\mc{D}\epsilon_3\qquad \textrm{and}\qquad \phi_9'\gamma_{\hat{r}}\epsilon_3=\mc{D}^*\epsilon_1
\end{equation}
with $\mc{D}=\mc{D}_1+i\mc{D}_2$ and
\begin{eqnarray}
\mc{D}_1&=&-2\Sigma^{-1}h_2(\cosh\phi_1\cosh\phi_4\sinh\phi_2+\sinh\phi_1\sinh\phi_3\sinh\phi_4)\times \nonumber \\
& &\cosh\phi_3\cosh\phi_9\sinh\phi_1,\nonumber \\
\mc{D}_2&=&h_1\Sigma^{-1}\cosh^2\phi_2\cosh^2\phi_4\sinh\phi_9-h_2\Sigma^{-1}\cosh^2\phi_3\cosh\phi_9\sinh^2\phi_1\nonumber \\
& &+h_2\Sigma^{-1}\cosh\phi_9(\cosh\phi_1\cosh\phi_4\sinh\phi_2+\sinh\phi_1\sinh\phi_3\sinh\phi_4)^2\, .
\end{eqnarray}
By repeating the same analysis, we find the BPS equation for $\phi_9$ of the form
\begin{equation}
\phi_9'=\frac{\eta}{|\mc{A}|}\textrm{Re}(\mc{A}^*\mc{D})
\end{equation}
together with an algebraic constraint
\begin{equation}
\textrm{Im}(\mc{A}^*\mc{D})=0\, .\label{con2}
\end{equation}
A similar constraint also arises in the analysis of BPS equations for Janus solutions from $N=4$ gauged supergravity with $SO(2)\times ISO(3)$ gauge group studied in \cite{Janus_5D_ISO3}. In that case, the constraint appears to obstruct the existence of supersymmetric Janus solutions interpolating between $AdS_5$ vacua. In the present case, this is not the case, and regular Janus solutions interpolating between $AdS_5$ vacua can be found. We also point out that the algebraic constraint \eqref{con2} is compatible with all the previously obtained BPS equations.
\\
\indent We now move to the gravitino variations along $AdS_4$ directions. By splitting the five-dimensional coordinates as $x^\mu=(x^\alpha,r)$ for $\alpha=0,1,2,3$, we find the following conditions from $\delta\psi_{\hat{\alpha}i}$ variation
\begin{eqnarray}
& &\left(A'-\frac{i}{\ell}\kappa_1 e^{-A}\right)\gamma_{\hat{r}}\epsilon_1=\mc{W}\epsilon_3\label{gravitino_eq1}\\
\textrm{and}\qquad & & \left(A'-\frac{i}{\ell}\kappa_3 e^{-A}\right)\gamma_{\hat{r}}\epsilon_3=\mc{W}^*\epsilon_1\label{gravitino_eq2}
\end{eqnarray}
with $\mc{W}=\mc{W}_1+i\mc{W}_2$ and
\begin{eqnarray}
\mc{W}_1&=&-\frac{1}{6}\Sigma^{-1}\cosh\phi_3\left\{\sqrt{2}\Sigma^{3}\left[(2g_2-g_1-g_1\cosh2\phi_1)\sinh\phi_3\sinh\phi_4\right.\right.\nonumber \\
& &\left.-g_1\cosh\phi_4\sinh2\phi_1\sinh\phi_2\right]-2h_2\sinh\phi_9(\cosh\phi_4\sinh2\phi_1\sinh\phi_2\nonumber \\
& &\left.\phantom{\sqrt{1}}+2\sinh^2\phi_1\sinh\phi_3\sinh\phi_4)\right\},\nonumber \\
\mc{W}_2&=&-\frac{1}{12}\Sigma^{-1}\sinh2\phi_1\sinh\phi_2\sinh\phi_3\sinh2\phi_4(2h_2\sinh\phi_9+\sqrt{2}g_1\Sigma^3)\nonumber \\
& &-\frac{1}{48}\Sigma^{-1}\left[4h_2\sinh^2\phi_1\sinh\phi_9+\sqrt{2}\Sigma^3(g_1-2g_2+g_1\cosh2\phi_1)\right]\times \nonumber \\
& &\cosh2\phi_3(\cosh2\phi_4-3)-\frac{1}{24}\Sigma^{-1}\cosh^2\phi_4\left[8h_1\cosh^2\phi_2\cosh\phi_9\phantom{\sqrt{1}}\right. \nonumber \\
& & \left. -4h_2\sinh\phi_9(\cosh2\phi_1-\cosh^2\phi_1\cosh2\phi_2)\right.\nonumber \\
& &\left. -2\sqrt{2}\Sigma^3(g_2+g_1\cosh2\phi_1-g_1\cosh2\phi_2\sinh^2\phi_1)\right].
\end{eqnarray}
It should be noted that $\mc{W}$ is related to the eigenvalue $\alpha$ via a relation $\mc{W}=-i\sqrt{\frac{2}{3}}\alpha^*$. 
\\
\indent In obtaining equations \eqref{gravitino_eq1} and \eqref{gravitino_eq2}, we have rewritten the covariant derivative of the Killing spinor $D_\alpha$ in terms of the covariant derivative on $AdS_4$ $\widetilde{\nabla}_\alpha$ as 
\begin{equation}
D_\alpha\epsilon_i=\widetilde{\nabla}_\alpha\epsilon_i-\frac{1}{2}A'\gamma_{r}\gamma_{\alpha}\epsilon_i
\end{equation}
and used the Killing spinor equations on $AdS_4$ of the form
\begin{equation}
\widetilde{\nabla}_\alpha\epsilon_i=\frac{i}{2\ell}\kappa_i\gamma_{r}\gamma_{\alpha}\epsilon_i
\end{equation}
with $\kappa_i=\pm 1$. We also note that the chirality matrix on $AdS_4$ is given by $\gamma_r=i\gamma_{\hat{0}}\gamma_{\hat{1}}\gamma_{\hat{2}}\gamma_{\hat{3}}$. Consistency between \eqref{gravitino_eq1} and \eqref{gravitino_eq2} requires $\kappa_3=-\kappa_1$.
\\
\indent Using the $\gamma_{\hat{r}}$ projector given in \eqref{gamma_r_proj} and writing $\kappa=\kappa_1=-\kappa_3$, we find the BPS equation for $A$ of the form
\begin{equation}
A'=\eta \frac{\textrm{Re}(\mc{W}\mc{A}^*)}{|\mc{A}|}\label{BPS_gravi}
\end{equation} 
together with another algebraic constraint
\begin{equation}
\frac{\kappa}{\ell}e^{-A}=-\eta\frac{\textrm{Im}(\mc{W}\mc{A}^*)}{|\mc{A}|}\, .\label{con1}
\end{equation}
Combining equations \eqref{gravitino_eq1} and \eqref{gravitino_eq2}, we can also derive the following relation
\begin{equation}
{A'}^2+\frac{1}{\ell^2}e^{-2A}=|\mc{W}|^2\, .
\end{equation}
We have verified that the algebraic constraint \eqref{con1} is compatible with all the BPS equations. In addition, it can also be verified that all the BPS equations together with the two algebraic constraints in \eqref{con2} and \eqref{con1} imply the second-ordered field equations.
\\
\indent We finally consider the gravitino variations along $r$ of the form
\begin{equation}
\delta\psi_{\hat{r}i}=\pd_r\epsilon_i+iQ{_{ri}}^j\epsilon_j+\frac{i}{\sqrt{6}}\Omega_{ij}A^{jk}_1\gamma_{\hat{r}}\epsilon_k\, .
\end{equation}
The composite connection can be computed from the coset representative and takes the form
\begin{equation}
{Q_{ri}}^j=Q(r)\textrm{diag}(-i,0,i,0)
\end{equation}
with
\begin{equation}
Q(r)=\cosh\phi_3\cosh\phi_4\sinh\phi_2\phi_1'+\sinh\phi_4\phi_3'\, .
\end{equation}
Using the explicit form of $A_1^{ij}$ and the definition of $\mc{W}$, we find
\begin{equation}
\epsilon_1'=\frac{1}{2}\mc{W}\gamma_{\hat{r}}\epsilon_3-iQ(r)\epsilon_1\qquad\textrm{and}\qquad  \epsilon_3'=\frac{1}{2}\mc{W}^*\gamma_{\hat{r}}\epsilon_1-iQ(r)\epsilon_3\, .
\end{equation}
Using equations \eqref{gravitino_eq1} and \eqref{gravitino_eq2}, we eventually obtain
\begin{eqnarray}
\epsilon_{1,3}'=\left[\frac{1}{2}\left(A'-\frac{i}{\ell}\kappa_{1,3}e^{-A}\right)-iQ(r)\right]\epsilon_{1,3}
\end{eqnarray}
which leads to the Killing spinors
\begin{equation}
\epsilon_{1,3}=e^{\frac{1}{2}A+i\Phi_{1,3}}\epsilon_{1,3}^{(0)}\, .
\end{equation}
In this equation, $\epsilon_{1,3}^{(0)}$ are two $r$-independent spinors, and the phase functions are defined by
\begin{equation}     
\Phi_{1,3}=-\int \left(\frac{\kappa_{1,3}}{2\ell}e^{-A}+Q(r)\right)dr
\end{equation}
with $\kappa_3=-\kappa_1$ as before. It should be noted that in the limit $\ell\rightarrow \infty$ and $\phi_1=\phi_3=0$, the Killing spinors reduce to those of standard flat domain walls $\epsilon_{1,3}=e^{\frac{A}{2}}\epsilon^{(0)}_{1,3}$.
\\
\indent We end this section by giving some comments on the two algebraic constraints \eqref{con2} and \eqref{con1}. The constraint \eqref{con2} dose not appear in the analysis of \cite{5D_N4_Janus} since there are no scalar fields that are not charged under $SO(2)_D$, $\phi_9$ and $\phi_2$ in the present case. However, a similar constraint has also arisen in \cite{Janus_5D_ISO3} in which $SO(2)$ invariant scalars do appear. Similar to the situation in \cite{Janus_5D_ISO3}, we cannot simply set $\phi_9=0$ since this is not consistent with the corresponding BPS equations. On the other hand, the algegraic constraint \eqref{con1} is related to the curvature of the domain wall. In particular, if the right hand side of \eqref{con1} vanishes, the corresponding field configurations cannot support the curvature of the $AdS_4$ slices as $\ell\rightarrow \infty$ in this case. Although the explicit form of \eqref{con1} is rather complicated, we can at least check that the right hand side of \eqref{con1} vanishes for $\phi_1=0$ and $\phi_3=0$. We also point out that setting $\phi_1=\phi_2=\phi_9=0$, we recover the BPS equations studied in \cite{5D_N4_Janus}. 
\\
\indent There is an interesting consistent truncation given by setting $\phi_3=\phi_4=0$ and $g_2=0$. In this case, the BPS equations as well as the corresponding solutions can be regarded as those of $N=4$ gauged supergravity with $SO(2)\times SO(3)\times SO(3)$ gauge group considered in \cite{5D_N4_flow_Davide} and \cite{5D_N4_flow}. In this case, the $N=2$ $AdS_5$ vacua disappear, but there are still two $N=4$ $AdS_5$ vacua given by critical points I and II. Since Janus solutions in this $N=4$ gauged supergravity has not been studied previously, we will separately consider the solutions in this truncation.  
%%%%%%%%%%%%%%%%%%%%%%%%%%%%%%%%%%%%%%%%%%%%%%%%%%%%%%%%%%%%%%%%%%%%%%%%%%%%%%%%%%%%%%%%%%%%%%%%%%%%%%%%%%%%%%%%%%%%%%%%%%%%%%%%%%%%%%%%%
\section{Supersymmetric Janus solutions}\label{Janus_solutions}
In this section, we look for supersymmetric Janus solutions to the BPS equations obtained in the previous section. We will consider only regular Janus solutions interpolating between $AdS_5$ vacua. The solutions holographically describe conformal interfaces within $N=2$ or $N=1$ SCFTs dual to $N=4$ or $N=2$ $AdS_5$ vacua. The solutions are called Janus interfaces if the dual SCFTs on both sides of the interfaces are the same. If the dual SCFTs on the two sides of the interfaces are different, the resulting solutions are called RG-flow interfaces between two different SCFTs. We also recall that only $\epsilon_1$ and $\epsilon_3$ correspond to the Killing spinors of the unbroken supersymmetry, and these spinors are subject to the projector \eqref{gamma_r_proj}. Therefore, the solutions to the BPS equations preserve only four supercharges corresponding to $N=1$ superconformal symmetry on the three-dimensional conformal interfaces. Furthermore, we will look for numerical Janus solutions due to the complexity of the relevant BPS equations.  
\\
\indent In order to find regular Janus solutions, we need to smoothly combine the two branches of solutions corresponding to each sign of $\eta$ in the BPS equations. The existence of these solutions also implies a turning point of the warp factor $A(r)$ such that $A'(r_0)=0$ for a particular value of $r=r_0$. Without loss of generality, we can choose $r_0=0$. The solutions must also be asymptotically $AdS_5$ with $A(r)\rightarrow \frac{r}{L}$ as $r\rightarrow \pm \infty$. This also implies that $A(r)$ attains a minimum at $r_0$. A general procedure to find the solutions is to choose a boundary condition at $r_0$ such that $A'(r_0)=0$ and $A''(r_0)>0$ and smoothly join the two branches of solutions on both sides of $r_0$. Alternatively, we can find the solutions without any branch cuts by solving the second-ordered field equations with the boundary conditions for first derivatives of various fields given by the BPS equations. 
\\
\indent Before giving explicit solutions, we discuss holographic implications of the solutions in the dual field theories in some detail. The regular Janus solutions considered here are asymptotic to an $AdS_5$ vacuum. Near this $AdS_5$ geometry, the five-dimensional metric can be written in terms of an $AdS_4$ slice as
\begin{equation}
ds^2=\frac{\cosh^2 u}{z^2}(-dt^2+dx_1^2+dx_2^2+dz^2)+du^2\, .\label{AdS4_AdS5}
\end{equation}
The directions $(t,x_1,x_2)$ lie along the three-dimensional conformal defect or interface while $z$ parametrizes the direction away from the interface. We have also taken the $AdS_5$ radius to be unity for convenience. The behavior of a scalar field near the $AdS_5$ geometry takes the form of
\begin{equation}
\phi\sim \hat{\phi}e^{-(4-\Delta)u}+\tilde{\phi}e^{-\Delta u}\label{scalar1}
\end{equation} 
for $\Delta\neq 2$, and
\begin{equation}
\phi\sim \hat{\phi}ue^{-2u}+\tilde{\phi}e^{-2u}\label{scalar2}
\end{equation} 
for $\Delta= 2$. At this stage, we cannot apply the usual interpretation of $\hat{\phi}$ and $\tilde{\phi}$ as dual to a source or a deformation and a vacuum expectation value since this is only valid for a flat- or Poincare-sliced $AdS_5$ geometry. Accordinly, we need to transform the metric \eqref{AdS4_AdS5} into a Poincare-sliced form.
\\
\indent  Recall that the $AdS_5$ metric in the Poincare-sliced form is given by
\begin{equation}
ds^2=\frac{1}{\xi^2}(-dt^2+dx_1^2+dx_2^2+dy^2+d\xi^2).\label{flat_AdS5}
\end{equation}
The solution for a scalar field near this geometry can be written as
\begin{eqnarray}
& &\phi\sim \phi_{(s)}\xi^{4-\Delta}+\phi_{(v)}\xi^\Delta,\qquad \Delta\neq 2,\nonumber \\
\textrm{or} \qquad & &\phi\sim \phi_{(s)}\xi^{2}\ln \xi+\phi_{(v)}\xi^2,\qquad \Delta= 2 \label{scalar3}
\end{eqnarray}
in which $\phi_{(s)}$ and $\phi_{(v)}$ are identified with a deformation and a vacuum expectation value of the operator dual to $\phi$, respectively. 
\\
\indent It can be easily verified that we can bring the metric given in \eqref{flat_AdS5} to the form \eqref{AdS4_AdS5} by the following coordinate transformation
\begin{equation}
y=z \tanh u\qquad \textrm{and}\qquad \xi=\frac{z}{\cosh u}\, .
\end{equation}  
As $u\rightarrow \infty$ or $\xi\rightarrow 0$, we find, to leading order,
\begin{equation}
e^{-u}=\frac{\xi}{2y}\, .
\end{equation}
We can then rewrite the behaviors of the scalar field in \eqref{scalar1} and \eqref{scalar2} as
\begin{eqnarray}
& &\phi\sim \frac{\hat{\phi}}{y^{4-\Delta}}\xi^{4-\Delta}+\frac{\tilde{\phi}}{y^\Delta}\xi^\Delta,\qquad \Delta\neq 2,\nonumber \\
\textrm{or} \qquad & &\phi\sim \frac{\hat{\phi}}{y^2}\xi^{2}\ln \xi+\frac{\tilde{\phi}}{y^2}\xi^2,\qquad \Delta= 2\, . \label{scalar4}
\end{eqnarray} 
By comparing \eqref{scalar4} with \eqref{scalar3}, we find the following identification of a deformation and a vacuum expectation value
\begin{equation}
\phi_{(s)}\sim \frac{\hat{\phi}}{y^{4-\Delta}}\qquad \textrm{and}\qquad \phi_{(v)}\sim \frac{\tilde{\phi}}{y^\Delta}\, .
\end{equation} 
This implies that the source term and the vacuum expectation value of the operator dual to $\phi$ acquire a $y$-dependence. Therefore, the presence of conformal interfaces involves position-dependent deformations and/or position-dependent vacuum expectation values. 

%%%%%%%%%%%%%%%%%%%%%%%%%%%%%%%%%%%%%%%%%%%%%%%%%%%%%%%%%%%%%%%%%%%
\subsection{Janus solutions from $SO(2)\times SO(3)\times SO(3)$ gauge group}
\begin{figure}
         \centering
         \begin{subfigure}[b]{0.3\textwidth}
                 \includegraphics[width=\textwidth]{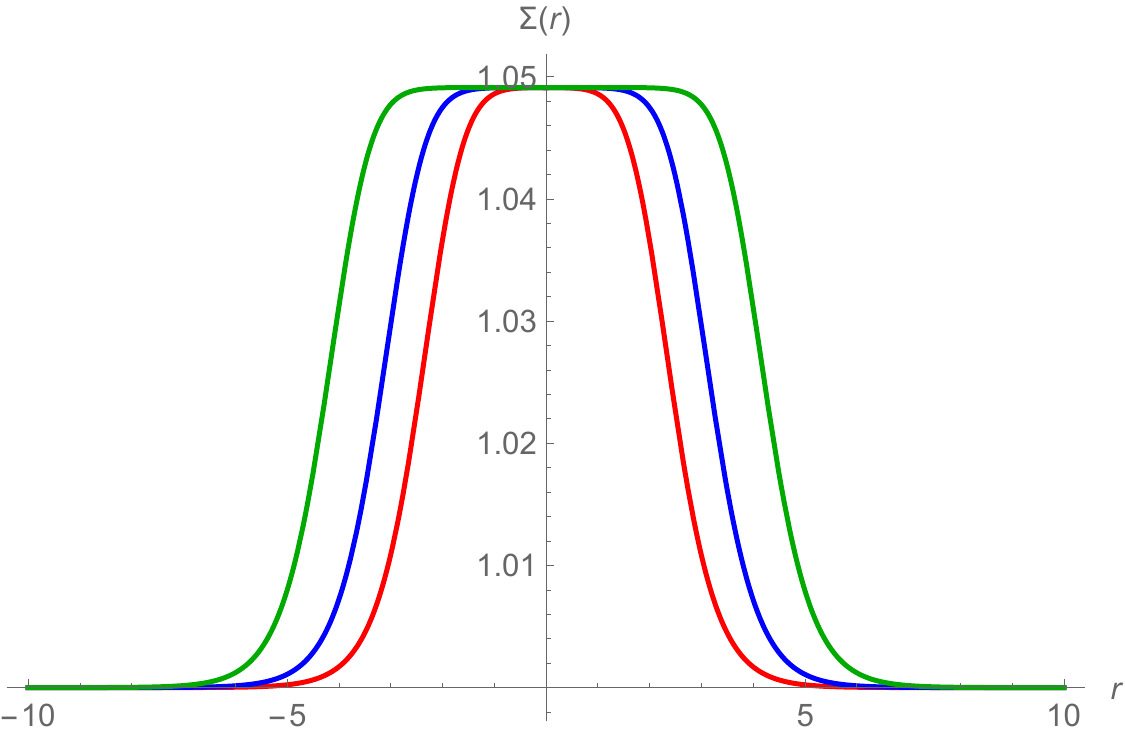}
                 \caption{Solutions for $\Sigma(r)$}
         \end{subfigure} \qquad
\begin{subfigure}[b]{0.3\textwidth}
                 \includegraphics[width=\textwidth]{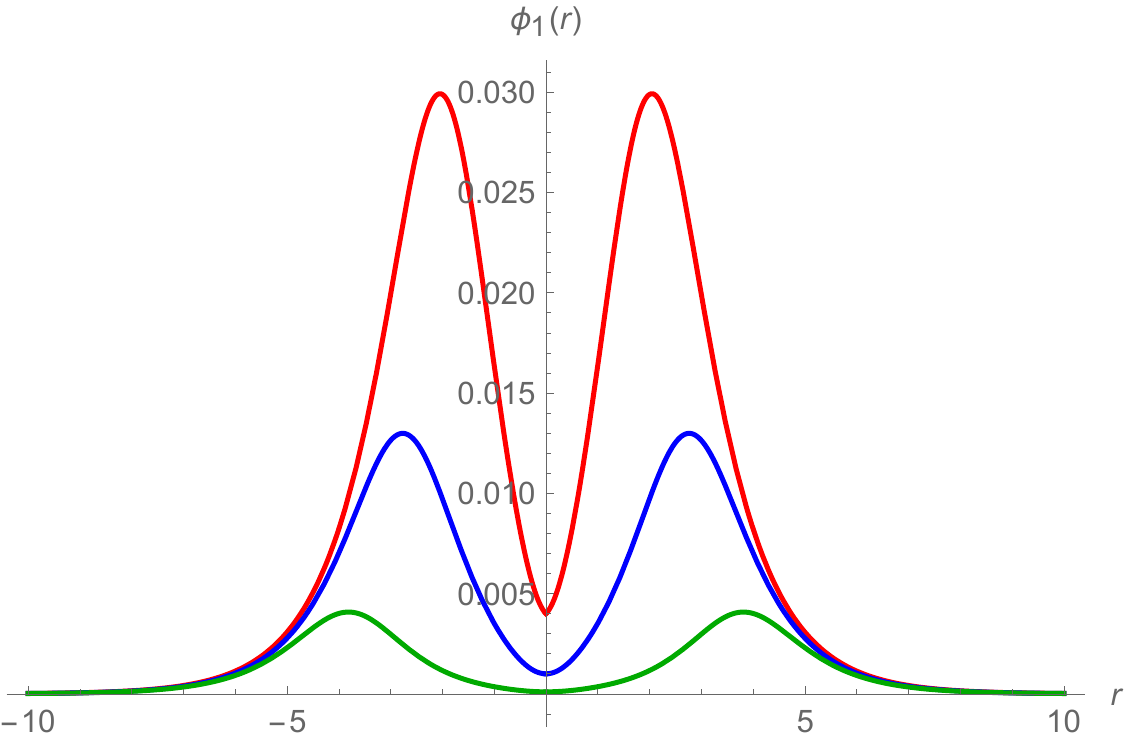}
                 \caption{Solutions for $\phi_1(r)$}
         \end{subfigure}
         \begin{subfigure}[b]{0.3\textwidth}
                 \includegraphics[width=\textwidth]{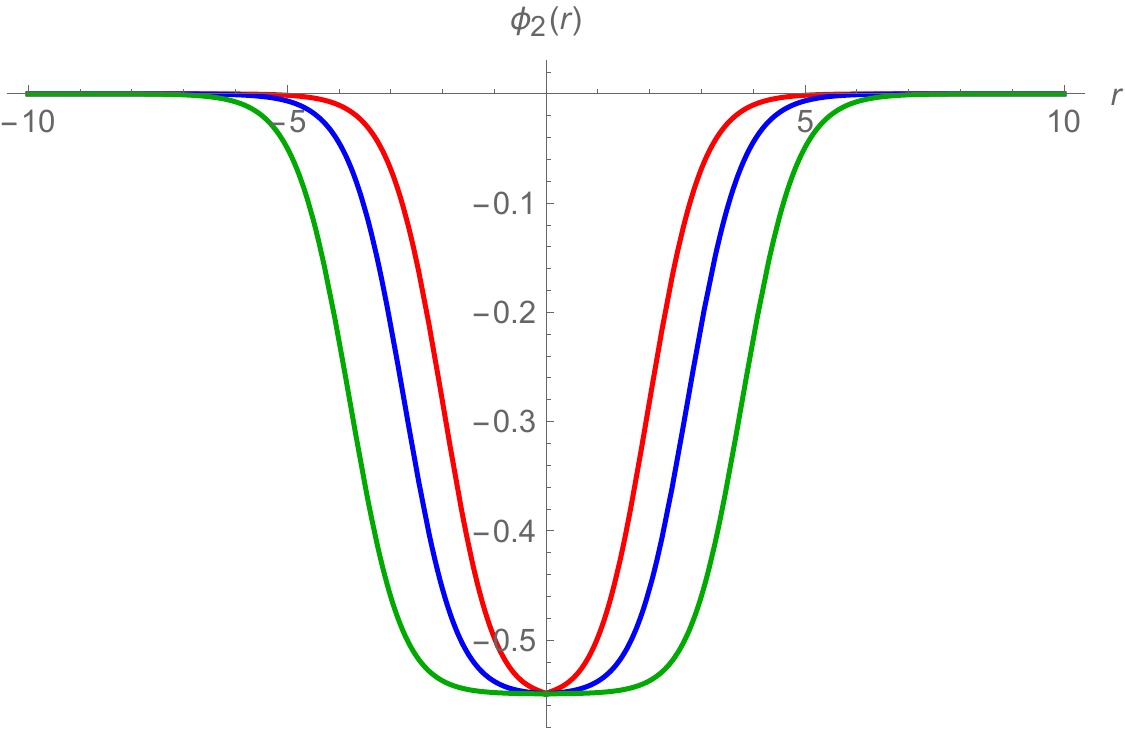}
                 \caption{Solutions for $\phi_2(r)$}
         \end{subfigure}\\ 
         \begin{subfigure}[b]{0.3\textwidth}
                 \includegraphics[width=\textwidth]{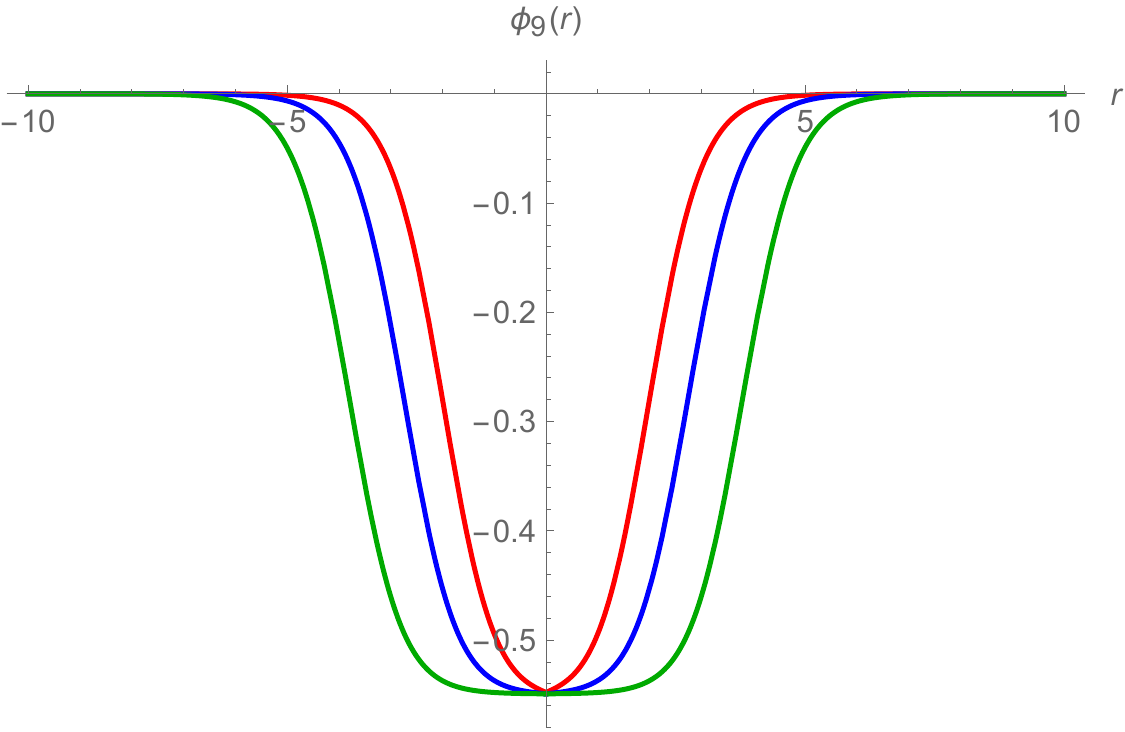}
                 \caption{Solutions for $\phi_9$}
         \end{subfigure} 
         \begin{subfigure}[b]{0.3\textwidth}
                 \includegraphics[width=\textwidth]{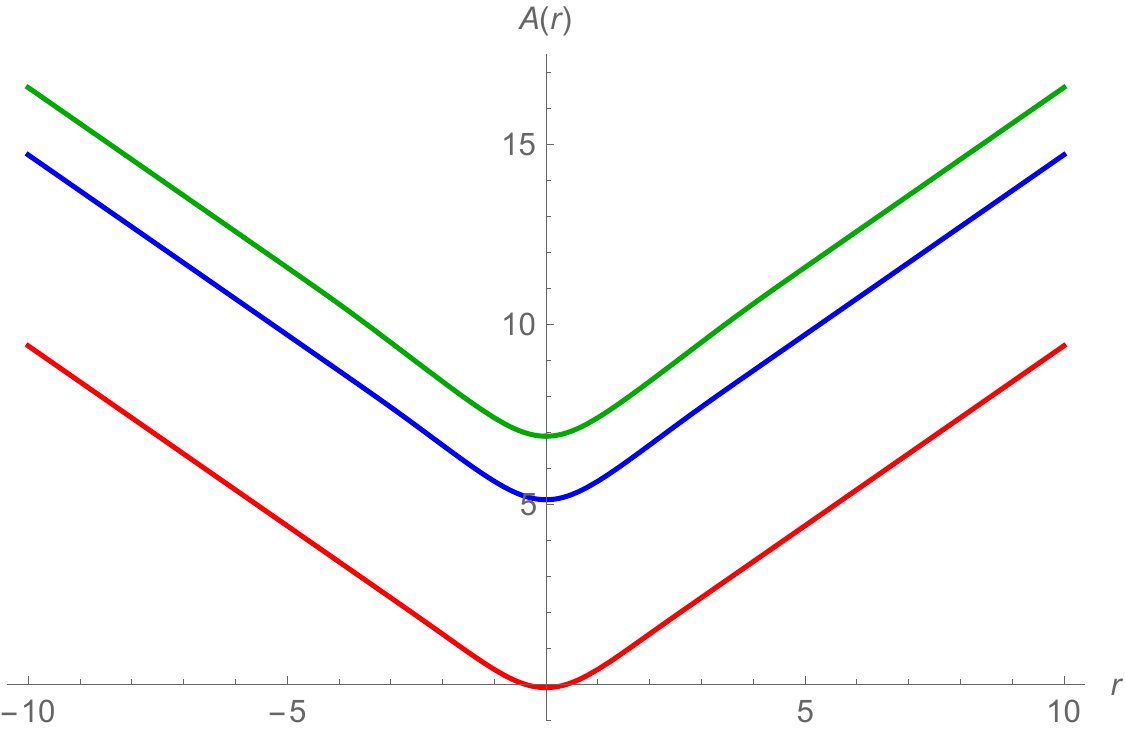}
                 \caption{Solutions for $A(r)$}
         \end{subfigure}
         \begin{subfigure}[b]{0.3\textwidth}
                 \includegraphics[width=\textwidth]{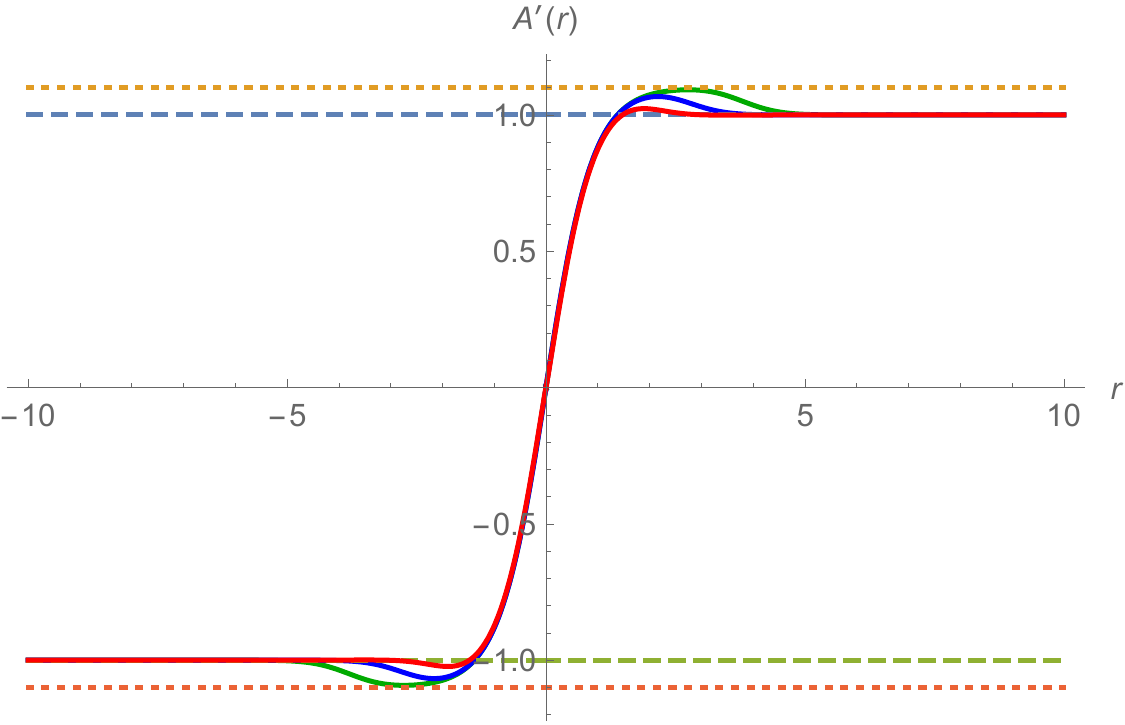}
                 \caption{Solutions for $A'(r)$}
         \end{subfigure}
         \caption{Examples of Janus solutions interpolating between $N=4$ $AdS_5$ critical point I (red) and between $N=4$ $AdS_5$ critical point II (green). The blue line corresponds to a solution interpolating between critical point I that flows close to critical point II. In these solutions, we have chosen the numerical values of $\ell=1$, $\kappa=-1$, $h_1=2$, $g_1=-\sqrt{2}$ and $h_2=4$.}\label{fig1}
 \end{figure}

\begin{figure}
        \centering
         \begin{subfigure}[b]{0.3\textwidth}
                 \includegraphics[width=\textwidth]{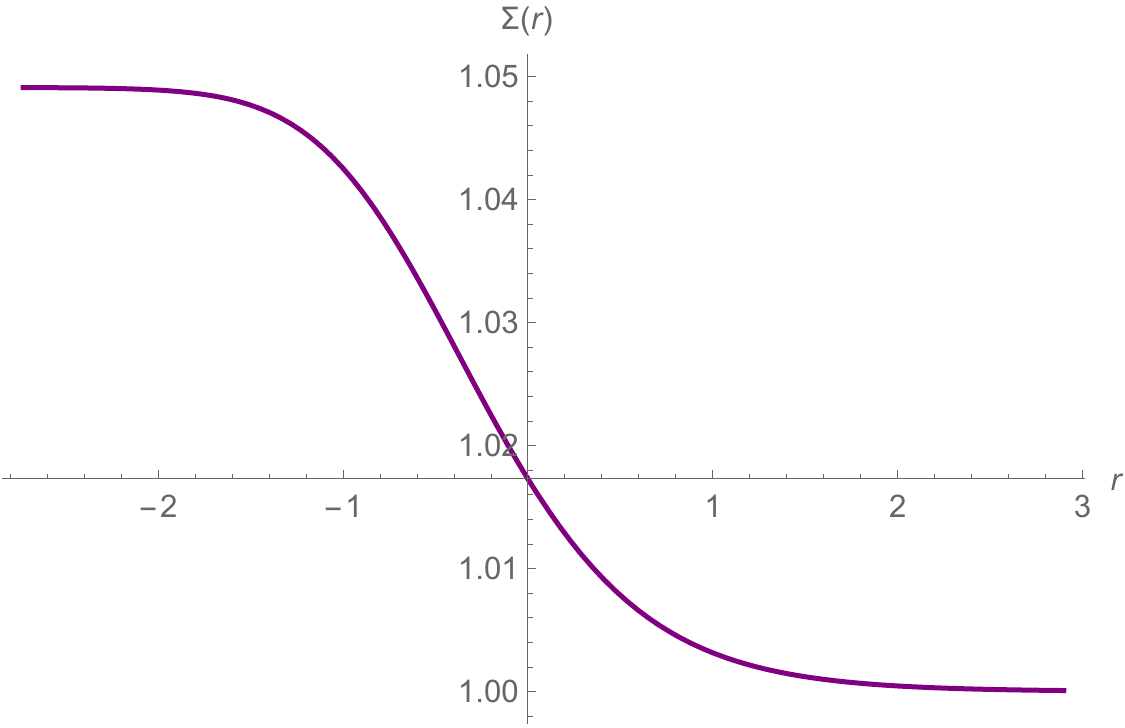}
                 \caption{Solution for $\Sigma(r)$}
         \end{subfigure} \qquad
\begin{subfigure}[b]{0.3\textwidth}
                 \includegraphics[width=\textwidth]{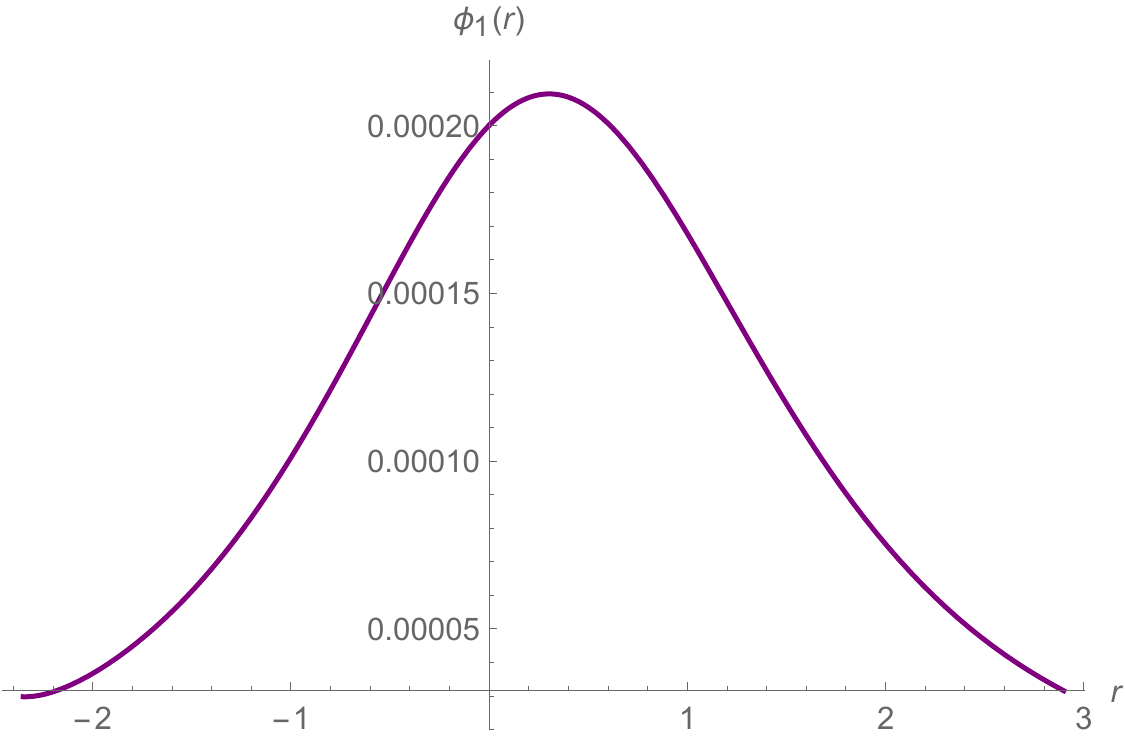}
                 \caption{Solution for $\phi_1(r)$}
         \end{subfigure}
         \begin{subfigure}[b]{0.3\textwidth}
                 \includegraphics[width=\textwidth]{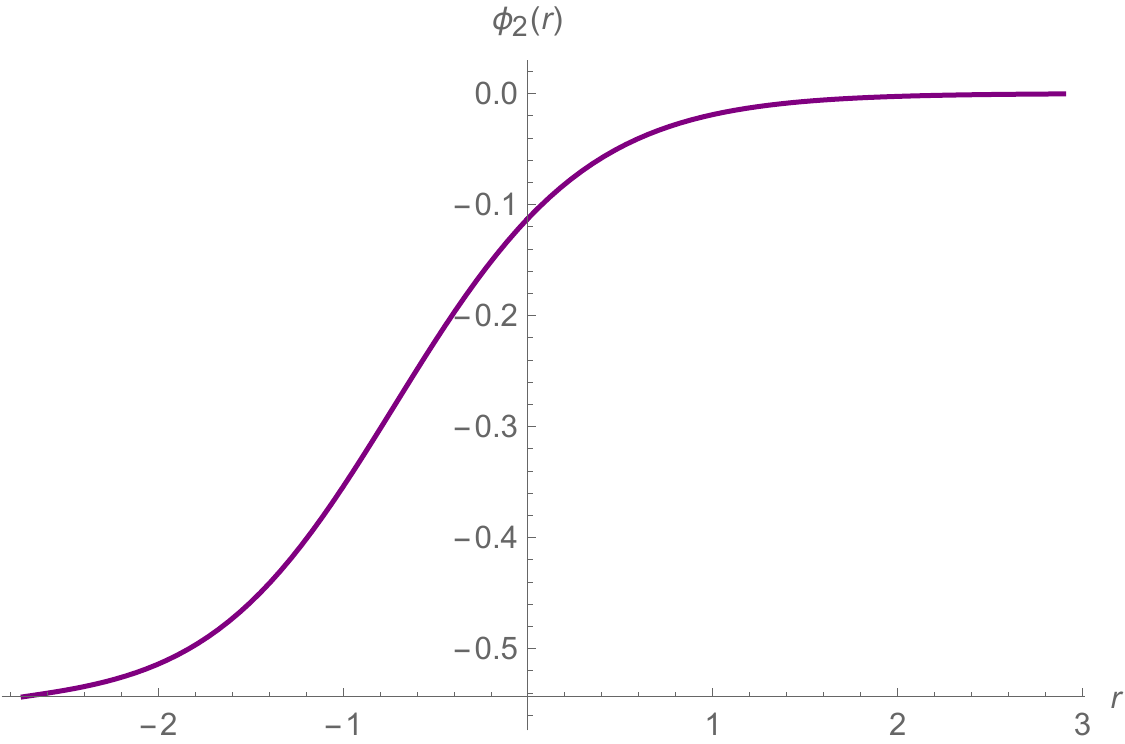}
                 \caption{Solution for $\phi_2(r)$}
         \end{subfigure}\\ 
         \begin{subfigure}[b]{0.3\textwidth}
                 \includegraphics[width=\textwidth]{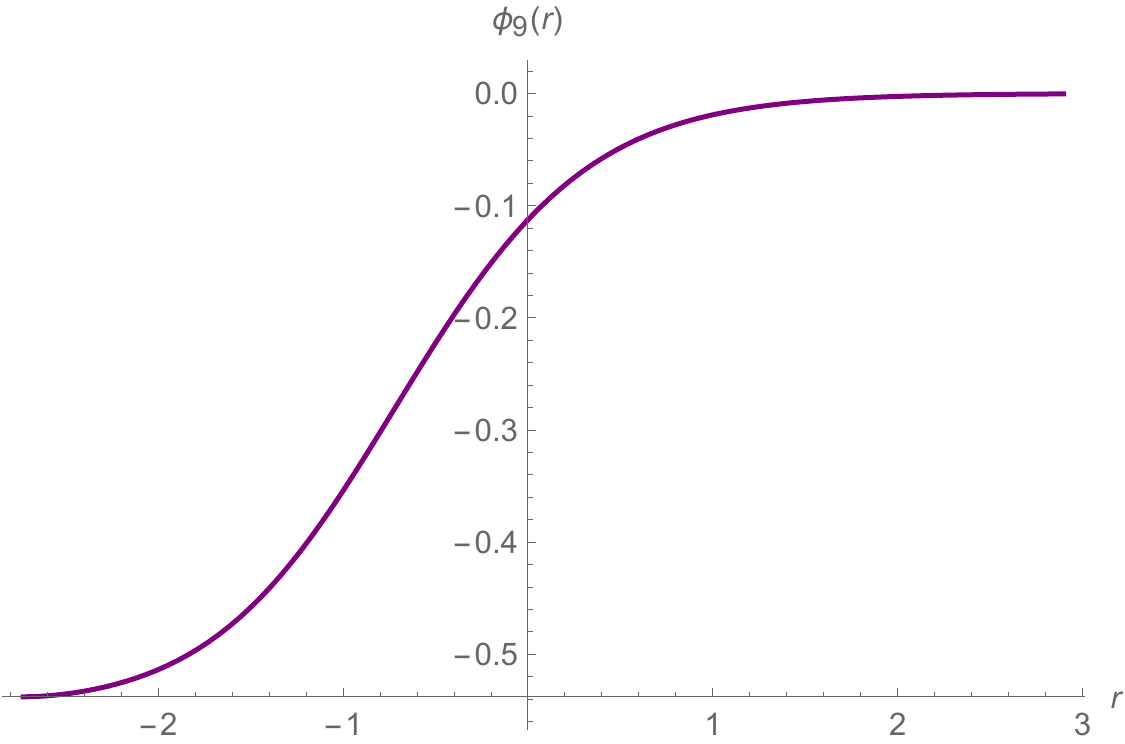}
                 \caption{Solution for $\phi_9$}
         \end{subfigure} 
         \begin{subfigure}[b]{0.3\textwidth}
                 \includegraphics[width=\textwidth]{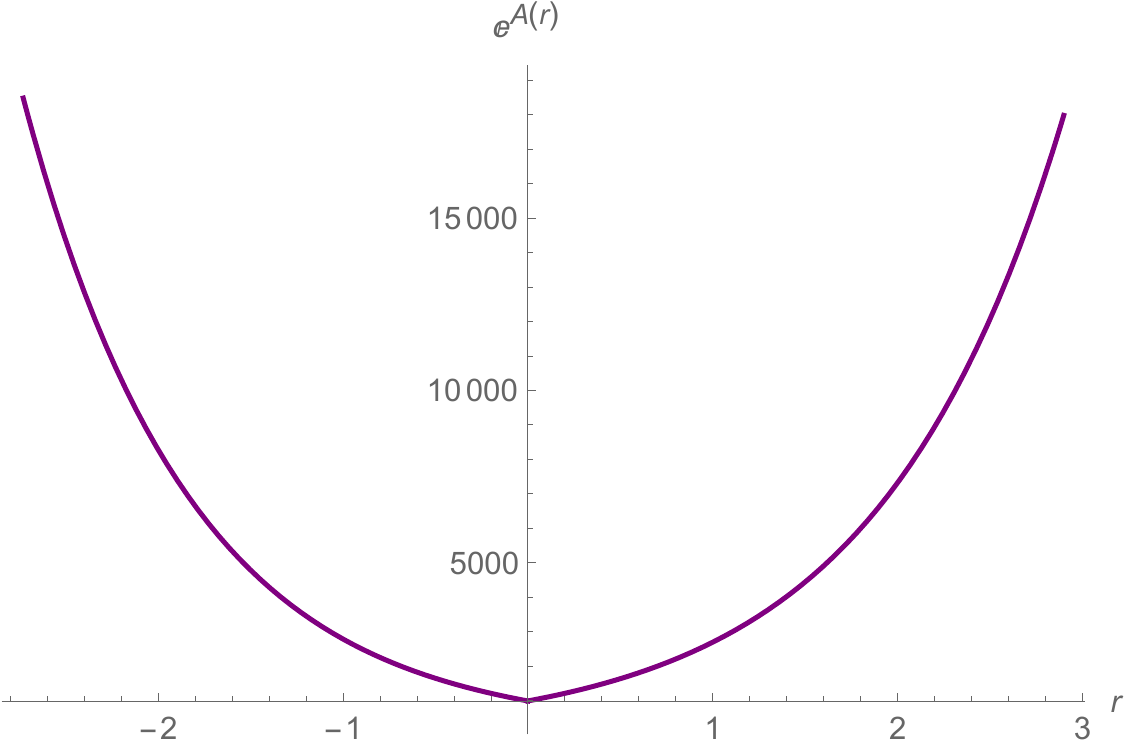}
                 \caption{Solution for $e^{A(r)}$}
         \end{subfigure}
         \begin{subfigure}[b]{0.3\textwidth}
                 \includegraphics[width=\textwidth]{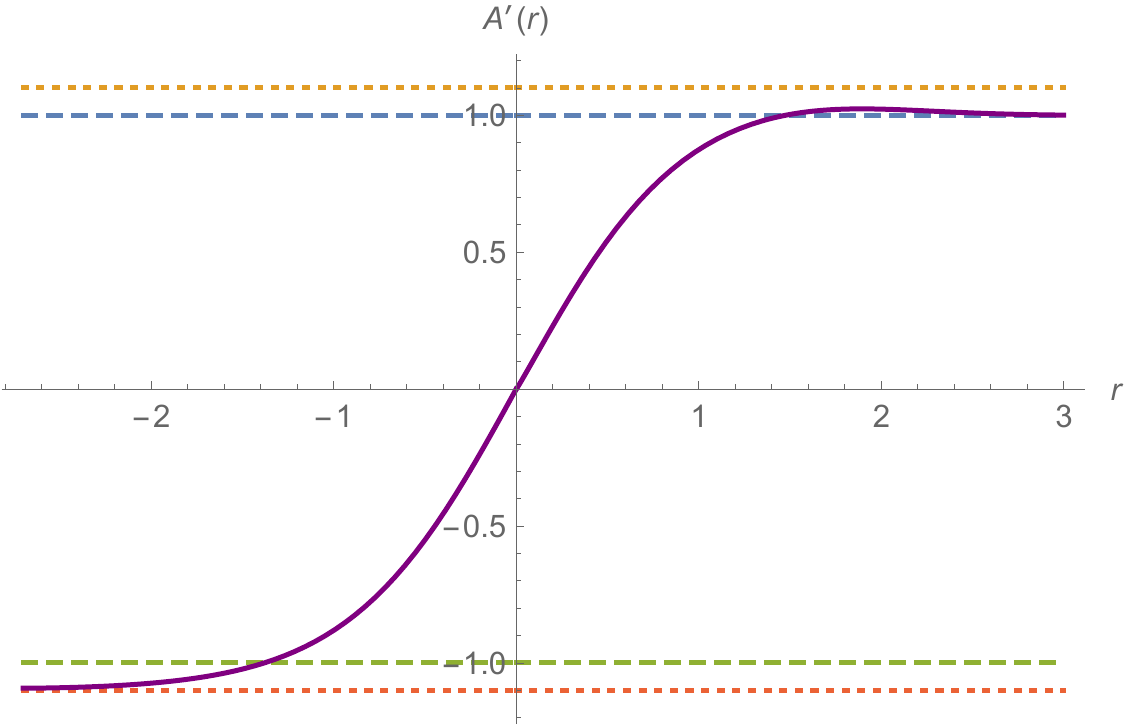}
                 \caption{Solution for $A'(r)$}
         \end{subfigure}
         \caption{An example of RG-flow interfaces interpolating between $N=4$ $AdS_5$ vacua given by critical points I and II with $\ell=1$, $\kappa=-1$, $h_1=2$, $g_1=-\sqrt{2}$ and $h_2=4$.}\label{fig2}
 \end{figure}

In this section, we consider Janus solutions in $N=4$ gauged supergravity with $SO(2)\times SO(3)\times SO(3)$ gauge group. This gauge group can be embedded in a smaller global symmetry $SO(5,3)$ arising from $N=4$ supergravity coupled to three vector multiplets. Accordingly, we will also truncate out all scalar fields from the remaining two vector multiplets namely $\phi_3$ and $\phi_4$ for simplicity. The corresponding BPS equations can be obtained from the general analysis in the previous section by setting $g_2=0$ and $\phi_3=\phi_4=0$. With this truncation, various expressions simplify considerably. 
\\
\indent The BPS equations take the form
\begin{eqnarray} 
\Sigma'&=&\eta\sqrt{\mc{A}_1^2+\mc{A}_2^2},\qquad \cosh\phi_2\phi_1'=\eta\frac{\textrm{Im}(\mc{A}\mc{B})}{|\mc{A}|},\qquad \phi_2'=\eta\frac{\textrm{Re}(\mc{A}\mc{B})}{|\mc{A}|},\nonumber \\
\phi_9'&=&\eta\frac{\textrm{Re}(\mc{A}^*\mc{D})}{|\mc{A}|},\qquad A'=\eta\frac{\textrm{Re}(\mc{A}^*\mc{W})}{|\mc{A}|}
\end{eqnarray}
together with two algebraic constraints
\begin{eqnarray}
\Sigma^3&=&-\frac{h_1h_2}{\sqrt{2}g_1(h_1\sinh\phi_9+h_2\cosh\phi_9)},\\
\textrm{and}\quad \frac{\kappa}{\ell}e^{-A}&=&-\frac{\eta}{3\sqrt{2}|\mc{A}|}g_1\Sigma^2\cosh^2\phi_2\sinh2\phi_1\sinh\phi_2(h_1\cosh\phi_9+h_2\sinh\phi_9).\nonumber \\
& & \label{con_trun2}
\end{eqnarray} 
All the functions appearing in the BPS equations are given by
\begin{eqnarray}
\mc{A}_1&=&\frac{1}{3}\sinh2\phi_1\sinh\phi_2(h_2\sinh\phi_9-\sqrt{2}g_1\Sigma^3),\nonumber \\
\mc{A}_2&=&\frac{1}{6}h_2\sinh\phi_9(\sinh^2\phi_1+\cosh2\phi_1-\cosh^2\phi_1\cosh2\phi_2)\nonumber \\
& &-\frac{1}{3}h_1\cosh^2\phi_2\cosh\phi_9-\frac{1}{6\sqrt{2}}g_1\Sigma^3(1+3\cosh2\phi_1-2\cosh2\phi_2\sinh^2\phi_1),\nonumber \\
\mc{B}_1&=&\frac{1}{4}\Sigma^{-1}\cosh\phi_2\sinh2\phi_1(2h_2\sinh\phi_9-\sqrt{2}g_1\Sigma^3),\nonumber \\
\mc{B}_2&=&\frac{1}{4}\Sigma^{-1}\sinh2\phi_2(2h_1\cosh\phi_9+2h_2\cosh^2\phi_1\sinh\phi_9-\sqrt{2}g_1\Sigma^3\sinh^2\phi_1),\nonumber \\
\mc{D}_1&=&-2h_2\Sigma^{-1}\sinh\phi_1\cosh\phi_1\sinh\phi_2,\nonumber \\
\mc{D}_2&=&h_1\Sigma^{-1}\cosh^2\phi_2\sinh\phi_9+h_2\Sigma^{-1}\cosh\phi_9(\cosh^2\phi_1\sinh^2\phi_2-\sinh^2\phi_1),\nonumber
\end{eqnarray} 
\begin{eqnarray}
\mc{W}_1&=&\frac{1}{6}\Sigma^{-1}\sinh2\phi_1\sinh\phi_2(\sqrt{2}g_1\Sigma^3+2h_2\sinh\phi_9),\nonumber \\
\mc{W}_2&=&\frac{1}{24}\Sigma^{-1}\left[4h_2\sinh^2\phi_1\sinh\phi_9+\sqrt{2}g_1\Sigma^{3}(1+\cosh2\phi_1)\right]\nonumber \\
& &+\frac{1}{12}\Sigma^{-1}\left[2h_2\sinh\phi_9(\cosh2\phi_1-\cosh^2\phi_1\cosh2\phi_2)-4h_1\cosh^2\phi_2\cosh\phi_9\right.\nonumber \\
& &\left.+\sqrt{2}g_1\Sigma^3(\cosh2\phi_1-\cosh2\phi_2\sinh^2\phi_1)\right]. 
\end{eqnarray}
From the algebraic constraint \eqref{con_trun2}, we immediately see that both $\phi_1$ and $\phi_2$ must be non-vanishing in order for Janus solutions in the form of $AdS_4$-sliced domain walls to exist. For $\phi_1=0$ and $\phi_9=\phi_2$, the scalar coset representative reduces to that studied in \cite{5D_N4_flow} with $SO(2)\times SO(3)_{\textrm{diag}}$ symmetry. In this case, there exist only a flat domain wall interpolating between the two $N=4$ $AdS_5$ vacua. Accordingly, no supersymmetric Janus solutions with $SO(2)\times SO(3)_{\textrm{diag}}$ symmetry exist. Furthermore, as previously mentioned, a truncation of $\phi_9$ cannot be consistently performed. Therefore, the BPS equations cannot be further simplified.  
\\
\indent  From scalar masses given in \cite{5D_N4_flow}, near critical point I, $\phi_2$ and $\phi_9$ are dual to operators of dimensions $\Delta=2$. The dilaton $\Sigma$ is also dual to a dimension-$2$ operator. As pointed out above, the existence of supersymmetric Janus solutions requires non-vanishing $\phi_1$. This scalar is dual to an operator of dimension $\Delta=3$. According to the AdS/CFT correspondence reviewed previously, the conformal interfaces should arise from a position-dependent deformation associated with a dimension-$3$ operator dual to $\phi_1$. By solving the linearized BPS equations as in \cite{6D_Janus}, we find that near critical point I
\begin{eqnarray}
& &\Sigma\sim C_\Sigma e^{-\frac{2r}{L_{\textrm{I}}}}-\frac{2}{L_{\textrm{I}}}C_1^2 r e^{-\frac{2r}{L_{\textrm{I}}}},\qquad \phi_1\sim C_1e^{-\frac{r}{L_{\textrm{I}}}},\nonumber \\
& &\phi_2\sim C_2e^{-\frac{2r}{L_{\textrm{I}}}}-\frac{2}{3}\frac{C_1^2}{L_{\textrm{I}}}re^{-\frac{2r}{L_{\textrm{I}}}},\qquad \phi_9\sim \frac{1}{2}h_2 L_{\textrm{I}}(3C_\Sigma-C_1^2)e^{-\frac{2r}{L_{\textrm{I}}}}
\end{eqnarray} 
with $C_{\Sigma,1,2}$ being constants. We have also chosen the $AdS_5$ critical point to appear as $r\rightarrow \infty$. From these equations, we see that there is a source term for a dimension-3 operator dual to $\phi_1$ corresponding to the constant $C_1$. Similarly, $C_2$ and $C_\Sigma$ are related to vacuum expectation values of operators of dimension $\Delta=2$ dual to $\phi_2$ and $\Sigma$. In addition, we note that these two operators also have a source term given by $C_1$. In the case of holographic RG flows studied in \cite{5D_N4_flow_Davide} and \cite{5D_N4_flow}, there is no deformation corresponding to $\phi_1$, see also \eqref{con_trun2} in which $\phi_1=0$ leads to a flat domain wall with $\ell\rightarrow \infty$. In that case, both of the operators dual to $\Sigma$ and $\phi_2$ do not have source terms, so the RG flows are driven by vacuum expectation values. We then see that turning on a deformation corresponding to $C_1$ leads to a conformal interface within the dual $N=2$ SCFT. On the other hand, there is no source term for an operator of dimension $\Delta=2$ dual to $\phi_9$, but this operator acquires a vacuum expectation value related to $C_\Sigma$ and $C_1$.  
\\
\indent Near critical point II, the leading behaviors of scalar fields are given as follows
\begin{eqnarray}
& &\Sigma\sim \tilde{C}_\Sigma e^{-\frac{2r}{L_{\textrm{II}}}},\qquad \phi_1\sim \tilde{C}_1e^{\frac{r}{L_{\textrm{II}}}},\nonumber \\
& &\phi_2-\phi_9\sim \alpha e^{-\frac{4r}{L_{\textrm{II}}}},\qquad \phi_9+2\phi_2\sim \beta e^{\frac{2r}{L_{\textrm{II}}}}
\end{eqnarray}
with constants $\tilde{C}_\Sigma$, $\tilde{C}_{1}$, $\alpha$ and $\beta$. We now see that the operator dual to $\phi_1$ becomes irrelevant with dimension $\Delta=5$ while $\phi_2$ and $\phi_9$ are dual to two linear combinations of a marginal ($\Delta=4$) and an irrelevant ($\Delta=6$) operators. In this case, there are source terms of operators of dimensions $\Delta=5$ and $\Delta=6$ respectively dual to $\phi_1$ and a linear combination $2\phi_2+\phi_9$. There are also non-vanishing vacuum expectation values for a dimension-$2$ operator dual to $\Sigma$ together with a vacuum expectation value of a marginal operator ($\Delta=4$) dual to a linear combination $\phi_2-\phi_9$.  
\\
\indent To find numerical solutions, we will choose numerical values for various parameters as follows 
\begin{eqnarray}
h_1=2,\qquad \ell=1,\qquad\kappa=-1,\qquad h_2=2h_1\, .
\end{eqnarray}
Recall that the value of $g_1$ is fixed by the requirement that $\Sigma=1$ at the $N=4$ $AdS_5$ vacuum with $SO(2)\times SO(3)\times SO(3)$ symmetry, critical point I. The value of $h_1$ is related to the $AdS_5$ radius at critical point I as $L=\frac{2}{h_1}$. We have chosen the value of $h_1=2$ in order to set the $AdS_5$ radius at critical point I to unity. Without loss of generality, we can also choose to work with the unit-radius $AdS_4$, $\ell=1$. The sign choices of $\kappa=\pm1$ also lead to physically equivalent solutions. On the other hand, the value of $h_2$ is arbitrary, but the existence of critical point II implies that $h_2>h_1$ for $h_1>0$.     
\\
\indent With these numerical values, we find the values of scalar fields at the two $N=4$ $AdS_5$ vacua as follows
\begin{eqnarray}
\textrm{Critical point I}&:&\quad \phi_1=\phi_2=\phi_9=0,\qquad \Sigma=1,\nonumber \\ 
\textrm{Critical point II}&:&\quad \phi_1=0,\qquad \phi_2=\phi_9=-0.5493,\qquad \Sigma=1.0491\, .
\end{eqnarray}
For convenience, we also note 
\begin{equation}
\frac{1}{L_{\textrm{I}}}=1.0\qquad \textrm{and}\qquad \frac{1}{L_{\textrm{II}}}=1.1
\end{equation}
for $L_{\textrm{I}}$ and $L_{\textrm{II}}$ being the $AdS_5$ radii of critial points I and II, respectively.
\\
\indent Some examples of Janus solutions are shown in figure \ref{fig1}. As in similar solutions in other dimensions, there is a solution interpolating between critical point I on both sides shown by the red line in figure \ref{fig1}. This solution describes a conformal interface within $N=2$ SCFT dual to the $N=4$ $AdS_5$ vacuum with $SO(2)\times SO(3)\times SO(3)$ symmetry. There are also solutions that flow close to critical point II on both sides. An example of this solutions is given by the blue line in figure \ref{fig1}. By tuning the boundary condition at the turning point, the solutions can stay very close to critical point II longer and longer as shown by the green line. Following the interpretation given in \cite{warner_Janus}, we expect these solutions to describe conformal interfaces between $N=2$ SCFTs with $SO(2)\times SO(3)_{\textrm{diag}}$ symmetry dual to critical point II. These $N=2$ conformal phases on both sides of the interface arise from holographic RG flows from the $N=2$ conformal phase dual to critical point I. By fine-tuning the boundary condition, we find a solution describing an RG-flow interface between $N=2$ SCFTs dual to critical points I and II on each side of the interface. The solution is similar to that given in \cite{5D_N4_Janus} and shown in figure \ref{fig2}.    

%%%%%%%%%%%%%%%%%%%%%%%%%%%%%%%%%%%%%%%%%%%%%%%%%%%%%%%%%%%%%%%%%%%
\subsection{Janus solutions from $SO(2)_D\times SO(3)\times SO(3)$ gauge group} 
\begin{figure}
         \centering
               \begin{subfigure}[b]{0.45\textwidth}
                 \includegraphics[width=\textwidth]{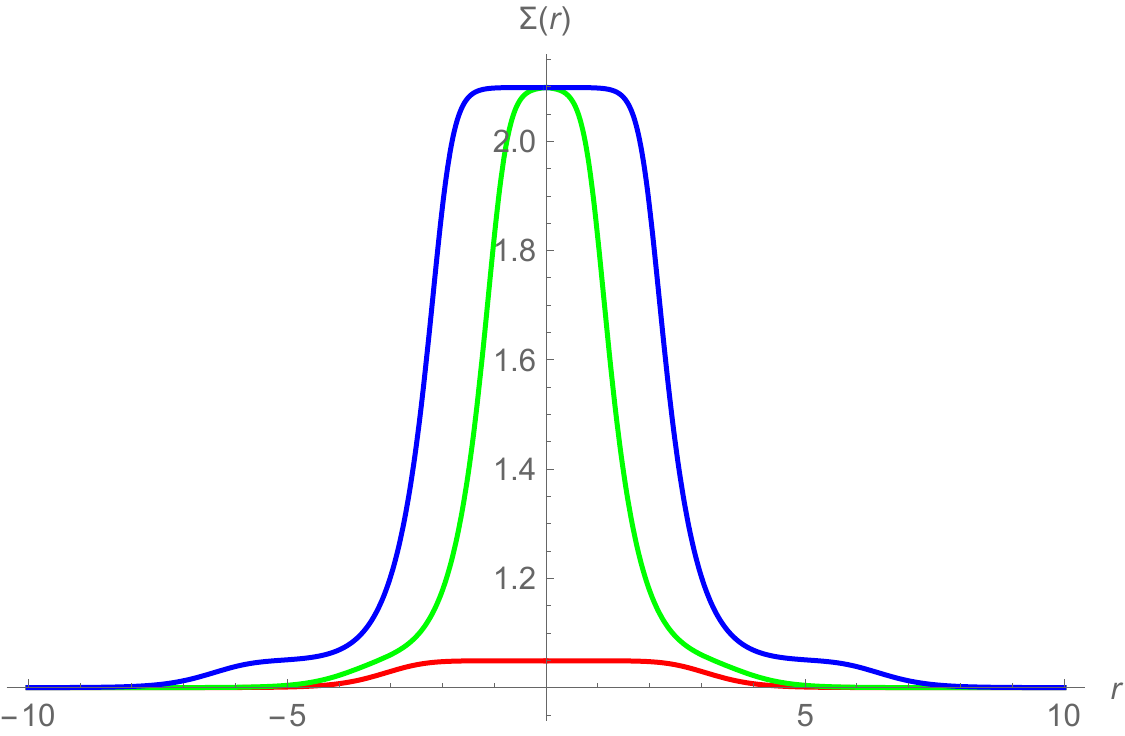}
                 \caption{Solutions for $\Sigma(r)$}
         \end{subfigure}
         \begin{subfigure}[b]{0.45\textwidth}
                 \includegraphics[width=\textwidth]{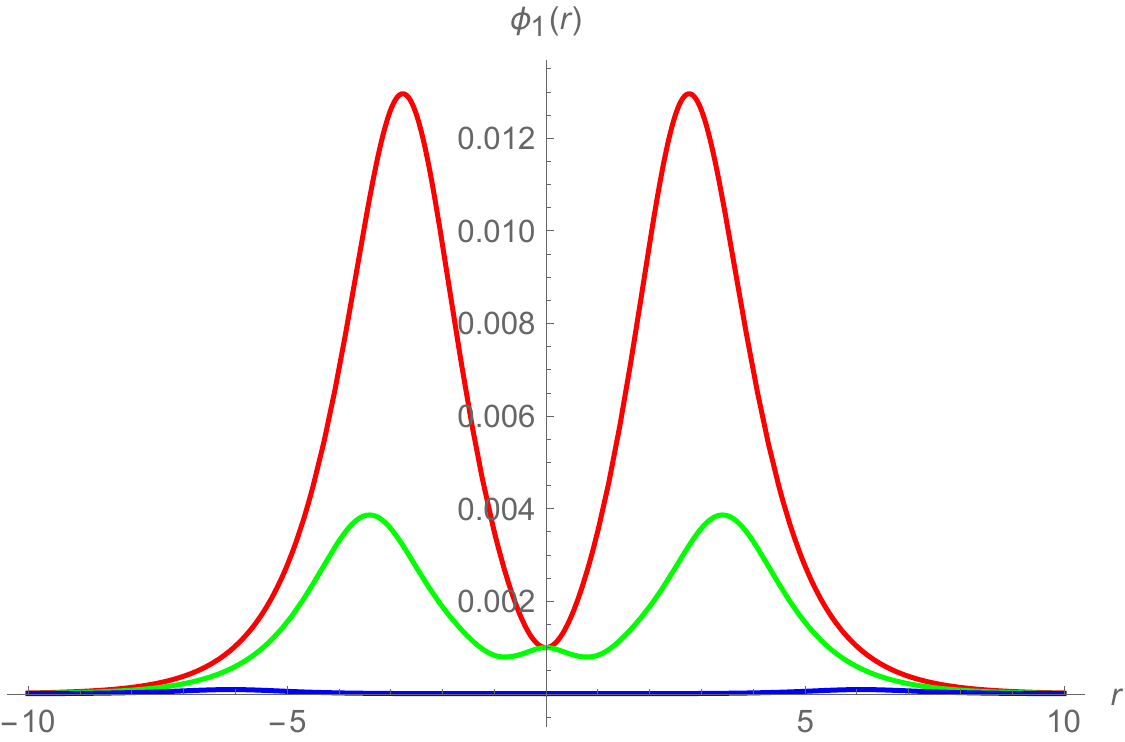}
                 \caption{Solutions for $\phi_1(r)$}
         \end{subfigure}\\
         \begin{subfigure}[b]{0.45\textwidth}
                 \includegraphics[width=\textwidth]{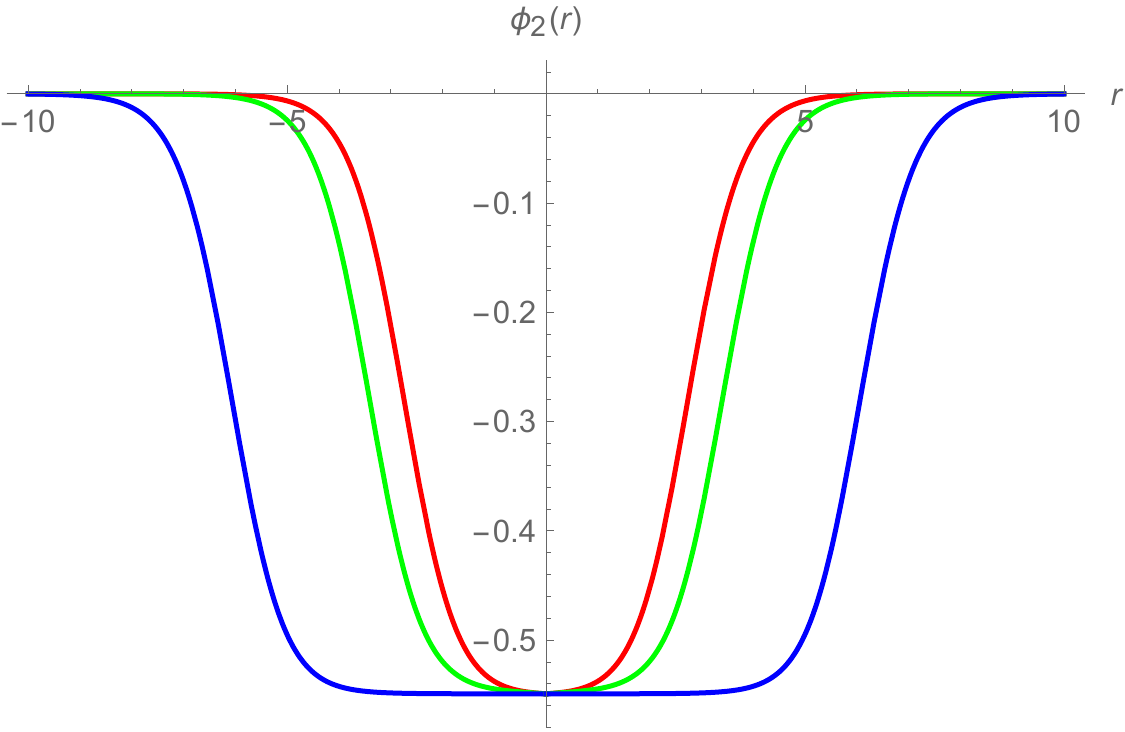}
                 \caption{Solutions for $\phi_2(r)$}
         \end{subfigure}
         \begin{subfigure}[b]{0.45\textwidth}
                 \includegraphics[width=\textwidth]{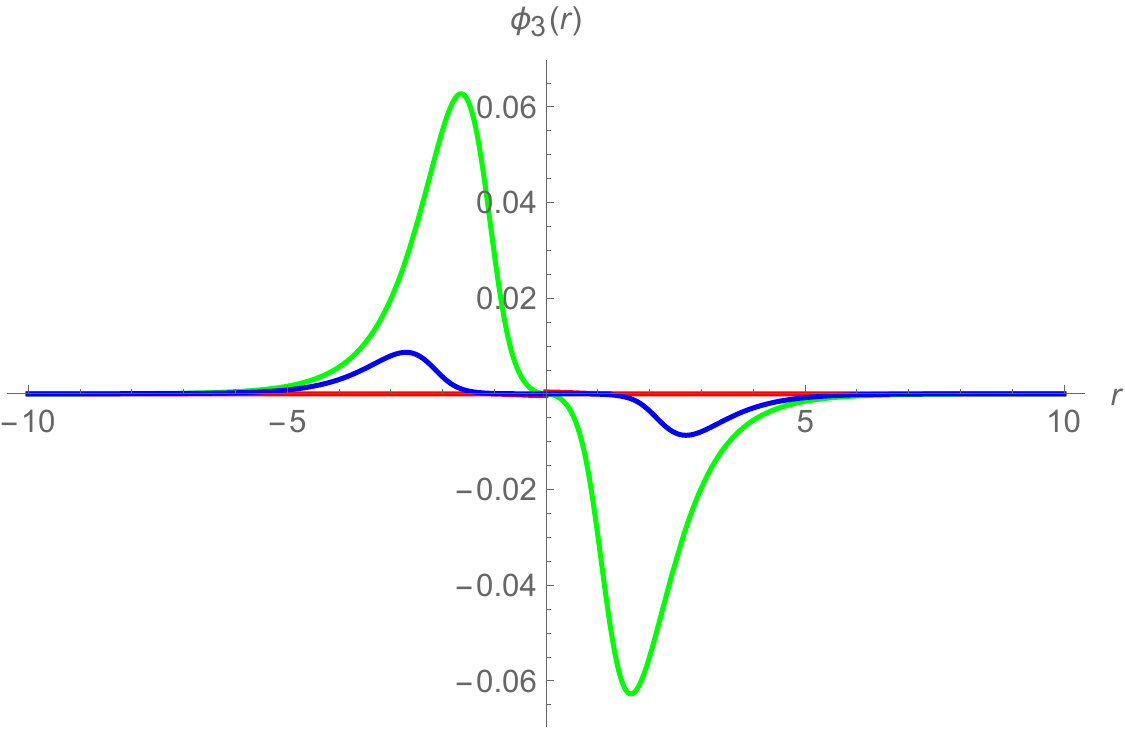}
                 \caption{Solutions for $\phi_3(r)$}
         \end{subfigure}\\
         \begin{subfigure}[b]{0.45\textwidth}
                 \includegraphics[width=\textwidth]{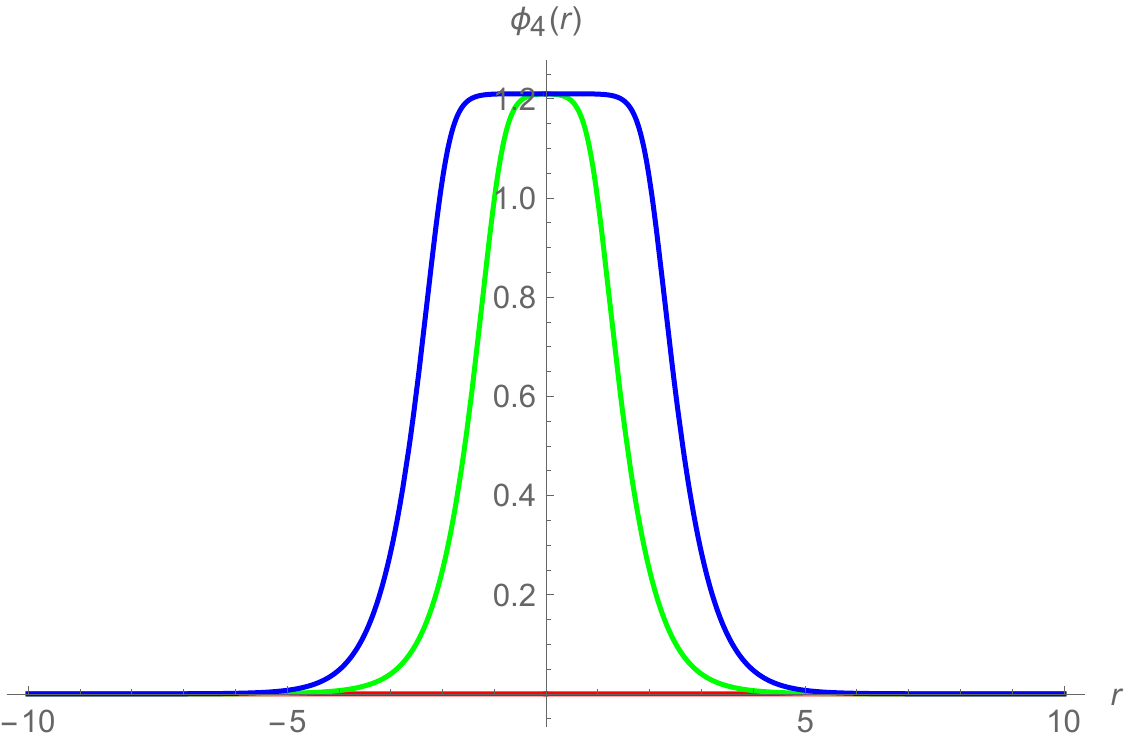}
                 \caption{Solutions for $\phi_4(r)$}
         \end{subfigure}
         \begin{subfigure}[b]{0.45\textwidth}
                 \includegraphics[width=\textwidth]{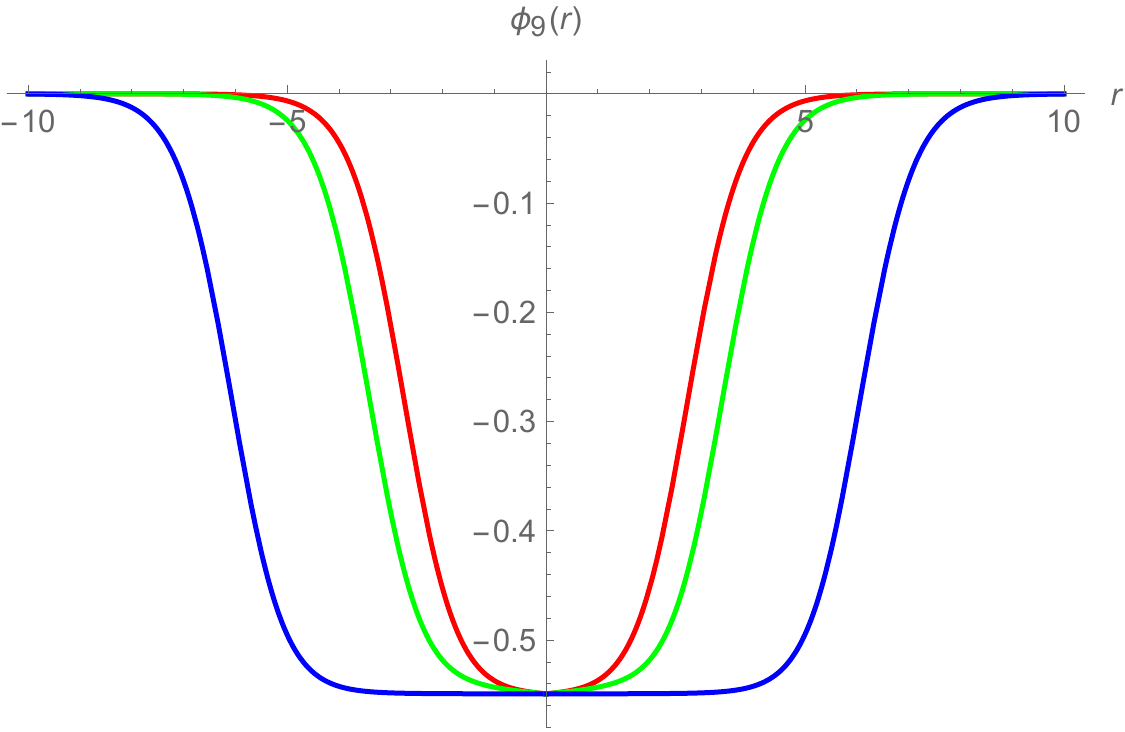}
                 \caption{Solutions for $\phi_9(r)$}
         \end{subfigure}\\
         \begin{subfigure}[b]{0.45\textwidth}
                 \includegraphics[width=\textwidth]{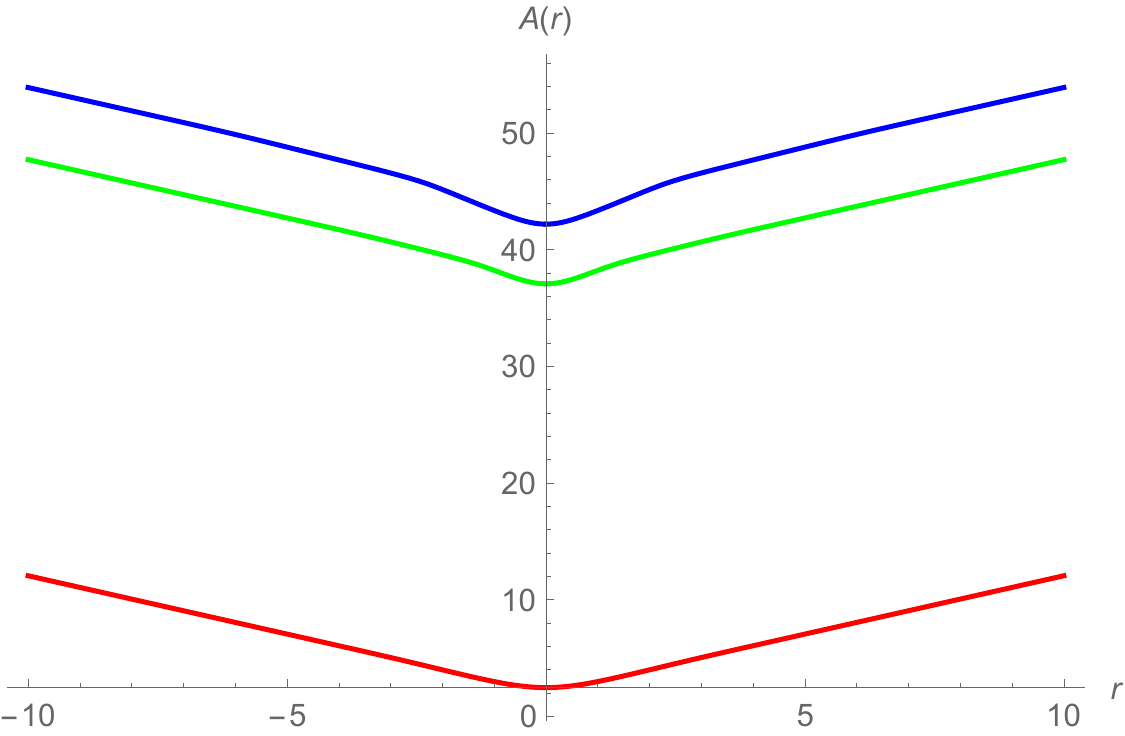}
                 \caption{Solutions for $A(r)$}
         \end{subfigure}
          \begin{subfigure}[b]{0.45\textwidth}
                 \includegraphics[width=\textwidth]{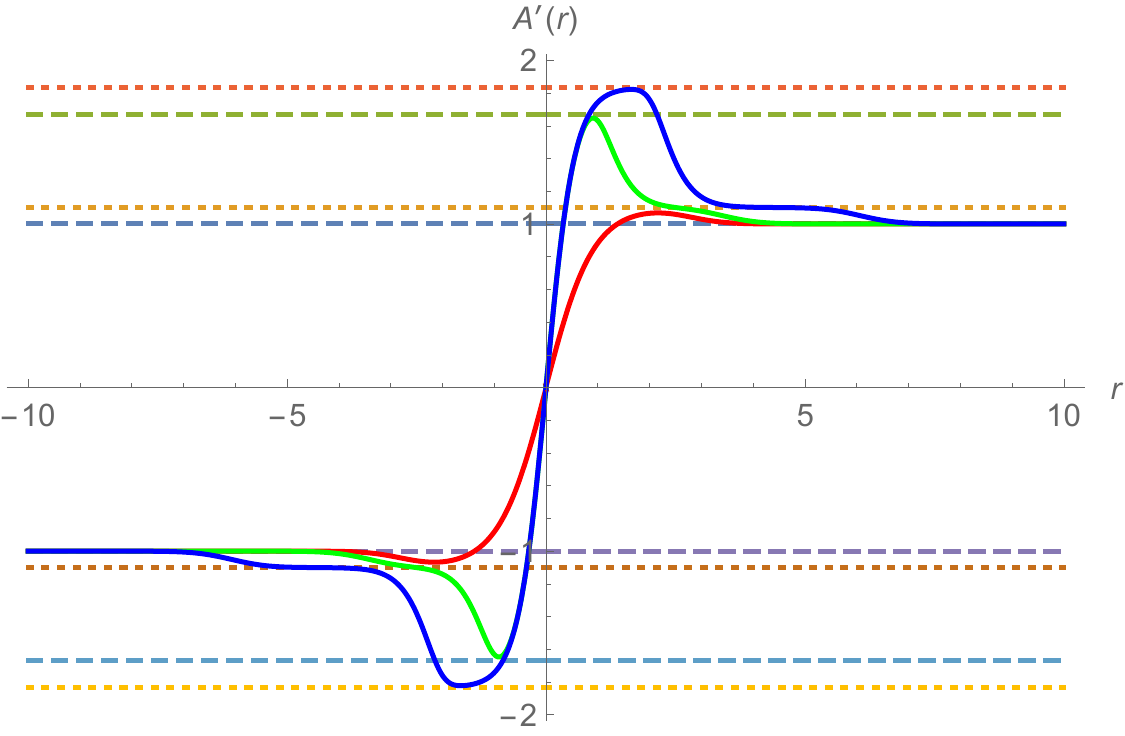}
                 \caption{Solutions for $A'(r)$}
         \end{subfigure}
         \caption{Examples of supersymmetric Janus solutions interpolating between $AdS_5$ vacua given by critical point II (red), III (green) and IV (blue).}\label{fig3}
 \end{figure}

 We now consider supersymmetric Janus solutions from $SO(2)_D\times SO(3)\times SO(3)$ gauge group. In this case, there are four supersymmetric $AdS_5$ vacua with $N=4$ and $N=2$ supersymmetries. In the numerical analysis, we will use the numerical values of various parameters as follows
\begin{eqnarray}
& &g_1=-\frac{h_1}{\sqrt{2}},\qquad h_2=2h_1,\qquad g_2=-\frac{1}{4}g_1,\nonumber \\
& &h_1=2,\qquad \ell=1,\qquad \kappa=-1\, .
\end{eqnarray} 
As in the previous case, the value of the coupling constant $h_2$ is a free parameter. Similarly, the value of $g_2$ is also arbitrary. Both of these coupling constants are chosen in such a way that critical points II, III and IV exist.
\\
\indent With these numerical values, the locations of all the four $AdS_5$ critical points are given by
\begin{itemize}
\item Critical point I:
\begin{eqnarray}
 \phi_1=\phi_2=\phi_3=\phi_4=\phi_9=0,\qquad \Sigma=1,\qquad L_{\textrm{I}}=1\, .
\end{eqnarray}    
\item Critical point II:
\begin{eqnarray}
 \phi_1&=&\phi_3=\phi_4=0,\qquad  \phi_2=\phi_9=-0.5493,\nonumber \\
  \Sigma&=&1.049,\qquad L_{\textrm{II}}=0.9086\, .
\end{eqnarray}    
\item Critical point III:
\begin{eqnarray}
\phi_1&=&\phi_2=\phi_3=\phi_9=0,\qquad \phi_4=1.2099,\nonumber \\ 
\Sigma&=&2,\qquad L_{\textrm{III}}=0.6000\, .
\end{eqnarray}    
\item Critical point IV:
\begin{eqnarray}
 \phi_1&=&\phi_3=0,\qquad \phi_2=\phi_9=-0.5493,\qquad \phi_4=1.2099,\nonumber \\
 \Sigma&=&2.0982,\qquad L_{\textrm{IV}}=0.5451\, .
\end{eqnarray}    
\end{itemize}
It is also useful to note all the values of the inverse $AdS_5$ radii at each critical point
\begin{equation}
\frac{1}{L_{\textrm{I}}}=1,\qquad \frac{1}{L_{\textrm{II}}}=1.1006,\qquad\frac{1}{L_{\textrm{III}}}=1.6667,\qquad\frac{1}{L_{\textrm{IV}}}=1.8344 \, .
\end{equation}
\indent Since the BPS equations in this case are much more complicated than those considered in the previous truncation, we refrain from giving the corresponding asymptotic behaviors near all of the four supersymmetric $AdS_5$ critical points here. In particular, the scaling dimensions of the dual operators depend on the values of the ratio between $g_1$ and $g_2$. In addition, various scalar modes are mixed in a complicated manners at all the four $AdS_5$ vacua. However, before giving numerical Janus solutions, it is useful to note some information on the operators dual to various scalar fields appearing in the BPS equations, see also scalar masses at all of the four supersymmetric $AdS_5$ vacua given in \cite{5D_flowII}. The parameter $\rho$ appearing in \cite{5D_flowII} is defined by
\begin{equation}
\rho=-\frac{2g_1}{g_2}\, .
\end{equation}
For a particular choice of the above numerical values for $g_1$ and $g_2$, we find $\rho=8$. At $N=4$ $AdS_5$ critical point I, the dilaton together with $\phi_2$ and $\phi_9$ are dual to operators of dimension $2$ while $\phi_1$ is dual to a dimension-$3$ operator. With $\rho=8$, $\phi_3$ is dual to a relevant operator of dimension $\frac{11}{4}$, and $\phi_4$ is dual to an operator of dimension $\frac{9}{4}$. At critical point II, the dilaton is still dual to a dimension-$2$ operator while $\phi_2$ and $\phi_9$ are dual to two combinations of a marginal operator of dimension $4$ and an irrelevant operator of dimension $6$. $\phi_1$ is also dual to an irrelevant operator of dimension $5$. On the other hand, $\phi_3$ and $\phi_4$ are still dual to operators of dimentions $\frac{11}{4}$ and $\frac{9}{4}$, respectively. 
\\
\indent At $N=2$ $AdS_5$ vacua, things are even more complicated due to the mixing of scalar fields in the mass matrices. At critical point III, the dilaton $\Sigma$ and $\phi_4$ are dual to two combinations of operators of dimensions $1+\sqrt{\frac{89}{5}}$ and $3+\sqrt{\frac{89}{5}}$. $\phi_1$ is dual to a dimension-$\frac{12}{5}$ operator while $\phi_2$ and $\phi_9$ are dual to two combinations of operators of dimensions $2$ and $\frac{17}{5}$. Finally, $\phi_3$ is dual to an operator of dimension $2+\sqrt{\frac{89}{5}}$. At critical point IV, $\Sigma$ and $\phi_4$ are again dual to two combinations of operators of dimensions $1+\sqrt{\frac{89}{5}}$ and $3+\sqrt{\frac{89}{5}}$. Similarly, $\phi_2$ and $\phi_9$ are dual to two combinations of operators of dimensions $1+\sqrt{\frac{17}{5}}$ and $3+\sqrt{\frac{17}{5}}$. $\phi_1$ and $\phi_3$ are dual respectively to operators of dimensions $\frac{13}{5}$ and $2+\sqrt{\frac{89}{5}}$.  
\\
\indent With four supersymmetric $AdS_5$ vacua, there are so many possible Janus solutions that we will not aim to give an exhaustive list. We will simply give some representative examples of interesting solutions that are different from the other settings with one or two $AdS_5$ vacua. There are a number of solutions describing conformal interfaces with holographic RG flows from critical point I to critical points II, III or IV on both sides of the interfaces. Examples of these solutions are shown in figure \ref{fig3}. The red line represents a solution describing an interface between $N=2$ SCFTs dual to critical point II on both sides. These SCFTs are in turn obtained from a holographic RG flow from the $N=2$ SCFT dual to $AdS_5$ critical point I. Similarly, solutions describing conformal interfaces between $N=1$ SCFTs dual to $AdS_5$ critical points III and IV are shown by the green and blue lines, respectively. The $N=1$ SCFTs on both sides are similarly generated by holographic RG flows from critical point I and II to critical point III or IV.  

\begin{figure}
         \centering
               \begin{subfigure}[b]{0.45\textwidth}
                 \includegraphics[width=\textwidth]{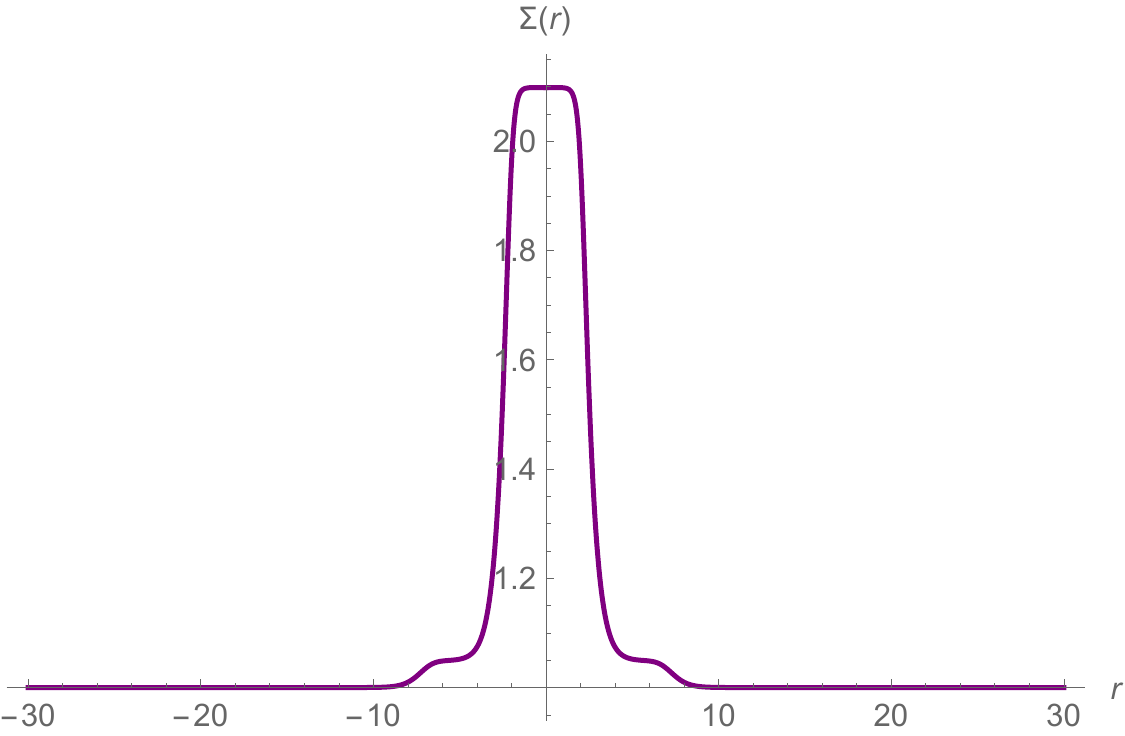}
                 \caption{Solution for $\Sigma(r)$}
         \end{subfigure}
         \begin{subfigure}[b]{0.45\textwidth}
                 \includegraphics[width=\textwidth]{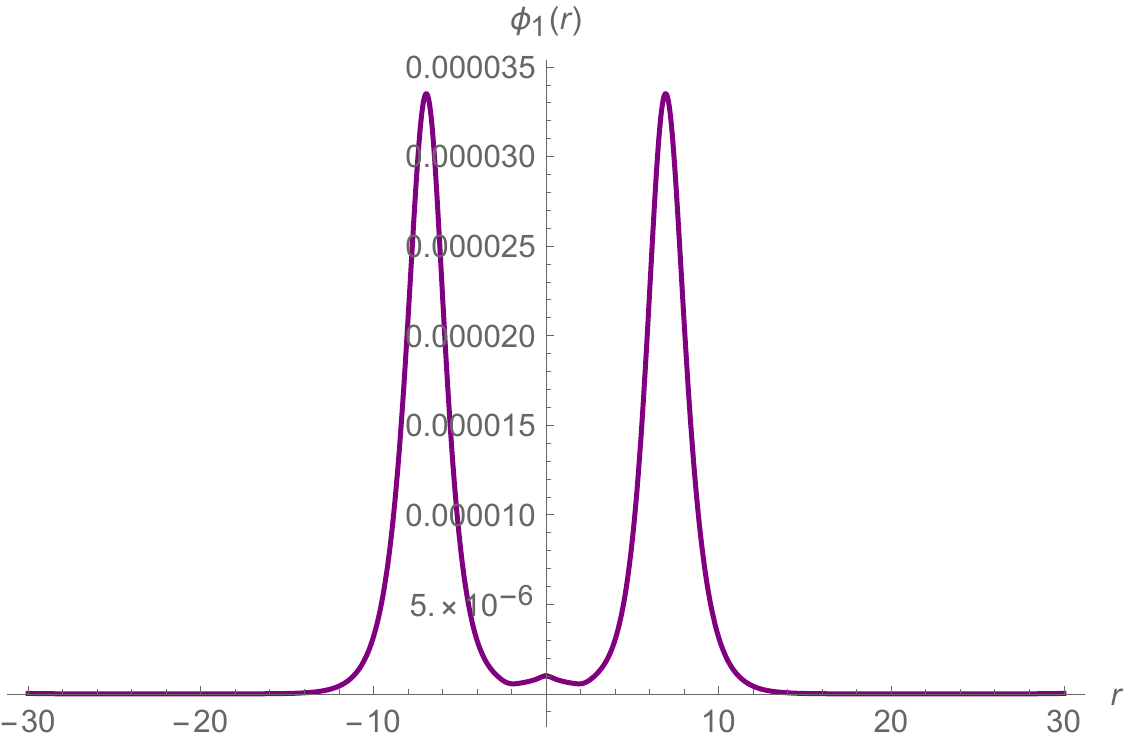}
                 \caption{Solution for $\phi_1(r)$}
         \end{subfigure}\\
         \begin{subfigure}[b]{0.45\textwidth}
                 \includegraphics[width=\textwidth]{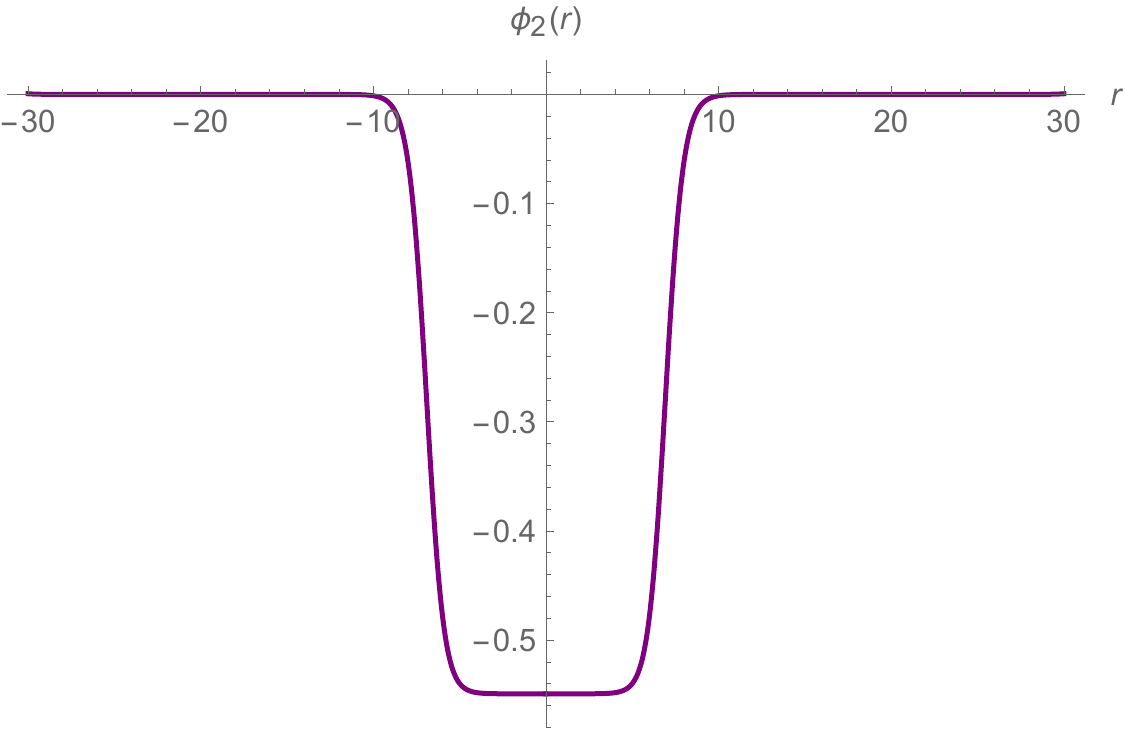}
                 \caption{Solution for $\phi_2(r)$}
         \end{subfigure}
         \begin{subfigure}[b]{0.45\textwidth}
                 \includegraphics[width=\textwidth]{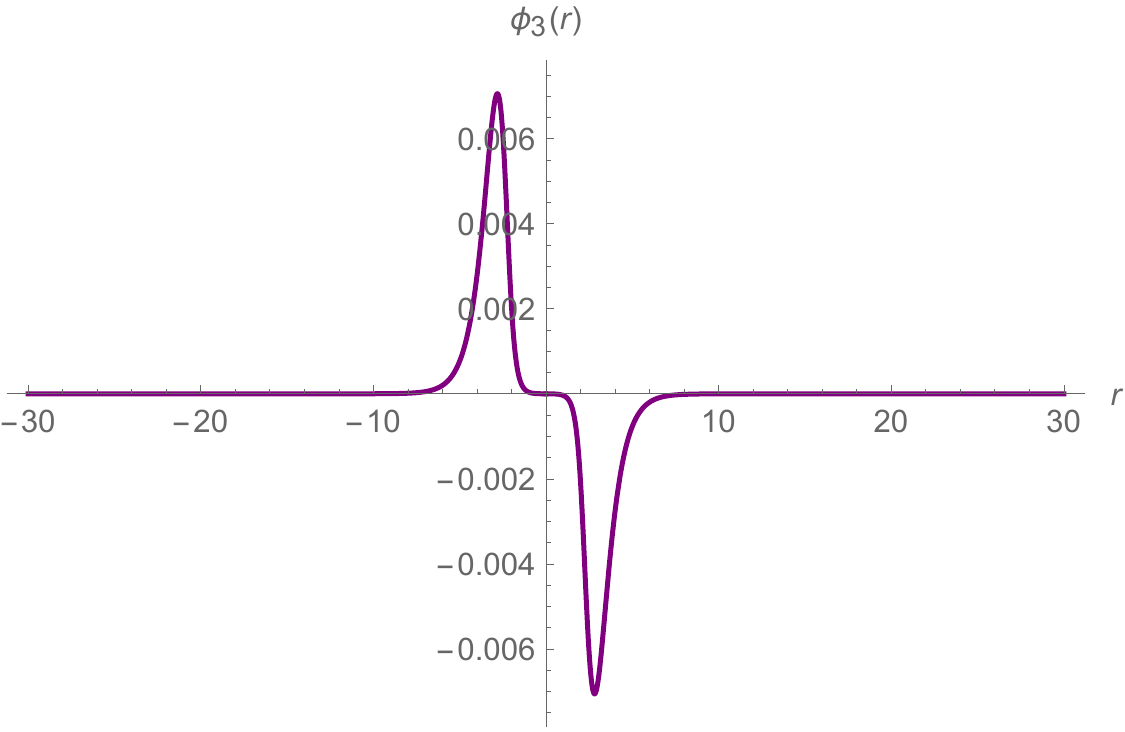}
                 \caption{Solution for $\phi_3(r)$}
         \end{subfigure}\\
         \begin{subfigure}[b]{0.45\textwidth}
                 \includegraphics[width=\textwidth]{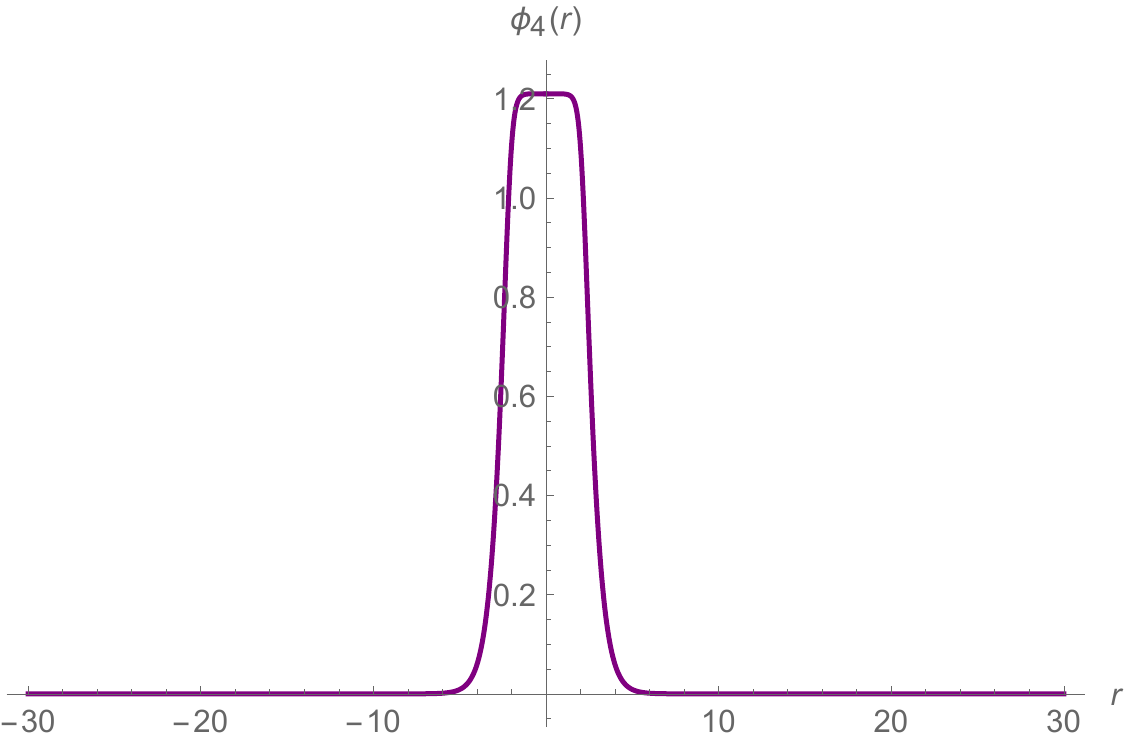}
                 \caption{Solution for $\phi_4(r)$}
         \end{subfigure}
         \begin{subfigure}[b]{0.45\textwidth}
                 \includegraphics[width=\textwidth]{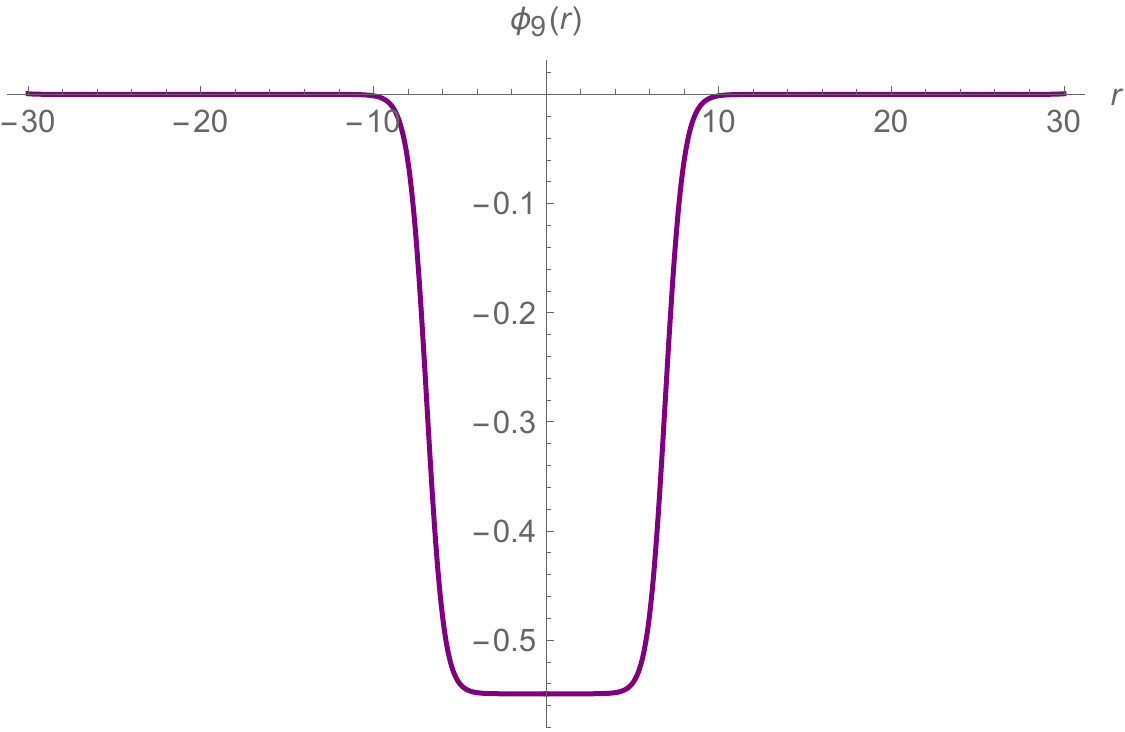}
                 \caption{Solution for $\phi_9(r)$}
         \end{subfigure}\\
         \begin{subfigure}[b]{0.45\textwidth}
                 \includegraphics[width=\textwidth]{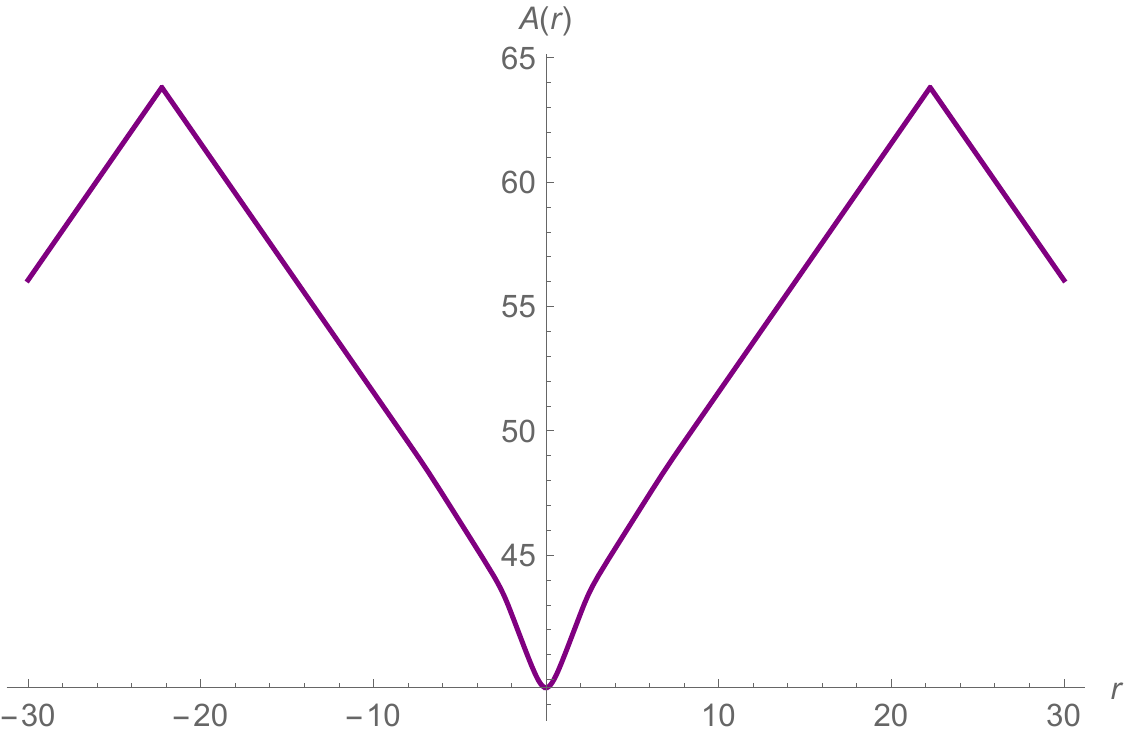}
                 \caption{Solution for $A(r)$}
         \end{subfigure}
          \begin{subfigure}[b]{0.45\textwidth}
                 \includegraphics[width=\textwidth]{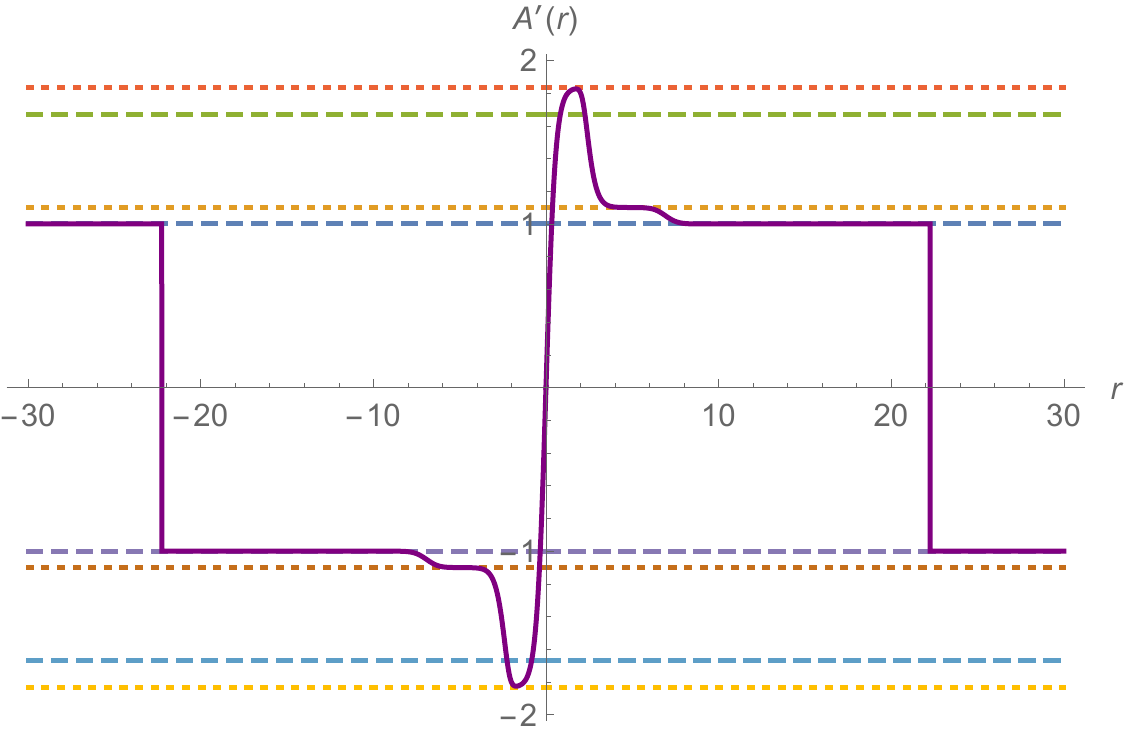}
                 \caption{Solution for $A'(r)$}
         \end{subfigure}
         \caption{An example of multi-Janus interfaces interpolating between $AdS_5$ vacua given by critical points I, II, III and IV. This solution describes three conformal interfaces, two interfaces between critical point I on the right and left of the origin and one interface between critical point IV at the origin.}\label{fig4}
 \end{figure}

\begin{figure}
         \centering
               \begin{subfigure}[b]{0.45\textwidth}
                 \includegraphics[width=\textwidth]{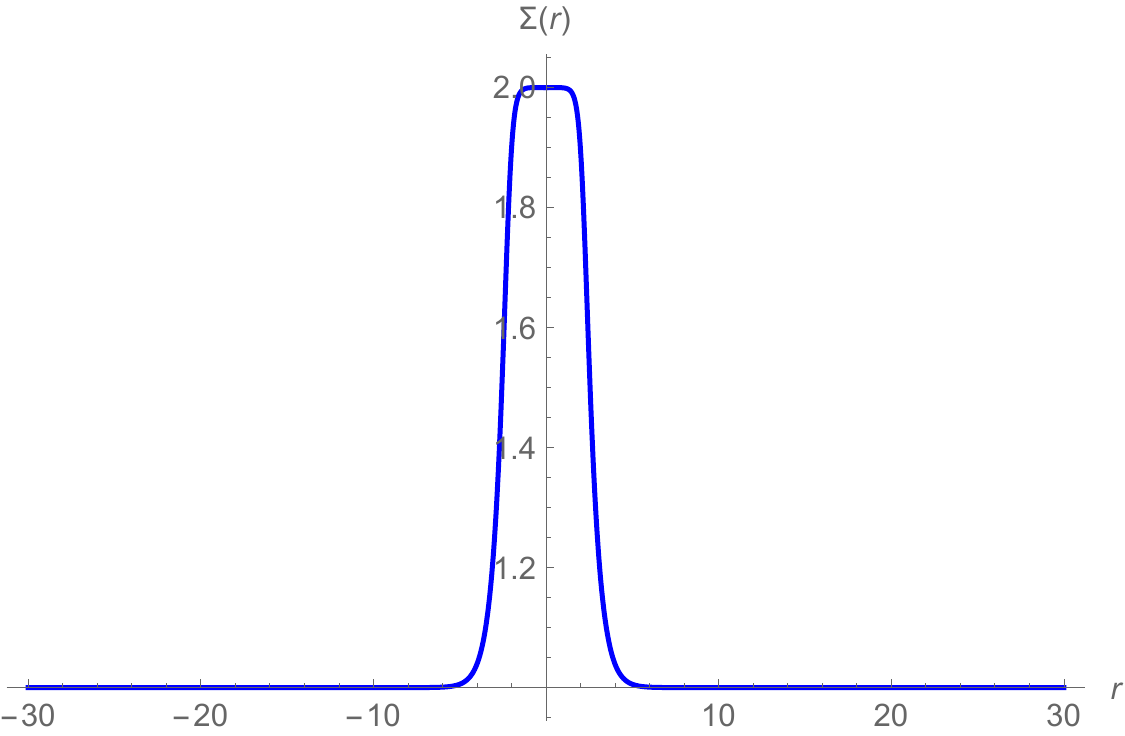}
                 \caption{Solution for $\Sigma(r)$}
         \end{subfigure}
         \begin{subfigure}[b]{0.45\textwidth}
                 \includegraphics[width=\textwidth]{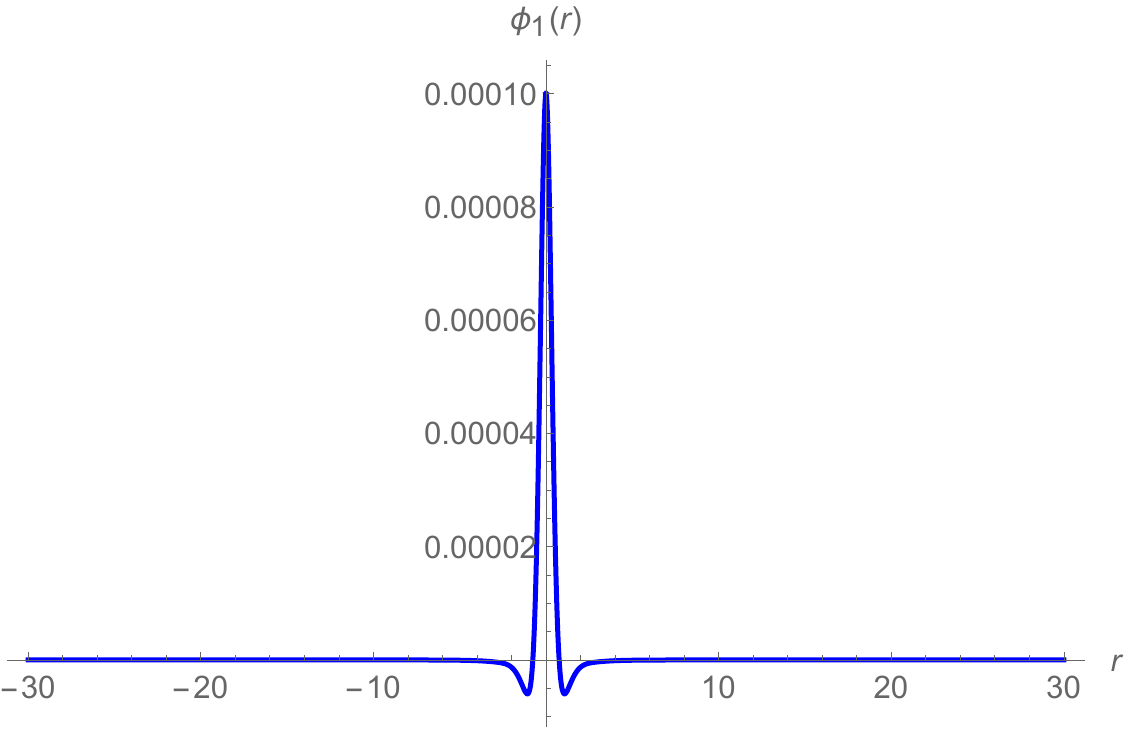}
                 \caption{Solution for $\phi_1(r)$}
         \end{subfigure}\\
         \begin{subfigure}[b]{0.45\textwidth}
                 \includegraphics[width=\textwidth]{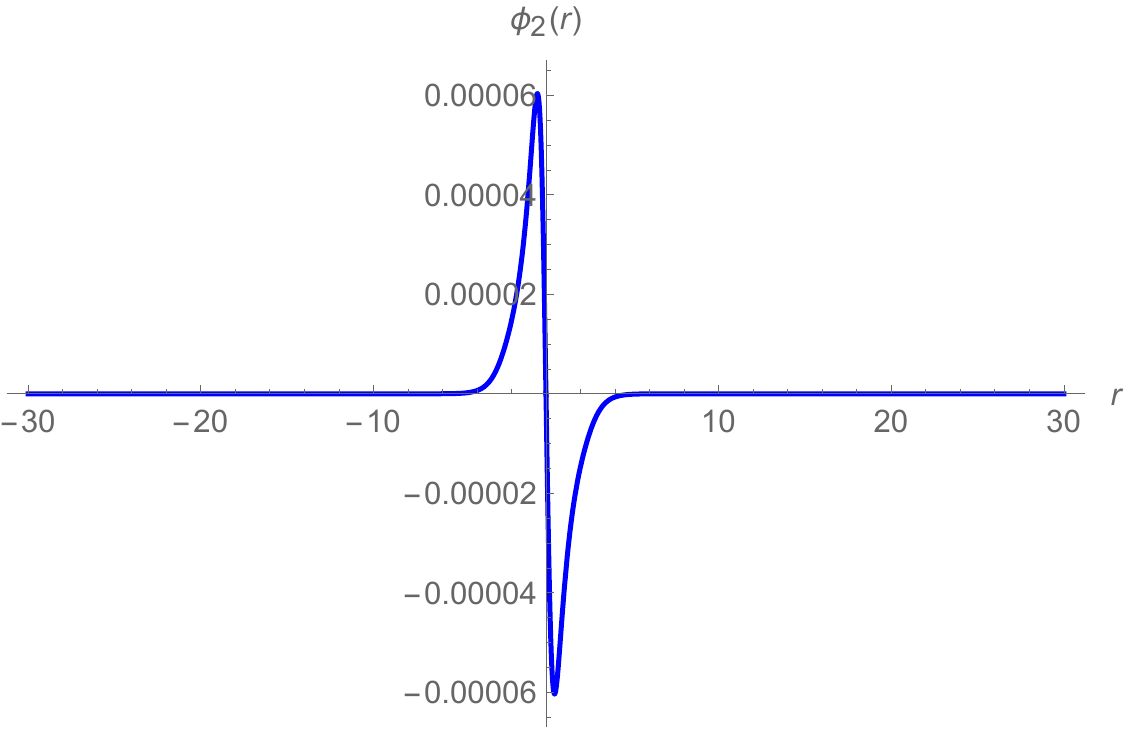}
                 \caption{Solution for $\phi_2(r)$}
         \end{subfigure}
         \begin{subfigure}[b]{0.45\textwidth}
                 \includegraphics[width=\textwidth]{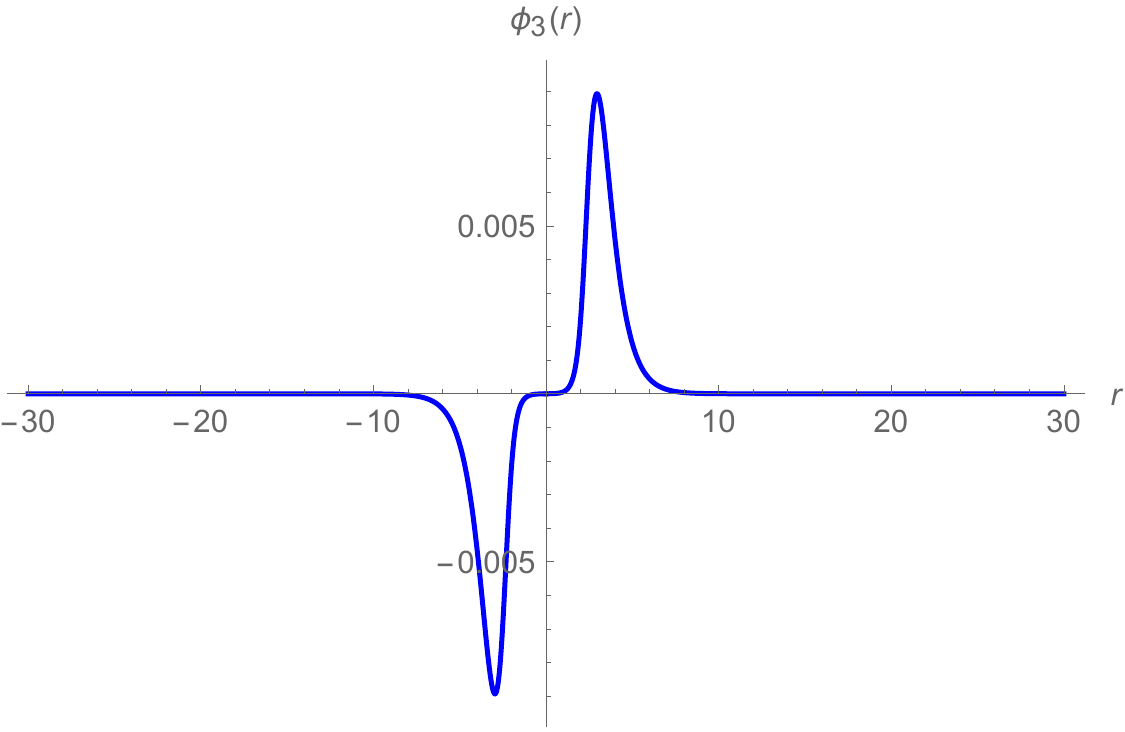}
                 \caption{Solution for $\phi_3(r)$}
         \end{subfigure}\\
         \begin{subfigure}[b]{0.45\textwidth}
                 \includegraphics[width=\textwidth]{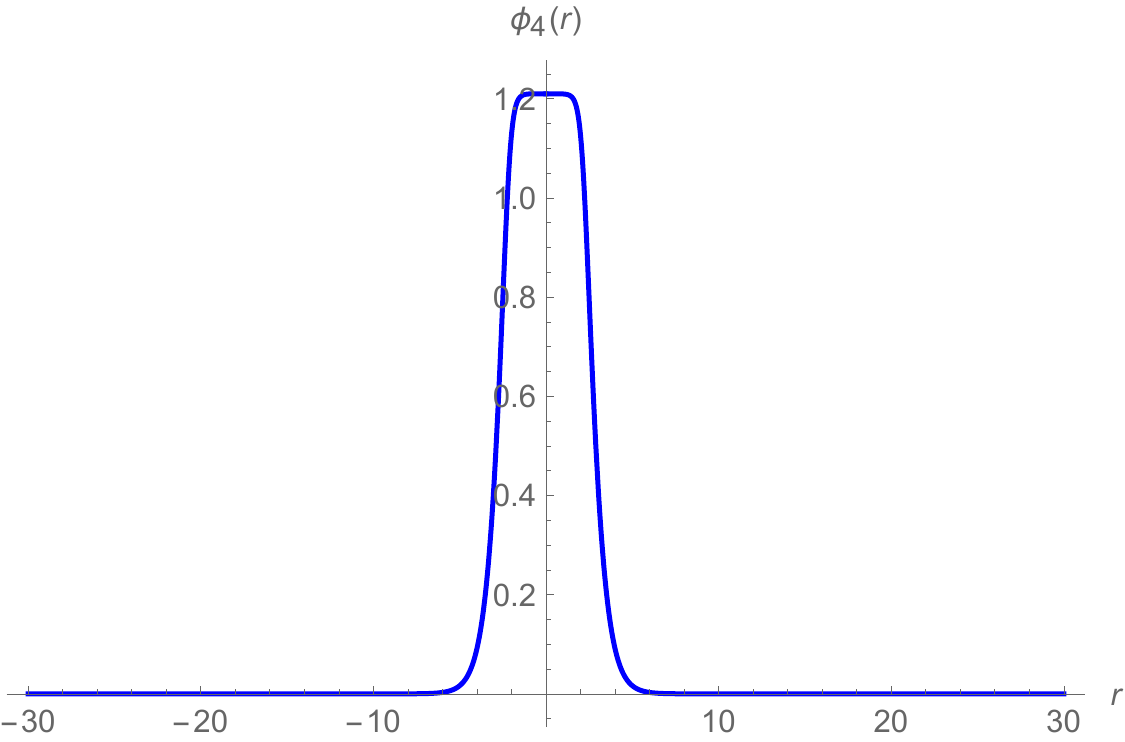}
                 \caption{Solution for $\phi_4(r)$}
         \end{subfigure}
         \begin{subfigure}[b]{0.45\textwidth}
                 \includegraphics[width=\textwidth]{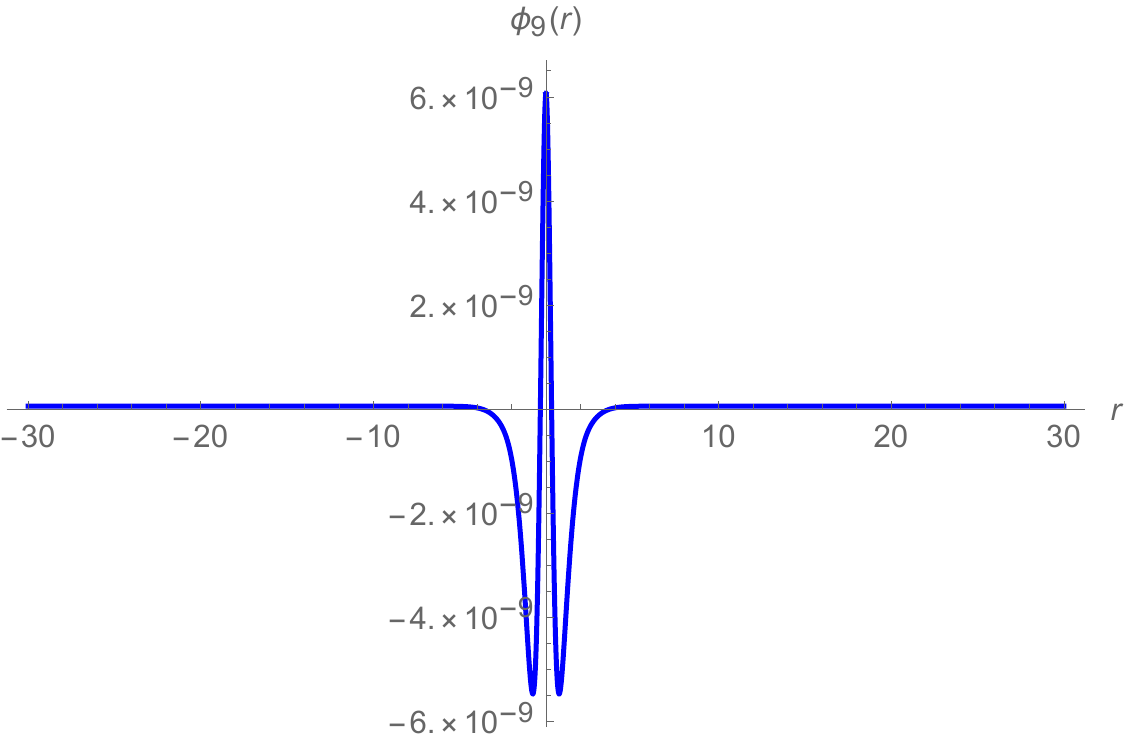}
                 \caption{Solution for $\phi_9(r)$}
         \end{subfigure}\\
         \begin{subfigure}[b]{0.45\textwidth}
                 \includegraphics[width=\textwidth]{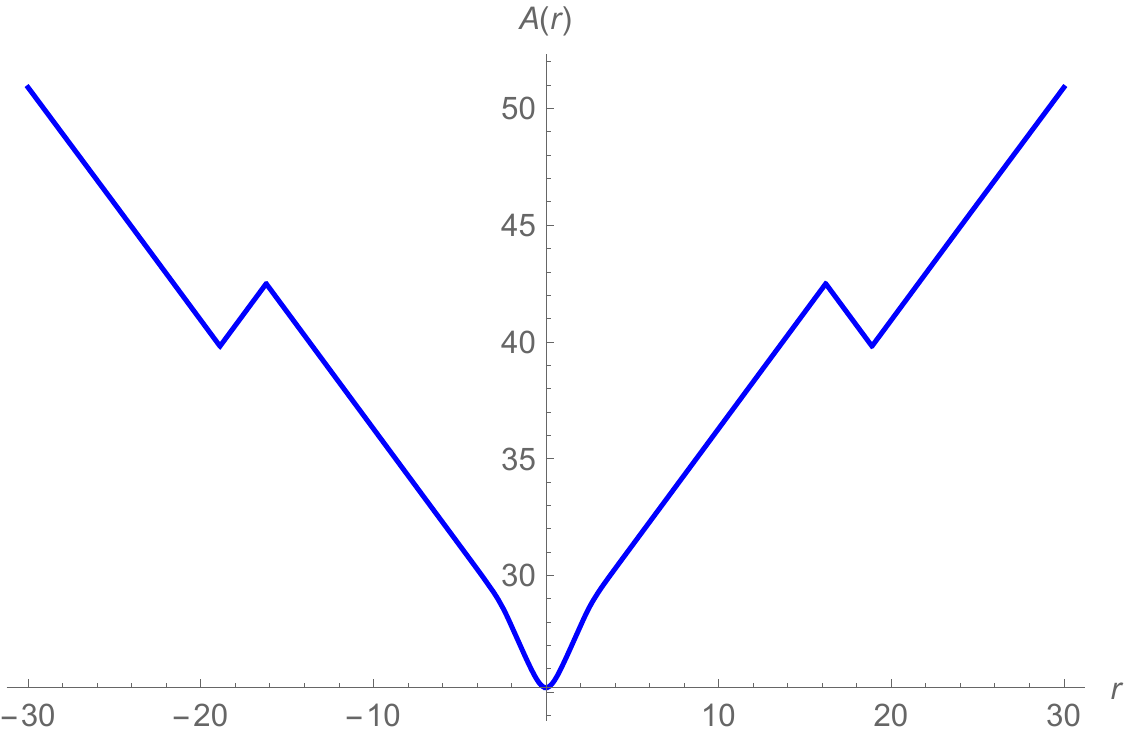}
                 \caption{Solution for $A(r)$}
         \end{subfigure}
          \begin{subfigure}[b]{0.45\textwidth}
                 \includegraphics[width=\textwidth]{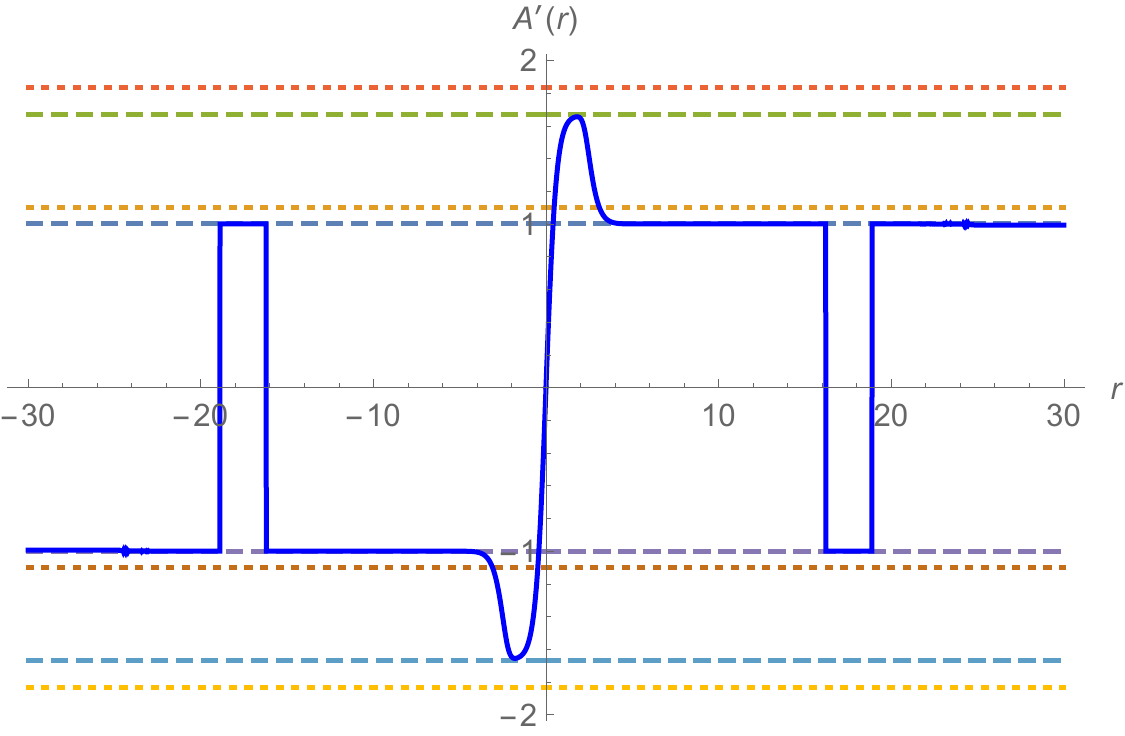}
                 \caption{Solution for $A'(r)$}
         \end{subfigure}
         \caption{An example of multi-Janus interfaces interpolating between $AdS_5$ vacua given by critical points I, II, III and IV. This solution describes five conformal interfaces, four interfaces between critical point I on the right and left of the origin and one interface between critical point III at the origin.}\label{fig5}
 \end{figure}

\begin{figure}
         \centering
               \begin{subfigure}[b]{0.45\textwidth}
                 \includegraphics[width=\textwidth]{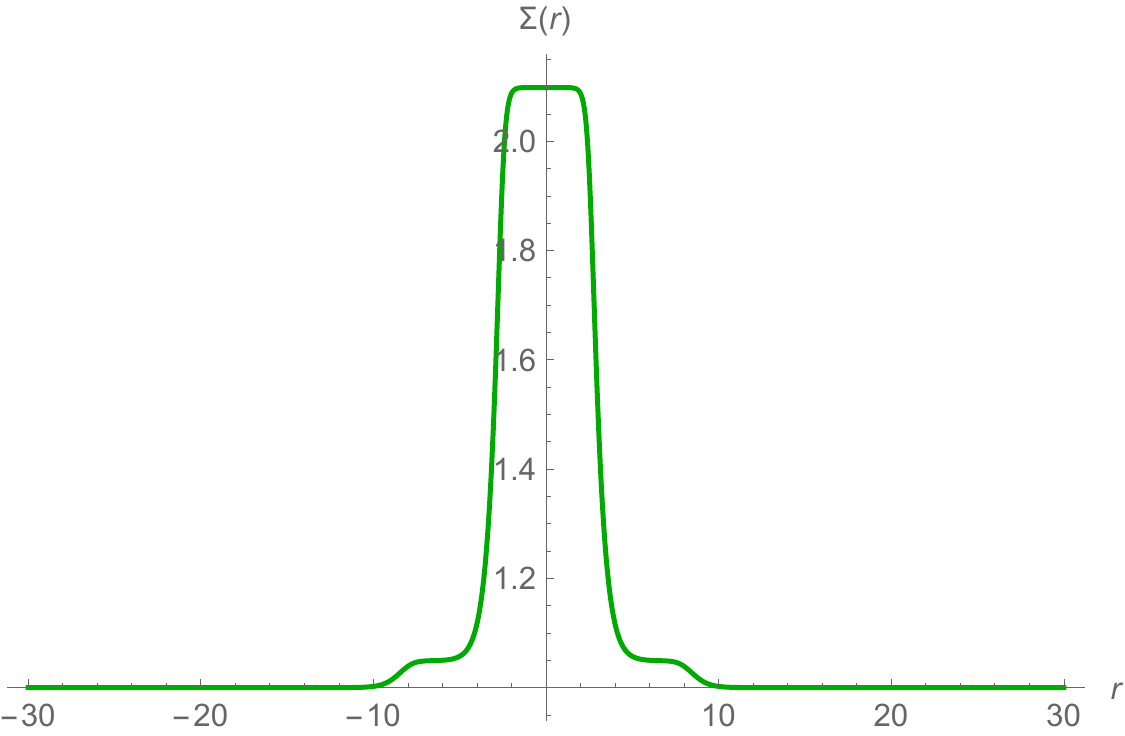}
                 \caption{Solution for $\Sigma(r)$}
         \end{subfigure}
         \begin{subfigure}[b]{0.45\textwidth}
                 \includegraphics[width=\textwidth]{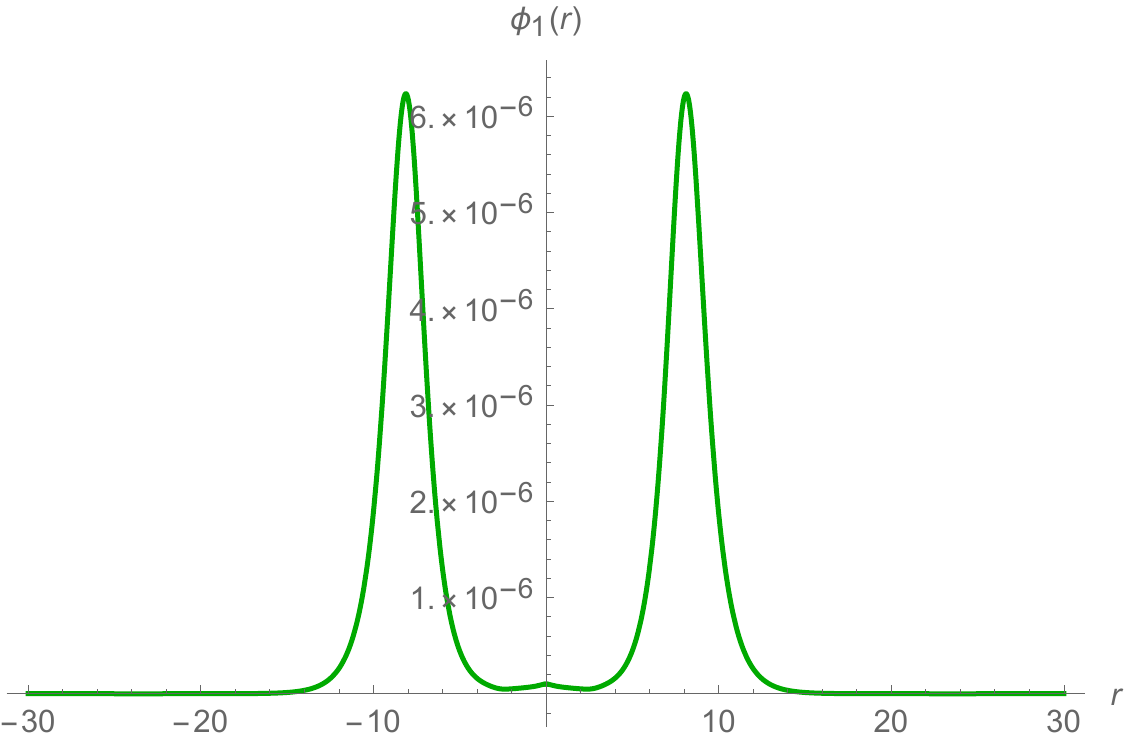}
                 \caption{Solution for $\phi_1(r)$}
         \end{subfigure}\\
         \begin{subfigure}[b]{0.45\textwidth}
                 \includegraphics[width=\textwidth]{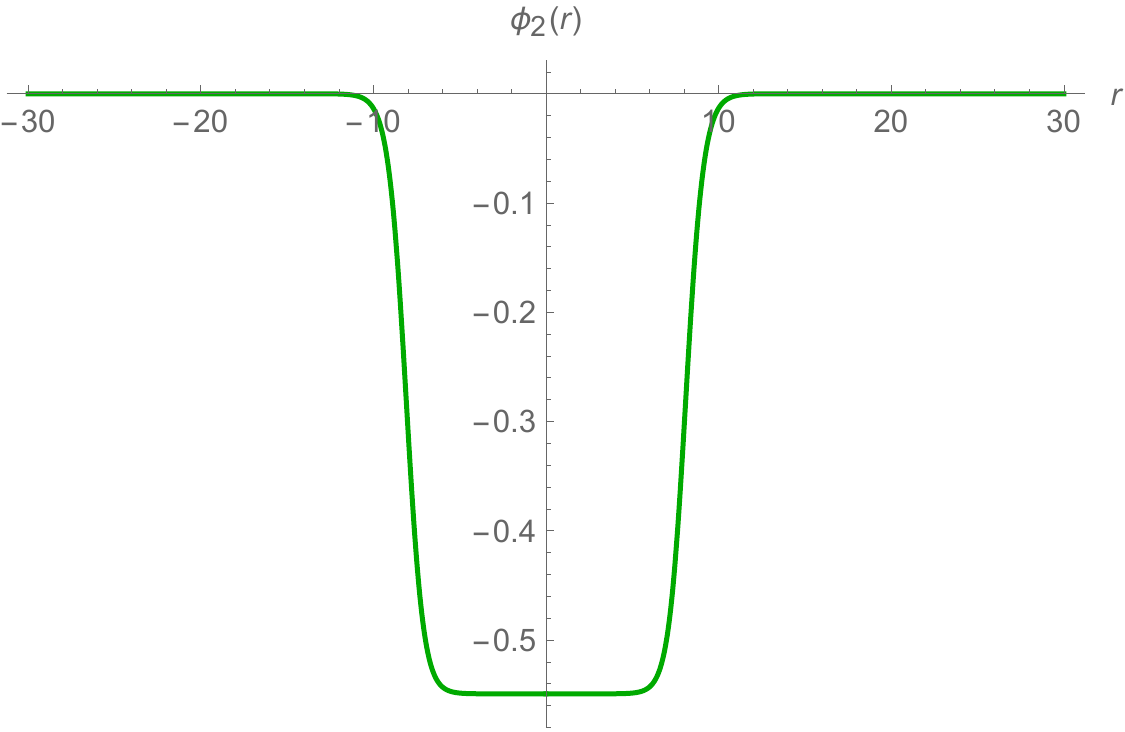}
                 \caption{Solution for $\phi_2(r)$}
         \end{subfigure}
         \begin{subfigure}[b]{0.45\textwidth}
                 \includegraphics[width=\textwidth]{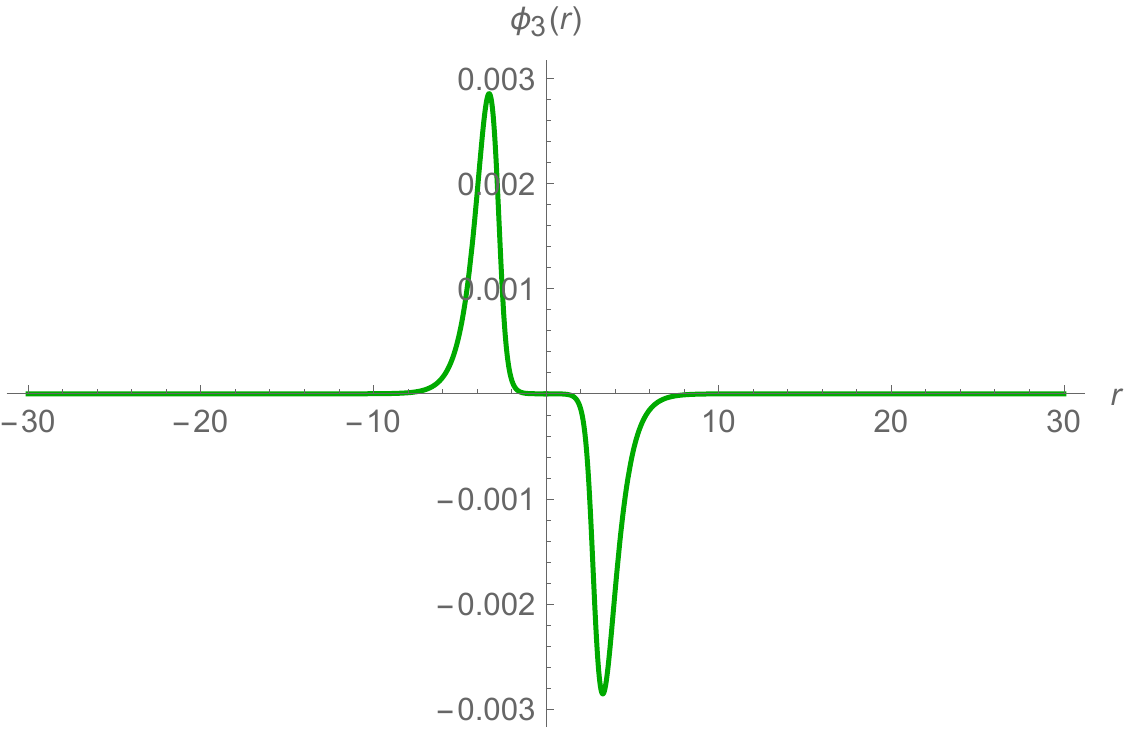}
                 \caption{Solution for $\phi_3(r)$}
         \end{subfigure}\\
         \begin{subfigure}[b]{0.45\textwidth}
                 \includegraphics[width=\textwidth]{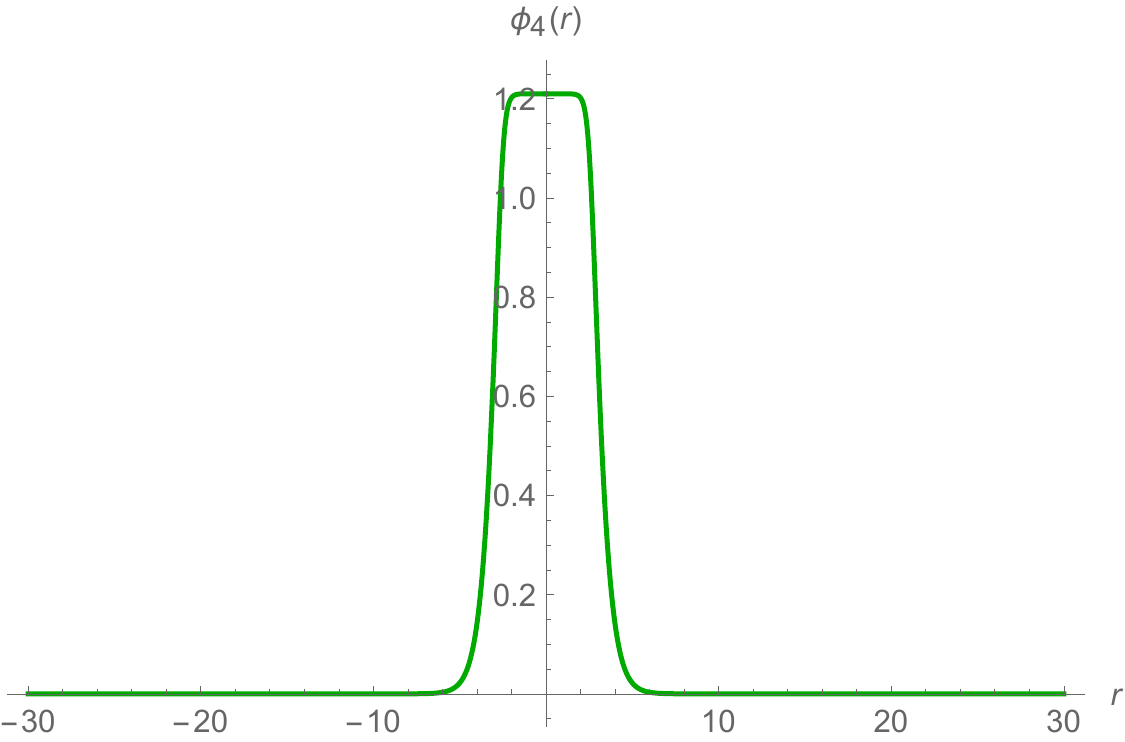}
                 \caption{Solution for $\phi_4(r)$}
         \end{subfigure}
         \begin{subfigure}[b]{0.45\textwidth}
                 \includegraphics[width=\textwidth]{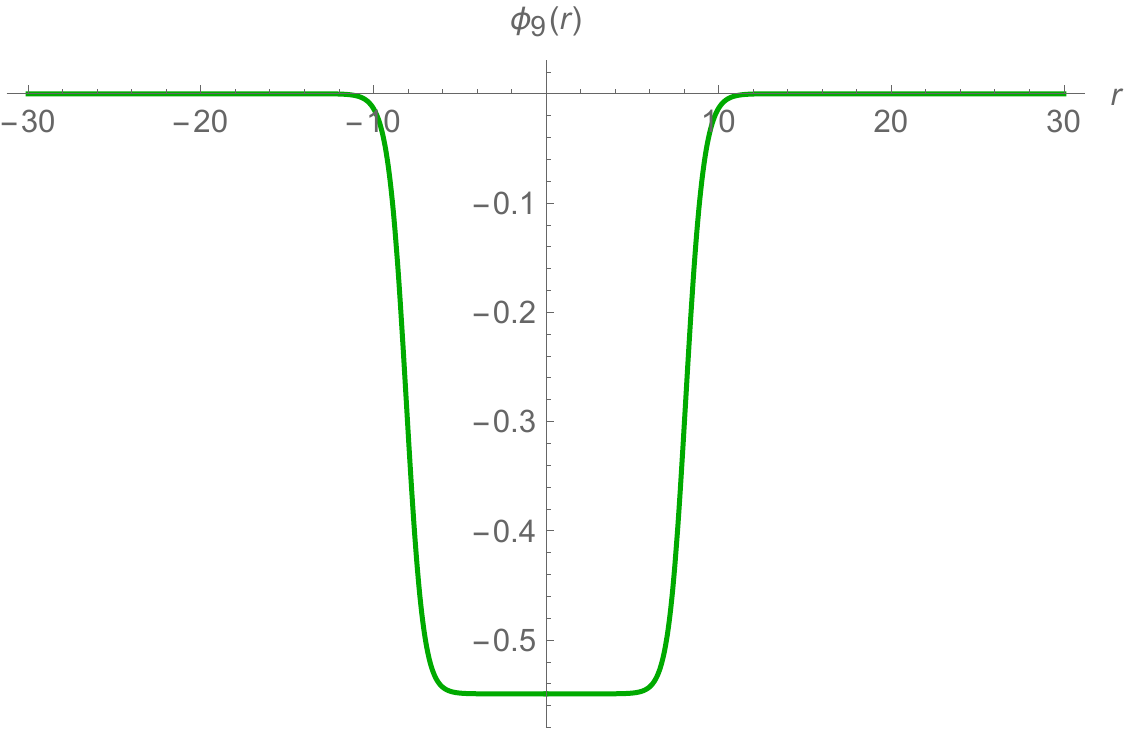}
                 \caption{Solution for $\phi_9(r)$}
         \end{subfigure}\\
         \begin{subfigure}[b]{0.45\textwidth}
                 \includegraphics[width=\textwidth]{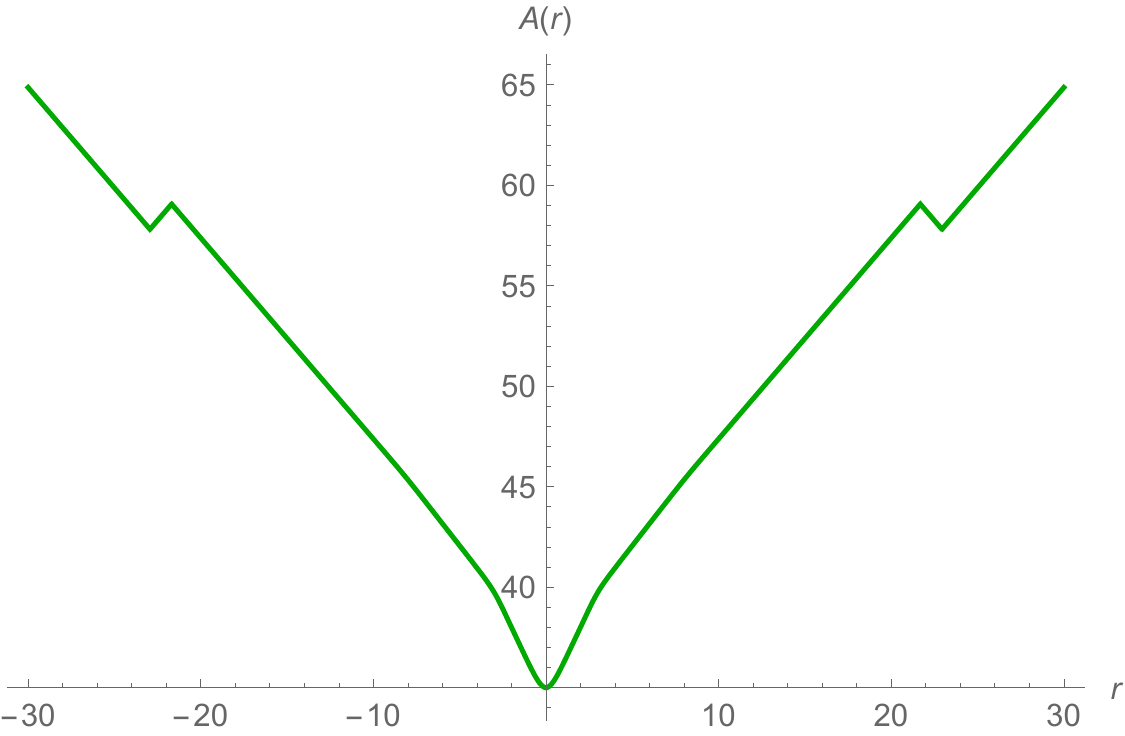}
                 \caption{Solution for $A(r)$}
         \end{subfigure}
          \begin{subfigure}[b]{0.45\textwidth}
                 \includegraphics[width=\textwidth]{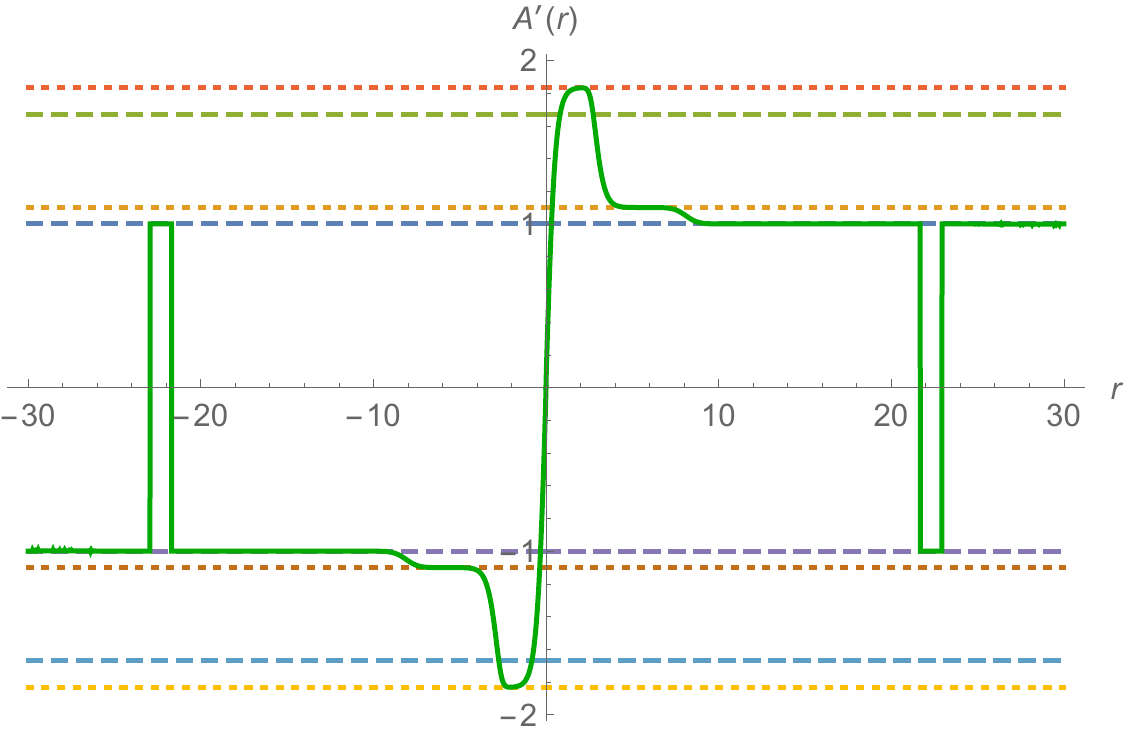}
                 \caption{Solution for $A'(r)$}
         \end{subfigure}
         \caption{An example of multi-Janus interfaces interpolating between $AdS_5$ vacua given by critical points I, II, III and IV. This solution describes five conformal interfaces, four interfaces between critical point I on the right and left of the origin and one interface between critical point IV at the origin.}\label{fig6}
 \end{figure}

\indent By fine-tuning the boundary conditions to make the turning point very close to the $N=2$ $AdS_5$ critical point III and IV, we find solutions describing multi-Janus interfaces. Examples of these solutions are shown in figures \ref{fig4}, \ref{fig5} and \ref{fig6}. In figure \ref{fig4}, the solution describes three conformal interfaces. The two interfaces on the left and right of the origin are conformal interfaces between $N=2$ SCFTs dual to critical point I while the interface at the origin corresponds to a conformal interface between $N=1$ SCFTs dual to critical point IV. There are also solutions describing five conformal interfaces shown in figures \ref{fig5} and \ref{fig6}. In these figures, four interfaces on the right and left of the origin are conformal interfaces between $N=2$ SCFTs dual to critical point I while the interface at the origin is a conformal interface between $N=1$ SCFTs dual to critical point III and IV, respectively. We also note that for solutions in figures \ref{fig4} and \ref{fig6} in which critical point IV appears on both side of the interface at the origin, the $N=1$ SCFT is generated by a holographic RG flow from the $N=2$ SCFTs dual to critical points I and II. On the other hand, the $N=1$ SCFT, dual to critical point III, on both sides of the interface at the origin in figure \ref{fig5} is obtained from a direct RG flow from the $N=2$ SCFT dual to critical point I. This is in agreement with the fact that there is no holographic RG flow from critical point II to critical point III, see \cite{5D_flowII}. 
\\
\indent All of these solutions are expected to holographically describe various conformal interfaces within $N=1$ and $N=2$ SCFTs induced by position-dependent deformations and vacuum expectation values of the dual operators with different scaling dimensions.

%%%%%%%%%%%%%%%%%%%%%%%%%%%%%%%%%%%%%%%%%%%%%%%%%%%%%%%%%%%%%%%%%%%%%%%%%%%%%%%%%%%%%%%%%%%%%%%%%%%%%%%%%%%%%%%%%%%%%%%%%%%%%%%%%%%%%%%%%
\section{Conclusions and discussions}\label{conclusion}
We have studied supersymmetric Janus solutions in five-dimensional $N=4$ gauged supergravity with $SO(2)\times SO(3)\times SO(3)$ and $SO(2)_D\times SO(3)\times SO(3)$ gauge groups. In the first gauge group, there are two $N=4$ supersymmetric $AdS_5$ vacua with $SO(2)\times SO(3)\times SO(3)$ and $SO(2)\times SO(3)_{\textrm{diag}}$ symmetries. We have found Janus solutions interpolating between the same $AdS_5$ vacua on both sides of the interfaces corresponding to conformal interfaces within the dual $N=2$ SCFTs. The existence of solutions interpolating between $N=2$ SCFTs with $SO(2)\times SO(3)\times SO(3)$ symmetry requires a non-vanishing scalar dual to a relevant operator of dimension $3$. The deformation of this operator also leads to source terms for the other two operators of dimension $\Delta=2$ dual to $\Sigma$ and $\phi_2$. The Janus solutions describing a conformal interface within the $N=2$ SCFT with $SO(2)\times SO(3)_{\textrm{diag}}$ symmetry imply the presence of a deformation involving an irrelevant operator of dimension $5$. The conformal interfaces are expected to arise from position-dependent sources and vacuum expectation values of these operators. For a particular choice of boundary conditions, we have also given an example of RG-flow interfaces interpolating between $SO(2)\times SO(3)\times SO(3)$ symmetric $AdS_5$ vacuum on one side of the interface and $SO(2)\times SO(3)_{\textrm{diag}}$ symmetric vacuum on the other side. This solution should correspond to an RG-flow interface between different $N=2$ SCFTs on the two sides.
\\
\indent For $SO(2)_D\times SO(3)\times SO(3)$ gauge group, the solution space has a very rich structure with many possible solutions to be considered. We have only given some representative examples of interesting solutions. The first class of solutions describes conformal interfaces within $N=2$ and $N=1$ SCFTs dual to $N=4$ and $N=2$ $AdS_5$ vacua given by critical points II, III or IV. These three conformal phases on both sides arise from holographic RG flows from $N=4$ critical point I on each side of the interfaces. We have also given some examples of solutions describing multiple conformal interfaces between the four $AdS_5$ critical points. We have explicitly given solutions corresponding to three and five conformal interfaces between critical point I and critical points III and IV. In these solutions, there are many scalar fields turned on, and these scalars are dual to relevant and irrelevant operators in the dual SCFTs. We also expect these solutions to describe conformal interfaces within $N=1$ and $N=2$ SCFTs arising from position-dependent sources and vacuum expectation values of these operators.   
\\
\indent As previously mentioned, unlike the solutions found in \cite{5D_N4_Janus} and \cite{Janus_5D_ISO3} in which the embedding in M-theory is known, the Janus solutions obtained from the $N=4$ gauged supergravities considered here have presently no known higher dimensional origins. However, these gauged supergravities provide an interesting framework to study Janus solutions in the presence of many $AdS_5$ vacua. We hopefully expect that the solutions given here would still be useful in the holographic study of conformal interfaces within $N=1$ and $N=2$ SCFTs in four dimensions and could also lead to some insights into different aspects of interfaced SCFTs. Finding an embedding of the two $N=4$ gauged supergravities in ten or eleven dimensions is clearly a very interesting task. This could be used to uplift solutions found in this paper to string/M-theory and eventually leads to interpretations of the corresponding conformal interfaces in terms of various brane configurations. Identifying possible dual $N=1$ and $N=2$ SCFTs dual to all the four supersymmetric $AdS_5$ vacua and deformations leading to the aforementioned interfaces is also of particular interest. We hope to shed some light on these issues in future works.
%%%%%%%%%%%%%%%%%%%%%%%%%%%%%%%%%%%%%%%%%%%%%%%%%%%%%%%%%%%%%%%%%%%%%%%%
%\\
%\textbf{Data Availability Statement:} No data associated in the manuscript


\begin{thebibliography}{99}
\bibitem{defect_review} N. Andrei et al. , ``Boundary and Defect CFT: Open Problems and Applications'', J. Phys. A: Math. Theor. 53 453002, arXiv: 1810.05697.
\bibitem{maldacena} J. M. Maldacena, ``The large $N$ limit of superconformal field theories and supergravity'', Adv. Theor. Math. Phys. \textbf{2} (1998) 231-252, arXiv: hep-th/9711200.
\bibitem{Gubser_AdS_CFT} S. S. Gubser, I. R. Klebanov and A. M. Polyakov, ``Gauge Theory Correlators from Non-Critical String Theory'', Phys. Lett. \textbf{B428} (1998) 105-114, arXiv: hep-th/9802.109.
\bibitem{Witten_AdS_CFT} E. Witten, ``Anti De Sitter Space and holography'', Adv. Theor. Math. Phys. \textbf{2} (1998) 253-291, arXiv: 9802150.
\bibitem{Bak_Janus} D. Bak, M. Gutperle and S. Hirano, ``A Dilatonic Deformation of $AdS_5$ and its Field Theory Dual'', JHEP 05 (2003) \textbf{072}, arXiv: hep-th/0304129.
\bibitem{5D_Janus_CK} A. Clark and A. Karch, ``Super Janus'', JHEP 10 (2005) \textbf{094}, arXiv: hep-th/0506265.
\bibitem{Freedman_Janus} A. B. Clark, D. Z. Freedman, A. Karch and M. Schnabl, ``Dual of the Janus solution: An interface conformal field theory'', Phys. Rev. \textbf{D71} (2005)
066003, arXiv: hep-th/0407073.
\bibitem{DHoker_Janus} E. D' Hoker, J. Estes and M. Gutperle, ``Interface Yang-Mills, supersymmetry, and Janus'', Nucl. Phys. \textbf{B753} (2006) 16, arXiv: hep-th/0603013.
\bibitem{Witten_Janus} D. Gaiotto and E. Witten, ``Janus Configurations, Chern-Simons Couplings, And The thetaAngle in N=4 Super Yang-Mills Theory'', JHEP 1006 (2010) 097, arXiv: 0804.2907.
\bibitem{Freedman_Holographic_dCFT} O. DeWolfe, D. Z. Freedman and H. Ooguri, ``Holography and Defect Conformal Field Theories'', Phys. Rev. \textbf{D66} (2002) 025009, arXiv: hep-th/0111135.
\bibitem{5D_Janus_Suh} M. W. Suh, ``Supersymmetric Janus solutions in five and ten dimensions'', JHEP 09 (2011) \textbf{064}, arXiv: 1107.2796.
\bibitem{5D_Janus_DHoker1} E. D'Hoker, J. Estes and M. Gutperle, ``Ten-dimensional supersymmetric Janus solutions'', Nucl. Phys. \textbf{B757} (2006) 79, arXiv: hep-th/0603012.
\bibitem{5D_Janus_DHoker2} E. D'Hoker, J. Estes and M. Gutperle, ``Exact half-BPS Type IIB interface solutions. I. Local solution and supersymmetric Janus'', JHEP 06 (2007) \textbf{021}, arXiv: 0705.0022.
\bibitem{5D_Janus_DHoker3} E. D'Hoker, J. Estes and M. Gutperle, ``Exact half-BPS Type IIB interface solutions. II: Flux solutions and multi-Janus'', JHEP 06 (2007) \textbf{022}, arXiv: 0705.0024.
\bibitem{Bobev_5D_Janus1} N. Bobev, F. F. Gautason, K. Pilch, M. Suh, J. van Muiden, ``Janus and J-fold Solutions from Sasaki-Einstein Manifolds'', Phys. Rev. \textbf{D100} (2019) 081901, arXiv: 1907.11132.
\bibitem{Bobev_5D_Janus2} N. Bobev, F. F. Gautason, K. Pilch, M. Suh, J. van Muiden, ``Holographic Interfaces in $N=4$ SYM: Janus and J-folds'', JHEP 05 (2020) \textbf{134}, arXiv: 2003.09154.
\bibitem{RG-interface_Gauntlett} I. Arav, K. C. Matthew Cheung, J. P. Gauntlett, M. M. Roberts and C. Rosen, ``Superconformal RG interfaces in holography'', JHEP 11 (2020) \textbf{168}, arXiv: 2007.07891.
\bibitem{6D_Janus} M. Gutperle, J. Kaidi and H. Raj, ``Janus solutions in six-dimensional gauged supergravity'', JHEP 12 (2017) \textbf{018}, arXiv: 1709.09204.
\bibitem{6D_Janus_RG} P. Karndumri, ``Janus and RG-flow interfaces from matter-coupled $F(4)$ gauged supergravity'', Phys. Rev. \textbf{D111} (2025) 026013, arXiv: 2405.17169.
\bibitem{warner_Janus} N. Bobev, K. Pilch and N. P. Warner, ``Supersymmetric Janus Solutions in Four Dimensions'', JHEP 1406 (2014) 058, arXiv: 1311.4883.
\bibitem{ISO3_flow} P. Karndumri, ``Holographic RG flows and Janus interfaces from ISO(3)$\times$U(1) $F(4)$ gauged supergravity'', Phys. Rev. \textbf{D111} (2025) 026022, arXiv: 2409.20151.
\bibitem{N3_Janus} P. Karndumri, ``Supersymmetric Janus solutions in four-dimensional $N=3$ gauged supergravity'', Phys. Rev. \textbf{D93} (2016) 125012, arXiv: 1604.06007.
\bibitem{N3_JanusII} P. Karndumri, ``New Janus interfaces from four-dimensional $N=3$ gauged supergravity'', Eur. Ohys. J. \textbf{C84} (2024) 1059, arXiv: 2408.00424.
\bibitem{N8_omega_Janus} P. Karndumri and C. Maneerat, ``Supersymmetric Janus solutions in $\omega$-deformed $N=8$ gauged supergravity'', Eur. Phys. J. \textbf{C81} (2021) 801, arXiv: 2012.15763.
\bibitem{N4_Janus} P. Karndumri, ``Holographic RG flows and Janus solutions from matter-coupled $N=4$ gauged supergravity'', Eur. Phys. J. \textbf{C81} (2021) 520, arXiv: 2102.05532.
\bibitem{ISO7_Janus} P. Karndumri and C. Maneerat, ``Janus solutions from dyonic $ISO(7)$ maximal gauged supergravity'', JHEP 10 (2021) \textbf{117}, arXiv: 2108.13398.
\bibitem{N4_omega_Janus} T. Assawasowan and P. Karndumri, ``New supersymmetric Janus solutions from $N=4$ gauged supergravity'', Phys. Rev. \textbf{D105} (2022) 106004, arXiv: 2203.03413.
\bibitem{Guarino_Janus_M} A. Anabalon, M. Chamorro-Burgos and A. Guarino, ``Janus and Hades in M-theory'', JHEP 11 (2022) \textbf{150}, arXiv: 2207.09287.
\bibitem{Guarino_S_fold_Janus} A. Guarino and M. Suh, ``S-folds and $AdS_3$ flows from the D3-brane'', JHEP 11 (2022) \textbf{134}, arXiv: 2207.14015.
\bibitem{4D_Janus_from_11D} E. D'Hoker, J. Estes, M. Gutperle and D. Krym, ``Janus solutions in M-theory'', JHEP 06 (2009) \textbf{018}, arXiv: 0904.3313.
\bibitem{tri-sasakian-flow} P. Karndumri, ``Supersymmetric deformations of 3D SCFTs from tri-sasakian truncation'', Eur. Phys. J. C (2017) \textbf{77}, 130, arXiv: 1610.07983.
\bibitem{orbifold_flow} P. Karndumri and K. Upathambhakul, ``Supersymmetric RG flows and Janus from type II orbifold compactification'', Eur. Phys. J. C (2017) \textbf{77}, 455, arXiv: 1704.00538.
\bibitem{Minwoo_4DN8_Janus} M. Suh, ``Supersymmetric Janus solutions of dyonic $ISO(7)$-gauged $N=8$ supergravity'', JHEP 04 (2018) \textbf{109}, arXiv: 1803.00041.
\bibitem{Kim_Janus} N. Kim and S. J. Kim, ``Re-visiting Supersymmetric Janus Solutions: A Perturbative Construction'', Chin. Phys. \textbf{C44} (2020) 7, 073104, arXiv: 2001.06789.
\bibitem{N5_flow} P. Karndumri and C. Maneerat, ``Supersymmetric solutions from $N=5$ gauged supergravity'', Phys. Rev. \textbf{D101} (2020) 126015, arXiv: 2003.05889.
\bibitem{N6_flow} P. Karndumri and J. Seeyangnok, ``Supersymmetric solutions from $N=6$ gauged supergravity'', Phys. Rev. \textbf{D103} (2021) 066023, arXiv: 2012.10978.
\bibitem{3D_Janus_de_Boer} C. Bachas, J. de Boer, R. Dijkgraaf, and H. Ooguri, ``Permeable conformal walls and holography'', JHEP 06 (2002) \textbf{027}, arXiv:hep-th/0111210.
\bibitem{3D_Janus_Bachas} C. Bachas and M. Petropoulos, ``Anti-de-Sitter D-branes'', JHEP 02 (2001) \textbf{025}, arXiv:hep-th/0012234.
\bibitem{3D_Janus_Bak} D. Bak, M. Gutperle and S. Hirano, ``Three dimensional Janus and time-dependent black holes'', JHEP 02 (2007) \textbf{068}, arXiv: hep-th/0701108. \bibitem{half_BPS_AdS3_S3_ICFT} M. Chiodaroli, M. Gutperle and D. Krym, ``Half-BPS Solutions locally asymptotic to $AdS_3\times S^3$
and interface conformal field theories'', JHEP 02 (2010) \textbf{066}, arXiv: 0910.0466.
\bibitem{exact_half_BPS_string} M. Chiodaroli, E. D'Hoker, Y, Guo and M. Gutperle, ``Exact half-BPS string-junction solutions in six-dimensional supergravity'', JHEP 12 (2011) \textbf{086}, arXiv: 1107.1722.
\bibitem{multi_face_Janus} D. Bak and H. Min, ``Multi-faced Black Janus and Entanglement'', JHEP 03 (2014) \textbf{046}, arXiv: 1311.5259.
\bibitem{3D_Janus} K. Chen and M. Gutperle ``Janus solutions in three-dimensional N=8 gauged supergravity'', JHEP 05 (2021) \textbf{008}, arXiv: 2011.10154.
\bibitem{3D_Janus2} K. Chen, M. Gutperle and C. Hultgreen-Mena, ``Janus and RG-flow interfaces in three-dimensional gauged supergravity'', JHEP 03 (2022) \textbf{057}, arXiv: 2111.01839.
\bibitem{3D_Janus3} M. Gutperle and C. Hultgreen-Mena, ``Janus and RG interfaces in three-dimensional gauged supergravity II: General $\alpha$'', JHEP 08 (2022) \textbf{126}, arXiv: 2205.10398.
\bibitem{3D_Janus4} M. Gutperle and C. Hultgreen-Mena, ``Janus and RG-interfaces in minimal 3d gauged supergravity'', arXiv: 2412.16749.
\bibitem{5D_N4_curved_DW} M. Zagermann, ``$N=4$ ``Fake'' Supergravity'', Phys.Rev. \textbf{D71} (2005) 125007, arXiv: hep-th/0412081.
\bibitem{Romans_5DN4} L. J. Romans, ``Gauged $N=4$ supergravities in five dimensions and their magnetovac backgrounds'', Nucl. Phys. \textbf{B267} (1986) 433.
\bibitem{5D_N4_Janus} P. Karndumri, ``Janus and RG-flow interfaces from $5D$ $N=4$ gauged supergravity'', Phys. Rev. \textbf{D111} (2025) 12, 126020, arXiv: 2503.06704.
\bibitem{Janus_5D_ISO3} P. Karndumri, ``Holographic solutions from 5D $SO(2)\times ISO(3)$ $N=4$ gauged supergravity'', arXiv: 2510.08258.
\bibitem{5D_flowII} P. Karndumri, ``$AdS_5$ vacua and holographic RG flows from 5D $N=4$ gauged supergravity'', Eur. Phys. J. \textbf{C83} (2023) 164, arXiv: 2209.05270.
\bibitem{5D_N4_flow_Davide} N. Bobev, D. Cassani and H. Triendl, ``Holographic RG Flows for Four-dimensional $N=2$ SCFTs'', JHEP 06(2018)\textbf{086}, arXiv: 1804.03276.
\bibitem{5D_N4_flow} H. L. Dao and P. Karndumri, ``Holographic RG flows and $AdS_5$ black strings from 5D half-maximal gauged supergravity'', Eur. Phys. J. \textbf{C79} (2019) 137, arXiv: 1811.01608.
\bibitem{N4_gauged_SUGRA} J. Schon and M. Weidner, ``Gauged $N=4$ supergravities'', JHEP 05 (2006) \textbf{034}, arXiv: hep-th/0602024.
\bibitem{5D_N4_Dallagata} G. Dall’Agata, C. Herrmann, and M. Zagermann, ``General matter coupled $N=4$ gauged supergravity in five-dimensions'', Nucl. Phys. \textbf{B612} (2001) 123–150, arXiv: hep-th/0103106.
\bibitem{AdS5_N4_Jan} J. Louis, H. Triendl and M. Zagermann, ``$N = 4$ supersymmetric $AdS_5$ vacua and their moduli spaces'', JHEP 10 (2015) \textbf{083}, arXiv:1507.01623.
\bibitem{5D_N4_black_stringII} P. Karndumri, ``New supersymmetric $AdS_5$ black strings from 5D $N=4$ gauged supergravity'', Eur. Phys. J. \textbf{C83} (2023) 432, arXiv: 2211.07456.
\end{thebibliography}
\end{document}